\newif\ifhide

\newif\ifarxiv
\arxivtrue 

\ifarxiv
\documentclass[a4paper,twocolumn,10pt]{article}
\else
\documentclass[a4paper,twocolumn,10pt]{quantumarticle}
\fi

\pdfoutput=1

\usepackage[utf8]{inputenc}
\usepackage[english]{babel}
\usepackage[T1]{fontenc}

\usepackage{amsmath,amsfonts}
\usepackage{hyperref}
\usepackage{graphicx}
\usepackage{braket}
\usepackage{color}
\usepackage{url}

\renewcommand{\vec}[1]{\boldsymbol{#1}}
\newcommand{\mtx}[1]{\boldsymbol{\mathsf{#1}}}
\newcommand{\unit}[1]{\hat{\boldsymbol{#1}}}

\ifhide

\else

\fi

\begin{document}

\title{The Virtual Quantum Optics Laboratory}

\ifarxiv
\author{Brian R. La Cour\thanks{blacour@arlut.utexas.edu}, Maria Maynard, Parth Shroff, Gabriel Ko, and Evan Ellis}
\date{Applied Research Laboratories, The University of Texas at Austin\\ P.O. Box 8029, Austin, TX 78713-8029}
\else
\author{Brian R. La Cour}
\email{blacour@arlut.utexas.edu}
\author{Maria Maynard}
\author{Parth Shroff}
\author{Gabriel Ko}
\author{Evan Ellis}
\affiliation{Applied Research Laboratories, The University of Texas at Austin, P.O. Box 8029, Austin, TX 78713-8029}
\fi

\maketitle

\ifarxiv
{\bfseries
\else
\begin{abstract}
\fi
We present a web-based software tool, the Virtual Quantum Optics Laboratory (VQOL), that may be used for designing and executing realistic simulations of quantum optics experiments.  A graphical user interface allows one to rapidly build and configure a variety of different optical experiments, while the runtime environment provides unique capabilities for visualization and analysis.  All standard linear optical components are available as well as sources of thermal, coherent, and entangled Gaussian states.  A unique aspect of VQOL is the introduction of non-Gaussian measurements using detectors modeled as deterministic devices that ``click'' when the amplitude of the light falls above a given threshold.  We describe the underlying theoretical models and provide several illustrative examples.  We find that VQOL provides a a faithful representation of many experimental quantum optics phenomena and may serve as both a useful instructional tool for students as well as a valuable research tool for practitioners.
\ifarxiv
}
\else
\end{abstract}
\fi


\section{Introduction}

The subject of quantum mechanics is notoriously difficult to understand and teach \cite{Muller2002,Scholz2020}.  In part, this stems from the complex mathematics often required to describe quantum systems, but the many conceptual difficulties of the subject pose what is arguably an even greater challenge.  Ideally, one would go to the laboratory and discover quantum phenomena first hand, but this is not always possible.  A potential remedy is to use high-fidelity simulation tools with which one may explore the quantum world.  Such an approach can indeed be of great practical and pedagogical benefit, but simulators must strike the right balance between simplicity and realism while giving users ample freedom to explore.  Simulations that are overly idealized or abstract may rob one of the deep insights that may be gained when working with real experimental observations.

The choice of physical systems is also of great importance.  We believe the physics of \emph{light} provides an excellent approach with which to introduce and study quantum concepts.  In using light, one can take advantage of the many classical concepts that translate directly to their quantum counterparts.  The polarization of classical light, for example, provides a direct analogy (\textit{sans} normalization) to a photon polarization state, which is itself one particular representation of a quantum bit or \emph{qubit}.  Furthermore, classical linear optical devices, such as beam splitters, phase shifters, and wave plates, translate directly to general unitary transformations.  Even nonlinear optical phenomena, such as parametric down conversion, can be introduced in classical terms.

The one important point of departure in these analogies lies in measurement.  Classical Gaussian measurements of light are typically in the form of time-varying intensities, but quantum light may also be measured as non-Gaussian detection events or ``clicks'' of, say, an avalanche photodiode.  Although quantum phenomena can certainly be found in intensity measures, as is done in heterodyne and homodyne detection, discrete detection events are essential for exhibiting the particle-like behavior of light. Thus, the difference between these two types of measurements can be used to encapsulate the differences between classical and quantum light  \cite{Bartlett2012}.

These considerations have led to our development of a novel quantum optics simulation tool we call the Virtual Quantum Optics Laboratory (VQOL).  VQOL is a unique marriage of classical and quantum optics based on two simple modeling principles:  First, we formally treat the vacuum modes of quantum optics as real (rather than virtual) random electromagnetic radiation corresponding to the zero-point field.  Second, detectors are modeled as deterministic devices that ``click'' when the intensity of incident light, including contributions from the vacuum, falls above a predefined threshold.  These two ingredients have been shown to be capable of reproducing many of the phenomena one observes in real quantum optics experiments \cite{LaCour&Williamson2020,LaCour&Yudichak2021a,LaCour&Yudichak2021b}.

There are, of course, a plethora of existing resources for simulating quantum optics, ranging from pedagogical games to sophisticated research tools \cite{Qubit4Matlab2008,QuTiP2013}.  We shall highlight a few that we believe are particularly relevant and contrast them with VQOL.

The first is Strawberry Fields by Xanadu, an open-source programming architecture for simulating continuous-variable photonic quantum computers \cite{StrawberryFields2019}.  Simulator backends allow users to construct quantum states through a sequence of unitary gates and apply both Gaussian (e.g., homodyne) and non-Gaussian (e.g., photon counting) measurements.  VQOL is similar to Strawberry Fields in using only Gaussian states, either thermal, coherent, or entangled, as light sources.  It differs in its focus on experimental design, vice abstract gate operations, and in treating detectors as nonideal devices.

QuantumLab by the University of Erlangen-Nuremberg is a web-based resource that uses Adobe Flash-Player to illustrate various real-world quantum optics experiments \cite{QuantumLabUEN}.  The experiments are all performed in a real laboratory and conducted under various configurations and parameter settings.  Users can select from among these options and visualize the real (not simulated) data that were actually observed.  QuantumLab is an excellent resource for demonstrating real-world quantum optics phenomena, but it lacks the flexibility provided by VQOL to allow users to design and conduct their own experiments.

Finally, Quantum Lab by Quantum Flytrap is a colorful web-based tool that allows users to design and run their own custom optics experiments \cite{QuantumLEGO,QuantumGame}.  A virtual optics table with a drag-and-drop pallet of components allows users to design and run a variety of optical experiments.  Lasers are modeled as simple, on-demand single-photon sources, while non-linear crystals provide a source of idealized entangled photon pairs.  Detectors are treated as having perfect efficiency and no dark counts, with random detection events that strictly follow the Born rule.  Quantum Lab is similar to VQOL in providing an open-ended user interface to design and conduct experiments.  VQOL, however, differs in its use of more realistic light sources and non-ideal detectors.  The advantage of our approach, we believe, lies in its ability to provide a more faithful representation of what is actually observed in the laboratory.

With these preliminary considerations, we summarize the organization of the paper as follows.  In Section \ref{sec:software} we provide a brief description of the VQOL software from a user interface perspective.  Sections \ref{sec:model} and \ref{sec:components} discusses the modeling assumptions and component descriptions underlying VQOL, with particular emphasis on departures from standard classical optics necessitated by our explicit use of vacuum modes.  Section \ref{sec:examples} gives several examples of experiments that may be performed with VQOL.  We discuss the pedagogical and research applications in Section \ref{sec:discussion} and summarize our conclusions in Section \ref{sec:conclusion}.


\section{Software Description}
\label{sec:software}

\begin{figure}[ht]
\centering
\scalebox{1.0}{\includegraphics[width=\columnwidth]{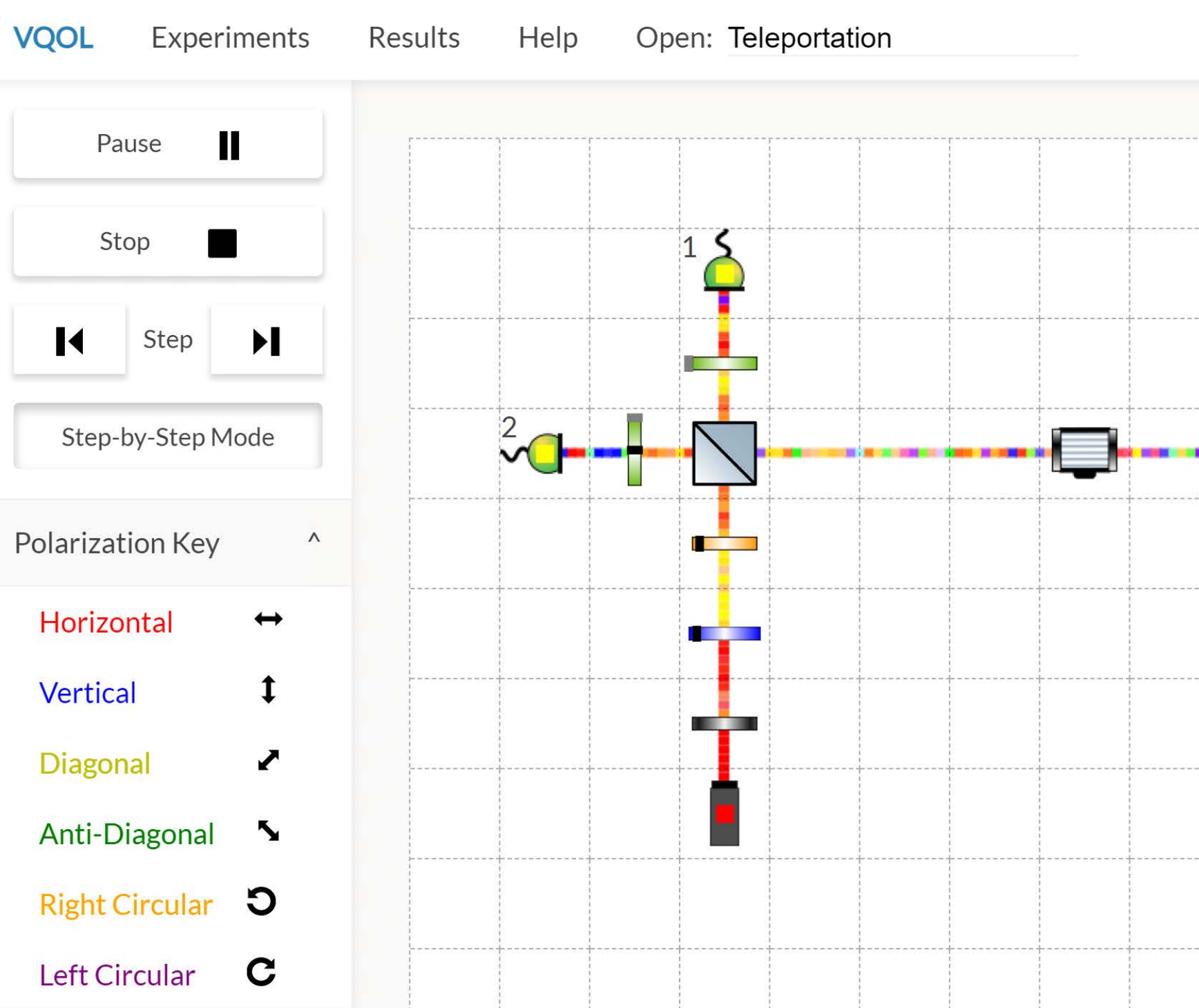}}
\caption{(Color online) VQOL quantum teleportation experiment.  Colors indicate different polarization states, as indicated in the Polarization Key.}
\label{fig:VQOL_example}
\end{figure}

This paper describes version 1.0 of the Virtual Quantum Optics Laboratory.  VQOL may be used online via the universal resource locator (URL) \texttt{https://www.vqol.org} or run locally from a downloadable client.  Using either a text editor or graphical interface, the user may define an experiment by placing (and optionally  orienting, and configuring) various component devices on a gridded virtual optics table.  Figure \ref{fig:VQOL_example} illustrates the interface and shows an implementation of a quantum teleportation experiment.  Sample experiments are provided, but the user is free to design their own.  

\begin{figure}[ht]
\scalebox{0.95}{\includegraphics[width=\columnwidth]{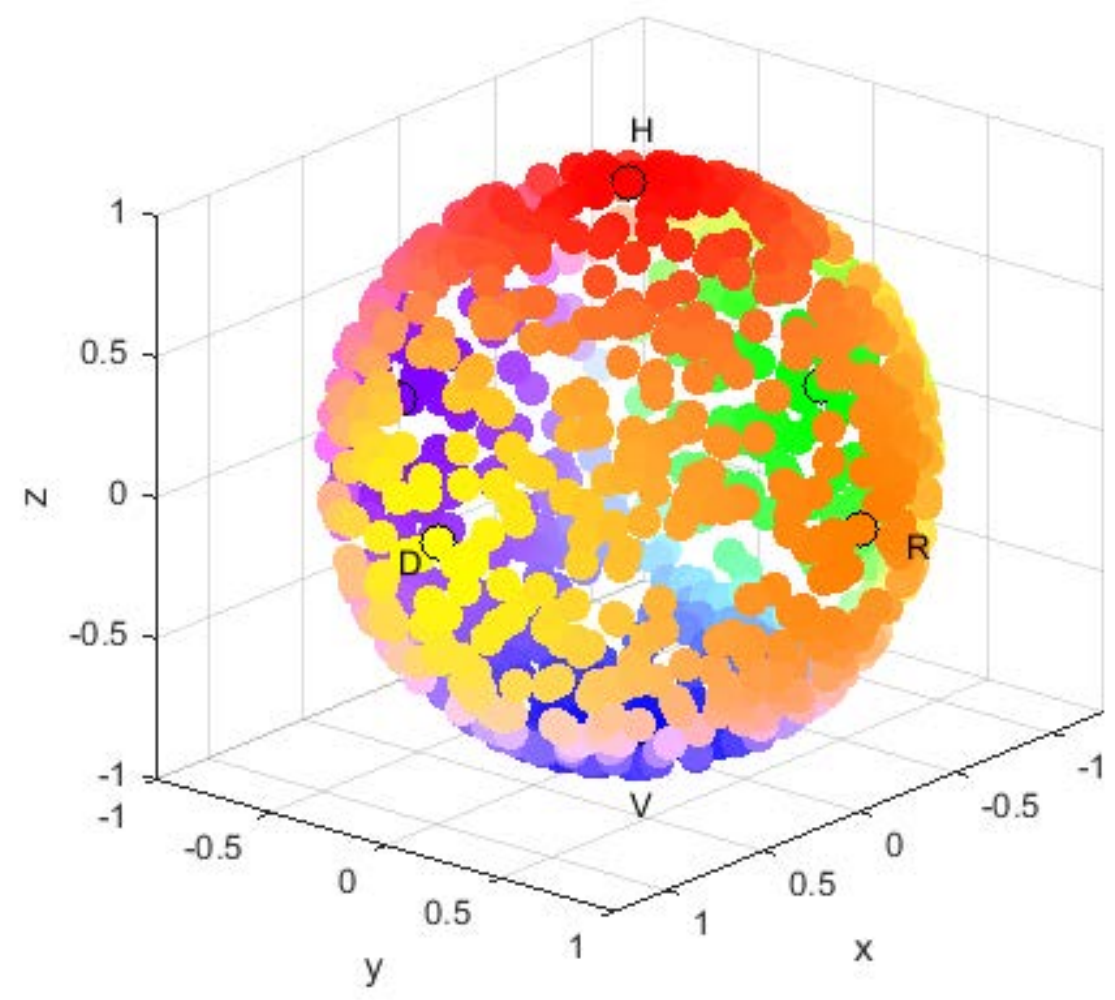}}
\caption{(Color online) Colors used by VQOL for various polarization states, as represented on the Bloch sphere.}
\label{fig:Bloch}
\end{figure}

Since polarization figures heavily in many quantum optics experiments, a design decision was made to use color to represent polarization.  Although the use of color to represent polarization can be misleading, as it should properly be construed as indicating wavelength (or frequency), we have found this causes no great confusion and is preferably to a two-dimensional projection of the electric field, which may be visually ambiguous.  The six main polarization states (and corresponding colors) are horizontal (red), vertical (blue), diagonal (yellow), anti-diagonal (green), right circular (orange), and left circular (violet).  Other, general elliptical polarizations are show in a color that is a blend of these six.  An illustration of these various colors, as seen on the Bloch sphere (or, equivalently, the Poincar\'{e} sphere) is given in Fig.\ \ref{fig:Bloch}.

Control buttons allow the user to start, pause, or stop the experiment.  There are also options to run the experiment in slow motion, step forward, or step back.  (Stepping back and then forward produces the same outcome.)  Experiments run for a default time of 1\,ms (in simulation time, not real time), but this may be changed in the graphical interface or by editing the variable \texttt{num\_seconds}. VQOL operates in time segments of $\Delta t = 1\,\mu\mathrm{s}$, so the default time of 1 ms corresponds to 1000 time samples.  The actual runtime of an experiment will depend on the local machine and the complexity of the experiment; a 1-ms experiment typically runs in about 15\,s of real time.  Setting the time to exactly $1\,\mu\mathrm{s}$ gives something that \emph{looks} like a single photon, though this is not how VQOL actually models quantum light.

A sample experiment for measuring the power of light light passing through  a polarizer rotated $30^\circ$ is the following:
\begin{verbatim}
# Malus's law experiment
num_seconds = 1e-3
offline_mode = False
Laser, x=1, y=1, orientation=0
Polarizer, x=3, y=1, orientation=0, angle=30
PowerMeter, x=5, y=1
\end{verbatim}
Note that comments may be added within the experiment using the hash symbol (\texttt{\#}).  

Another optional variable, \texttt{offline\_mode}, may be used to control the graphics display.  By default, this variable is set to \texttt{False}.  If set to \texttt{True}, VQOL will run the simulation without producing any animation graphics.  This can be useful for doing long simulations in which the user only wants the final results of the experiment.  In either case, once the experiment is completed the results are displayed on the screen and stored in a comma separated value (csv) file for later analysis.

One may use JavaScript\textsuperscript{\texttrademark} directly within the web-based editor to automate the simulation of multiple experimental scenarios.  To initiate the interpreter, the experiment editor should begin with \texttt{<JS>}.  An example for an experiment investigating Malus's law is shown below.
\begin{verbatim}
<JS>
// Malus's law experiment
for (let theta=0; theta<=90; theta+=5){
  runSimulation(
  "num_seconds = 1e-6\n" +
  "Laser, x=1, y=1, orientation=0\n" +
  "Polarizer, x = 3, y=1, orientation=0," +
  " angle=" + theta + "\n" +
  "PowerMeter, x=5, y=1"
  )
}
\end{verbatim}
The script runs 19 separate experiments, each 1-$\mu$s in duration, with polarizer angles of $0^\circ, 5^\circ, \ldots, 90^\circ$.  The ``\verb|//|'' symbol indicates a single-line comment.  The ``\texttt{+}'' symbols perform string concatenation, while the ``\verb|\n|'' symbol implements a newline character.  The results are stored in a set of folders that are downloaded as a single ZIP-formatted file.


\section{Model Description}
\label{sec:model}

In VQOL, light may travel in only one of four directions in the plane of the optics table: right ($\rightarrow$), left ($\leftarrow$), up ($\uparrow$), and down ($\downarrow$).  Out-of-plane or non-rectilinear directions are currently not supported.  Each $\Delta t$ time segment has a polarization given by a dimensionless Jones vector of the form
\begin{equation}
\vec{a} = \begin{pmatrix} a_H \\ a_V \end{pmatrix} \; ,
\end{equation}
where $a_H$ and $a_V$ are the complex horizontal and vertical polarization components, respectively.

For monochromatic light traveling, say, to the right with Jones vector $\vec{a}$, the electric field at the coordinate $(x,y)$ and time $t$ is given by
\begin{equation}
\vec{E}(x,y,t) \propto -(a_H \unit{y} + a_V \unit{z}) \, e^{i(k x - \omega t)} + \mathrm{c.c.} \; ,
\end{equation}
where $\omega$ is the angular frequency of the light, $k = \omega/c$ is the wavenumber, $c$ is the speed of light in a vacuum, and ``c.c.'' denotes the complex conjugate of the term to the left.  Note that VQOL uses a right-handed coordinate system such that the upper left corner is the origin, $\unit{x}$ points to the right, $\unit{y}$ points down, and $\unit{z}$ points into the optics table.

Classically, the Jones vector of the vacuum is simply the zero vector.  Quantum mechanically, we describe the vacuum by the Fock state $\ket{0}_H \otimes \ket{0}_V$, where $\ket{0}_H$ and $\ket{0}_V$ are the vacuum states of the horizontal and vertical polarization modes, respectively, and $\otimes$ denotes the tensor product.  More generally, the Wigner function of a thermal state for a given wave vector mode is given by
\begin{equation}
W_T(a_H, a_V) = \frac{1}{\pi^2 \sigma_T^4} \exp\left( -\frac{|a_H|^2+|a_V|^2}{\sigma_T^2} \right) \; ,
\end{equation}
where $\sigma_T > 0$ is given by Planck's second theory of black body radiation.  Specifically,
\begin{equation}
\sigma_T^2 = \frac{1}{e^{\hbar\omega/(k_{\rm B} T)} - 1} + \frac{1}{2} \; ,
\end{equation}
where $T$ is the absolute temperature (in Kelvin), $k_{\rm B}$ is Boltzmann's constant, and $\hbar$ is Planck's constant (divided by $2\pi$).  For $T = 0$, we define $\sigma_0^2 = 1/2$ as the limit of $\sigma_T^2$ as $T \to 0$.  At optical wavelengths and room temperature, $\sigma_T \approx \sigma_0$, which is taken to be the default in VQOL.  In quantum mechanical terms, $\sigma_T^2 \hbar \omega$ is the modal energy for a given wave vector and polarization, so each vacuum mode has an associated modal energy of $\frac{1}{2} \hbar \omega$.

In VQOL, a vacuum state is treated as having a random Jones vector of the form $\sigma_0 \vec{z}$, where
\begin{equation}
\vec{z} = \begin{pmatrix} z_H \\ z_V \end{pmatrix}
\end{equation}
is a standard complex Gaussian random vector, for which $z_H$ and $z_V$ are independent standard complex Gaussian random variables.  (We say that $z$ is a standard complex Gaussian random variable if $\mathsf{E}[z]=0$, $\mathsf{E}[z^2]=0$, and $\mathsf{E}[|z|^2]=1$, where $\mathsf{E}[\cdot]$ denotes the expectation value of a random variable.)  The probability density function of the Jones vector is therefore identical to the Wigner function of the vacuum state, by the optical equivalence theorem \cite{Cahill1969}.  Note that each vacuum mode has a random modal energy that is exponentially distributed with a mean value of $\frac{1}{2}\hbar\omega$.

In classical optics, a monochromatic plane wave may be represented by the Jones vector 
\begin{equation}
\vec{\alpha} = \begin{pmatrix} \alpha_H \\ \alpha_V \end{pmatrix} \; ,
\end{equation}
where $\alpha_H$ and $\alpha_V$ are complex numbers specifying the amplitude and phase of the horizontal and vertical polarization components.  Similarly, in quantum optics a monochromatic plane wave is described by the separable coherent state $\ket{\alpha_H} \otimes \ket{\alpha_V}$.  In VQOL, a mathematically equivalent representation is obtained by simply adding the vacuum components, resulting in the Jones vector
\begin{equation}
\vec{a} = \vec{\alpha} + \sigma_0 \vec{v} = \begin{pmatrix} \alpha_H + \sigma_0 z_H \\ \alpha_V + \sigma_0 z_V \end{pmatrix} \; .
\end{equation}
This, then, provides a model for laser light that includes classical light as a limiting case when the vacuum fluctuations can be neglected.

VQOL also provides a source of entangled light, modeled as a pair of random Jones vectors whose joint probability density function is identical to the Wigner function of a multi-mode Gaussian squeezed vacuum state and, hence, is Gaussian as well.  The random vectors are defined in terms of a pair of random vectors in a manner described in Sec.\ \ref{sec:components}.

The ability to model coherent and squeezed vacuum states as complex Gaussian random variables follows directly from the Gaussian nature of the quantum states themselves.  In this sense, they may be deemed classical.  (We note, however, that entangled Gaussian states need not admit a positive Glauber-Sudarshan $P$ function representation and, in this sense, are sometimes deemed non-classical.)  Non-Gaussian measurements are therefore essential to provide a mechanism for observing nonclassical behavior.

In VQOL we model detectors as devices that produce a detection event (or ``click'') when the amplitude of either the horizontal or vertical mode falls above a given threshold $\gamma \ge 0$.  Thus, a detection corresponds to the event
\begin{equation}
D = \Bigl\{ |a_H| > \gamma ~\mbox{ or }~ |a_V| > \gamma \Bigr\} \; .
\end{equation}
This particular definition of a detection event was chosen for its mathematical simplicity.  Using instead the magnitude of the Jones vector may be more physically motivated but would yield similar results.

For coherent light, the probability of event $D$ occurring is given by
\begin{equation}
\Pr[D] = 1 - P_{\gamma,\sigma_0}(\alpha_H) \, P_{\gamma,\sigma_0}(\alpha_V) \; ,
\label{eqn:PrD}
\end{equation}
where
\begin{equation}
P_{\gamma,\sigma}(\alpha)  = 1 - Q_1(\sqrt{2}|\alpha|/\sigma, \sqrt{2}\gamma/\sigma)
\end{equation}
and $Q_1(\cdot,\cdot)$ is the Marcum Q function \cite{Marcum1950}.  As we shall see, this result is different from the quantum probability of $1-\exp(-|\alpha_H|^2-|\alpha_V|^2)$ associated with one or more photons in a coherent state, as our model \emph{implicitly} incorporates the non-idealities of dark counts and sub-unity detection efficiency.

\begin{figure}[ht]
\centering
\scalebox{1.0}{\includegraphics[width=\columnwidth]{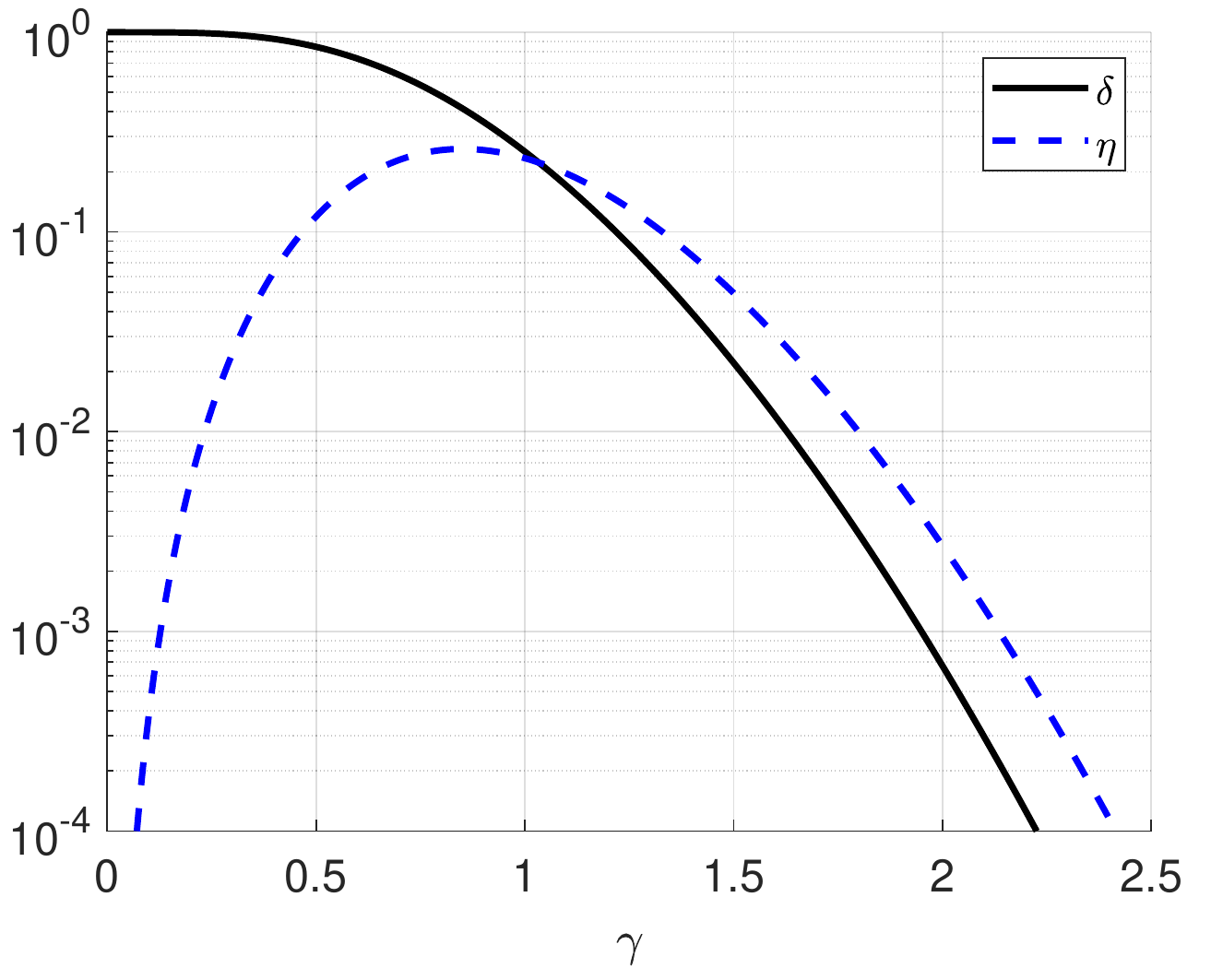}}
\caption{(Color online) Plot of dark count probability $\delta$ (black solid line) and nominal detection efficiency $\eta$ (blue dashed line) as a function of the threshold $\gamma$ for $\sigma_0^2 = 1/2$.  The parameter $\eta$ attains a maximum values of $0.26$ at $\gamma = 0.85$ and $\delta = 0.42$.}
\label{fig:threshold}
\end{figure}

To see this, note that we may approximate $P_{\gamma,\sigma}(\alpha)$, to second order in $|\alpha|$ and $\gamma$, as
\begin{equation}
P_{\gamma,\sigma}(\alpha) = 1 - e^{-\gamma^2/\sigma^2} \left[ 1 + \frac{\gamma^2}{\sigma^4} |\alpha|^2 + \mathcal{O}(|\alpha|^4) \right] \; .
\label{eqn:approximation}
\end{equation}
Thus, to lowest order,
\begin{equation}
\Pr[D] \approx \delta + \eta \left( |\alpha_H|^2 + |\alpha_V|^2 \right) \; ,
\end{equation}
where
\begin{equation}
\delta = 1 - \left( 1 - e^{-\gamma^2/\sigma_0^2} \right)^2
\label{eqn:DCP}
\end{equation}
is the probability of a dark count and
\begin{equation}
\eta = e^{-\gamma^2/\sigma_0^2} \left( 1 - e^{-\gamma^2/\sigma_0^2} \right) \frac{\gamma^2}{\sigma_0^4}
\label{eqn:eta}
\end{equation}
is the nominal detection efficiency.  In this way we can see that VQOL reproduces the Born rule in the limit of weak light (for fixed $\gamma$) and under the non-ideal, albeit realistic, conditions of a nonzero dark count rate and imperfect detection efficiency.  Note that it is not possible within the above parameterization to achieve both an arbitrarily low dark count and a detection efficiency arbitrarily close to one.  (See Figure \ref{fig:threshold}.)  Thus, VQOL is not capable of modeling an ideal detector.

\section{Description of Components}
\label{sec:components}


\subsection{Passive Optical Components}

\begin{figure}[ht]
\centering
\scalebox{1.0}{\includegraphics[width=\columnwidth]{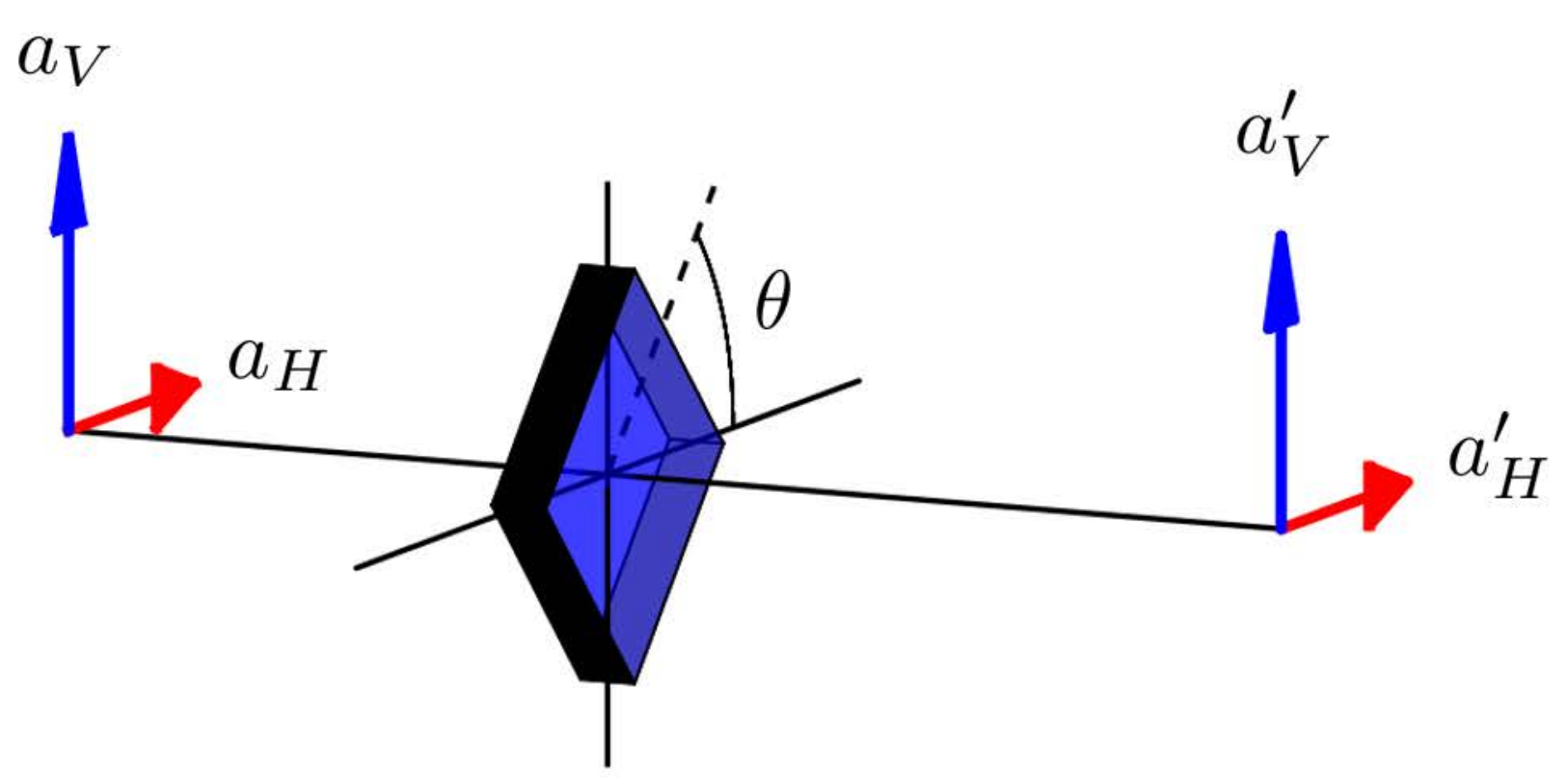}}
\caption{(Color online) Graphical illustration of the action of a wave plate with a fast-axis angle of $\theta$.  The input is shown on the left, while the output is shown on the right.}
\label{fig:HWP}
\end{figure}

Lossless optical components behave in accordance with their classical Jones matrix description.  For example, the action of a \textbf{half-wave plate} with a fast-axis angle of $\theta$ acting on a Jones vector $\vec{a} = (a_H,a_V)^\mathsf{T}$ is described by the Jones matrix transformation
\begin{equation}
\begin{pmatrix} a'_H \\ a'_V \end{pmatrix} = 
\begin{bmatrix} \cos(2\theta) & \sin(2\theta) \\ \sin(2\theta) & -\cos(2\theta) \end{bmatrix} \begin{pmatrix} a_H \\ a_V \end{pmatrix} \; .
\end{equation}
The angle $\theta$ is measured counterclockwise from the horizontal.  (Figure \ref{fig:HWP} provides a graphical illustration.)

Similarly, a \textbf{quarter-wave plate} with a fast-axis angle of $\theta$ has the Jones matrix
\begin{equation}
\begin{bmatrix} \cos^2\theta + i\sin^2\theta & (1-i) \cos\theta \sin\theta \\ (1-i) \cos\theta \sin\theta & \sin^2\theta + i\cos^2\theta \end{bmatrix} \; .
\end{equation}

Finally, a \textbf{phase delay} component may be used to apply a common phase shift $e^{i\phi}$ to both components.  This is represented by the Jones matrix
\begin{equation}
\begin{bmatrix} e^{i\phi} & 0 \\ 0 & e^{i\phi} \end{bmatrix} \; .
\end{equation}
The related \textbf{dephaser} component is a treated as a phase delay for which $\phi$ is a random variable that is uniformly distributed on the interval $[0,2\pi]$ and generated independently at each time step.

For longer delays, a \textbf{time delay} component may be used.  This component delays the light beam by an integer number of time steps, in units of $\Delta t$.  The default is zero (no delay).  A delay of 10 will set the light beam back by one whole grid space.  Physically, a time delay may be implemented with a coil of optical fiber, although in VQOL the time delay component is both amplitude and polarization preserving.

VQOL also offers the more general \textbf{rotator} and \textbf{phase retarder} components, whose Jones matrices are given by
\begin{equation}
\begin{bmatrix} \cos\theta & -\sin\theta \\ \sin\theta & \cos\theta \end{bmatrix} \; , 
\end{equation}
for $\theta \in [0^\circ,90^\circ]$, and
\begin{equation}
\begin{bmatrix} 1 & 0 \\ 0 & e^{i\phi} \end{bmatrix} \; ,
\end{equation}
respectively.

A general unitary may be represented by a rotator, two phase retarders, and a phase delay as follows:\cite{Jones1941a,Jones1941b}
\begin{equation}
\mtx{U} = 
e^{i\phi}
\begin{bmatrix} 1 & 0 \\ 0 & e^{i\phi} \end{bmatrix} 
\begin{bmatrix} \cos\theta & -\sin\theta \\ \sin\theta & \cos\theta \end{bmatrix}
\begin{bmatrix} 1 & 0 \\ 0 & e^{i\lambda} \end{bmatrix} \; .
\end{equation}
Taking $\theta = \frac{1}{2}\cos^{-1}(2u-1)$, where $u$ is uniformly distributed on the interval $[0,1]$, and taking $\phi, \lambda, \chi$ to be independent and uniformly distributed on the interval $[0,2\pi]$, the matrix $\mtx{U}$ becomes a Haar-distributed random matrix.  This is used to define a \textbf{depolarizer} in VQOL as a component that applies a random $\mtx{U}$ at each time step.

Lossy components are treated differently in VQOL from their classical counterparts.  In quantum optics, attenuation is treated as a partially transmitting beam splitter, where the second input port receives a vacuum mode.  Similarly, in VQOL a \textbf{neutral density filter} with an optical density of $d \ge 0$ acting on a Jones vector $\vec{a} = (a_H,a_V)^\mathsf{T}$ has the following output:
\begin{equation}
\begin{pmatrix} a'_H \\ a'_V \end{pmatrix} = 10^{-d/2} \begin{pmatrix} a_H \\ a_V \end{pmatrix} + \left( 1 - 10^{-d/2} \right) \sigma_0 \vec{z} \; ,
\end{equation}
where $\vec{z}$ is, again, an independent standard complex Gaussian random vector.  VQOL also offers a \textbf{beam blocker} component, which transforms an input Jones vector $\vec{a}$ into a vacuum state $\sigma_0 \vec{z}$.  This is equivalent to an NDF with $d \to \infty$.

In a similar manner, a general elliptical \textbf{polarizer} acting on the Jones vector $\vec{a}$ has the following output:
\begin{equation}
\begin{pmatrix} a'_H \\ a'_V \end{pmatrix} = \mtx{P}(\theta,\phi) \begin{pmatrix} a_H \\ a_V \end{pmatrix} + \Bigl[ \mtx{1}-\mtx{P}(\theta,\phi) \Bigr] \sigma_0 \vec{z} \; ,
\label{eqn:polarizer}
\end{equation}
where $\mtx{P}(\theta,\phi)$ is the standard polarizer Jones matrix,
\begin{equation}
\mtx{P}(\theta,\phi) = \begin{bmatrix}
\cos^2\theta & e^{-i\phi} \cos\theta \sin\theta \\ e^{i\phi} \cos\theta \sin\theta & \sin^2\theta
\end{bmatrix} \; ,
\end{equation}
and $\mtx{1}-\mtx{P}(\theta,\phi)$ is the complementary matrix projection.  Note that the presence of an additional vacuum term in Eqn.\ (\ref{eqn:polarizer}) invalidates the \emph{no-enhancement assumption} \cite{CH1974}.  In other words, it is possible that $|a'_H|^2 + |a'_V|^2 > |a_H|^2 + |a_V|^2$ for a given instance of $\vec{z}$.  This implies, in particular, that placing a polarizer in front of detector could generate a detection that, counterfactually, would not have occurred if the polarizer were not present.

\begin{figure}[ht]
\centering
\scalebox{1.0}{\includegraphics[width=\columnwidth]{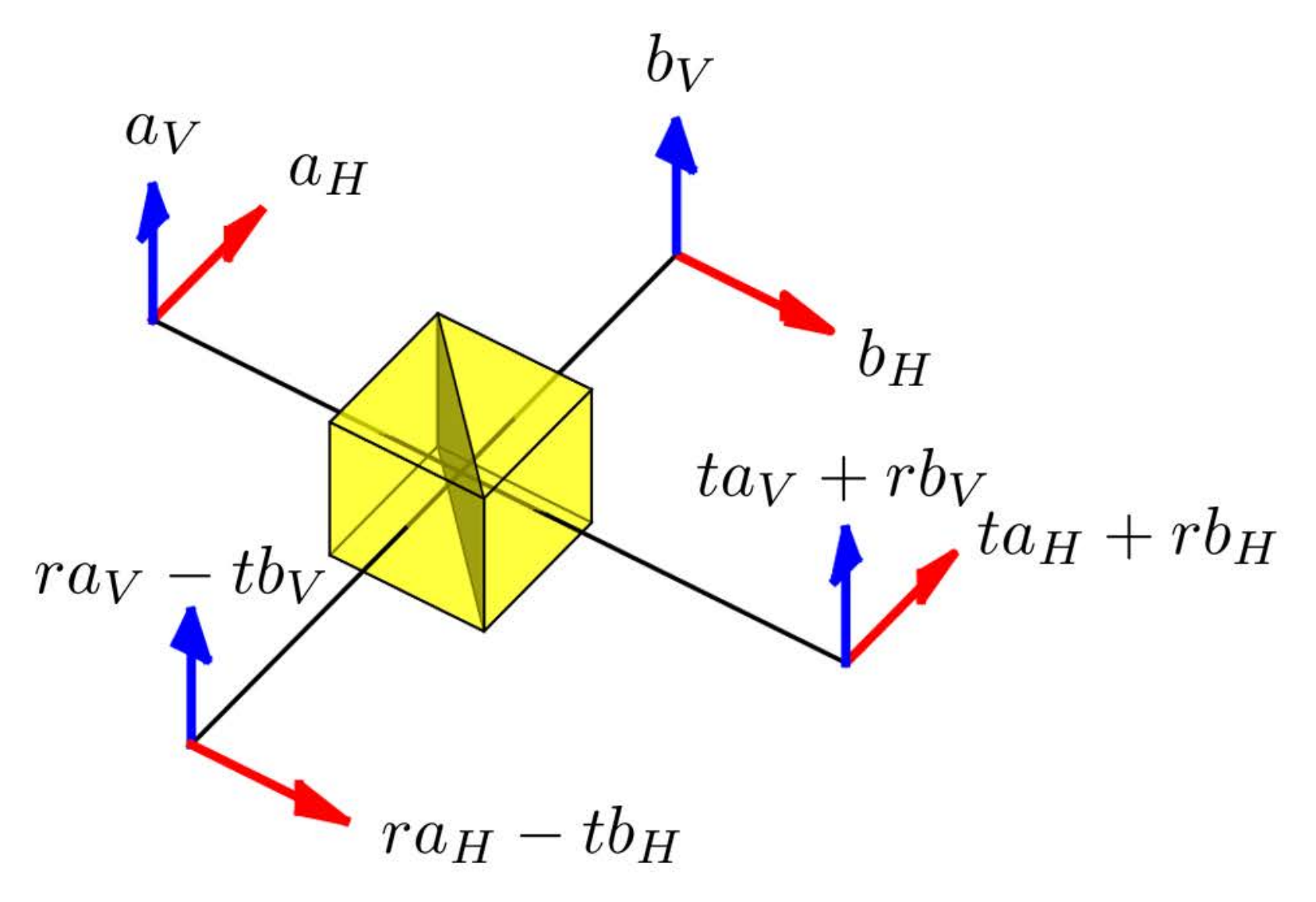}}
\caption{(Color online) Graphical illustration of the action of a beam splitter with transmission (reflection) coefficient $t$ ($r$).  The inputs are shown in the upper left and right, while the outputs are shown in the lower left and right.}
\label{fig:BS}
\end{figure}

A \textbf{beam splitter} in VQOL work much like a classical beam splitter.  Given the input Jones vectors $\vec{a} = (a_H, a_V)^\mathsf{T}$ and $\vec{b} = (b_H, b_V)^\mathsf{T}$, the output is given by
\begin{equation}
\begin{pmatrix} a'_H \\ a'_V \\ b'_H \\ b'_V \end{pmatrix} = \begin{pmatrix} t & 0 & r & 0 \\ 0 & t & 0 & r \\ r & 0 & -t & 0 \\ 0 & r & 0 & -t \end{pmatrix} \begin{pmatrix} a_H \\ a_V \\ b_H \\ b_V \end{pmatrix} \; ,
\end{equation}
where $r \in [0,1]$ is the reflection coefficient and $t = \sqrt{1-r^2}$ is the transmission coefficient.  (See Fig.\ \ref{fig:BS}.)  A \textbf{mirror} is a special case for which $r=1$.  Note that if a beam splitter is presented with only one input, the other is taken to be a pair of independent vacuum modes, represented by a random Jones vector $\sigma_0 \vec{z}$.

\begin{figure}[ht]
\centering
\scalebox{0.8}{\includegraphics[width=\columnwidth]{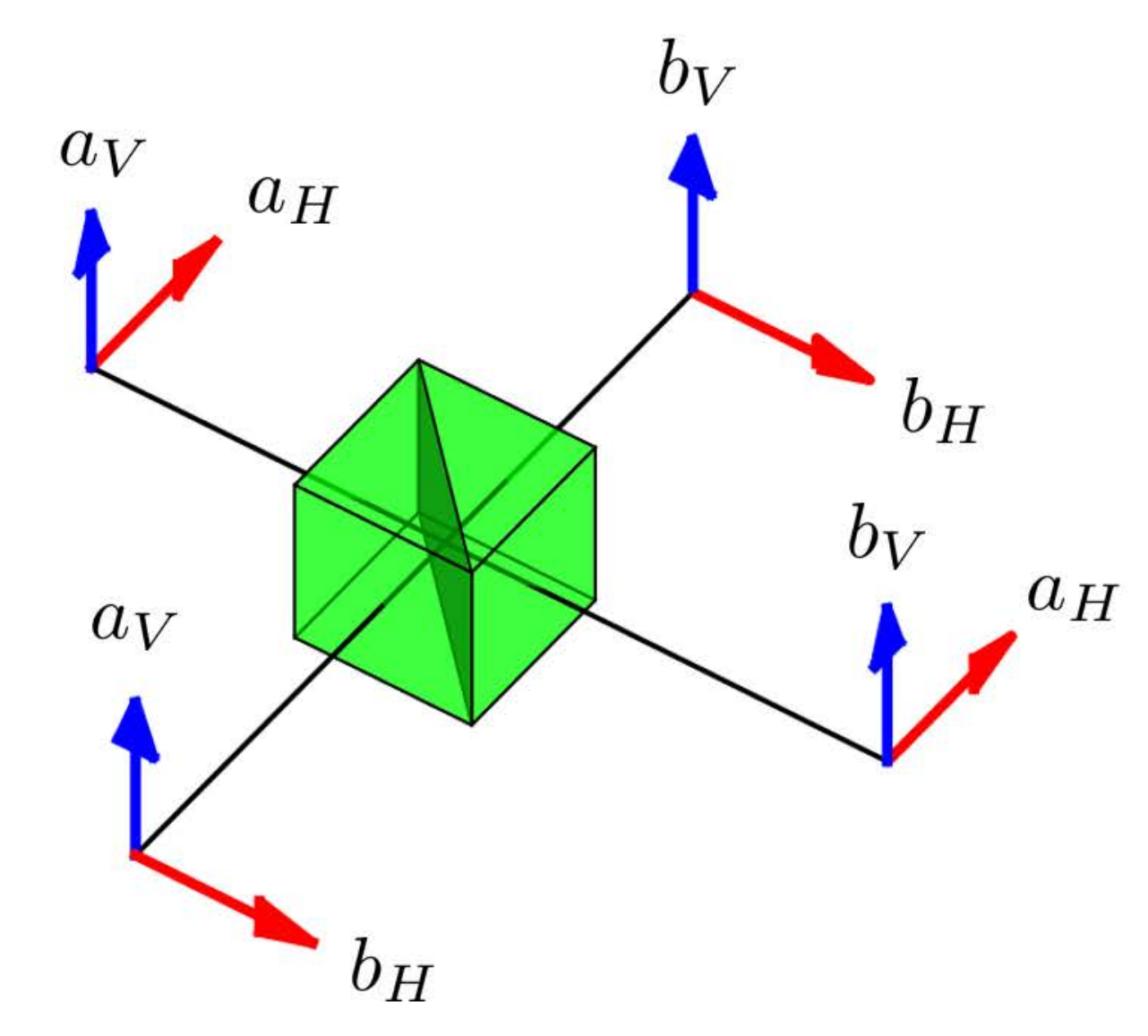}}
\caption{(Color online) Graphical illustration of the action of a polarizing beam splitter.  The inputs are in the upper left and right, while the outputs are in the lower left and right.}
\label{fig:PBS}
\end{figure}

A \textbf{polarizing beam splitter} (PBS) works similarly to a regular beam splitter.  Given two input Jones vectors, $\vec{a}$ and $\vec{b}$, the output is
\begin{equation}
\begin{pmatrix} a'_H \\ a'_V \\ b'_H \\ b'_V \end{pmatrix} = \begin{pmatrix} 1 & 0 & 0 & 0 \\ 0 & 0 & 0 & 1 \\ 0 & 0 & 1 & 0 \\ 0 & 1 & 0 & 0 \end{pmatrix} \begin{pmatrix} a_H \\ a_V \\ b_H \\ b_V \end{pmatrix} \; .
\label{eqn:PBShv}
\end{equation}
(See Fig.\ \ref{fig:PBS}.)  Again, if the PBS is presented with only one input, the other is taken to be a pair of independent vacuum modes.  Consequently, the light exiting each output port need not be restricted to only horizontal or vertical polarization, as reflected components of the opposite polarization will contribute from the vacuum modes.

VQOL offers three PBS variants for different polarization bases.  The default $H/V$ basis transmits horizontal light and reflects vertical light, as described by Eqn.\ (\ref{eqn:PBShv}).  An optional $D/A$ basis transmits diagonal light and reflects anti-diagonal light, while the $R/L$ basis option transmits right-circular light and reflects left-circular light.  Specifically, if $\vec{a} = (a_H, a_V)^\mathsf{T}$ and $\vec{b} = (b_H, b_V)^\mathsf{T}$ are the Jones vectors for the inputs to a $D/A$-basis PBS, then the output Jones vectors are
\begin{equation}
\begin{pmatrix} a'_H \\ a'_V \end{pmatrix} = \frac{a_H + a_V}{2} \begin{pmatrix} 1 \\ 1 \end{pmatrix} + \frac{b_H - b_V}{2} \begin{pmatrix} 1 \\ -1 \end{pmatrix}
\end{equation}
and
\begin{equation}
\begin{pmatrix} b'_H \\ b'_V \end{pmatrix} =  \frac{b_H + b_V}{2} \begin{pmatrix} 1 \\ 1 \end{pmatrix} + \frac{a_H - a_V}{2} \begin{pmatrix} 1 \\ -1 \end{pmatrix} \; .
\end{equation}
Similarly, for an $R/L$-basis PBS we have
\begin{equation}
\begin{pmatrix} a'_H \\ a'_V \end{pmatrix} = \frac{a_H - i a_V}{2} \begin{pmatrix} 1 \\ i \end{pmatrix} + \frac{b_H + i b_V}{2} \begin{pmatrix} 1 \\ -i \end{pmatrix}
\end{equation}
and
\begin{equation}
\begin{pmatrix} b'_H \\ b'_V \end{pmatrix} = \frac{b_H - i b_V}{2} \begin{pmatrix} 1 \\ i \end{pmatrix} + \frac{a_H + i a_V}{2} \begin{pmatrix} 1 \\ -i \end{pmatrix} \; .
\end{equation}

\subsection{Light Sources}

VQOL provides three different sources of light that are commonly available: a light-emitting diode for thermal light, a laser for coherent light, and a parametric downconversion source for entangled light.  All sources are assumed to have a coherence time of $\tau = \Delta t = 1\,\mu\mathrm{s}$, so an independent random realization of the ZPF is drawn every time step.  VQOL does not provide single-photon sources or indeed sources for any Fock states other than the vacuum.  Since these three light sources produce Gaussian states, they may be modeled classically by an equivalent Gaussian random vector, as described below.

A \textbf{light-emitting diode} (LED) is a source of incoherent thermal light.  In VQOL, an LED is treated as a monochromatic light source, with a wavelength of $\lambda = 496.61\,\mathrm{nm}$ ($603.68\,\mathrm{THz}$) and linewidth of that is perfectly collimated, appearing as a beam of constant width.  An LED is specified by the parameter \texttt{power}, in Watts, with a default value of $4\,\mathrm{mW}$.  The Jones vector of an LED is $\sqrt{\sigma^2+\sigma_0^2} \, \vec{z}$, where $\vec{z}$ is a standard complex Gaussian random vector and $\sigma \ge 0$ specifies the total power
\begin{equation}
P = (\sigma^2 + \sigma_0^2) \frac{\hbar\omega}{\Delta t} \; ,
\end{equation}
where $\Delta t = 1\,\mu\mathrm{s}$ is the length of each time step, corresponding to a passband of $1\,\mathrm{MHz}$, and $\omega = 2\pi c/\lambda$.  Note that the \texttt{power} parameter in VQOL is actually $P-P_0$, where $P_0 = \sigma_0^2 \hbar \omega / \Delta t = 2.0000 \times 10^{-13}\,\mathrm{W}$.  This is done to ensure that the vacuum modes persist even with the \texttt{power} parameter set to zero.

A \textbf{laser} in VQOL is a coherent monochromatic light source that is perfectly collimated.  The Jones vector of a laser is of the form $\vec{\alpha} + \sigma_0 \vec{z}$, where $\vec{\alpha} = \alpha \ket{\psi}$, $\alpha = 10^5$ (in dimensionless units), and $\ket{\psi}$ is a normalized polarization vector.  (Note that, although the use of ``ket'' notation is suggestive of a quantum state, here we use it solely to represent a normalized complex vector.)  Lasers may be configured so that $\ket{\psi}$ takes one of the following six standard forms:
\begin{subequations}
\begin{align}
\ket{H} &= \begin{pmatrix} 1 \\ 0 \end{pmatrix} \; , 
\quad\quad\;\;\: \ket{V} = \begin{pmatrix} 0 \\ 1 \end{pmatrix} \\
\ket{D} &= \frac{1}{\sqrt{2}} \begin{pmatrix} 1 \\ 1 \end{pmatrix} \; , 
\quad \ket{A} = \frac{1}{\sqrt{2}} \begin{pmatrix} 1 \\ -1 \end{pmatrix} \\
\ket{R} &= \frac{1}{\sqrt{2}} \begin{pmatrix} 1 \\ i \end{pmatrix} \; , \quad 
\ket{L} = \frac{1}{\sqrt{2}} \begin{pmatrix} 1 \\ -i \end{pmatrix}
\end{align}
\end{subequations}
The wavelength of a laser is the same as that of an LED, with the same default power as well.  The total power, $P$, is related to the average photon number, $\|\vec{\alpha}\|^2 = |\alpha_H|^2 + |\alpha_V|^2$, by
\begin{equation}
P = \left( \|\vec{\alpha}\|^2 + \sigma_0^2 \right) \frac{\hbar\omega}{\Delta t} \; .
\end{equation}
For the default \texttt{power} setting of $4\,\mathrm{mW}$ we have $\|\vec{\alpha}\|^2 = 10^{10}$, which corresponds to an average photon number of 10 billion per time segment $\Delta t$.  Using a neutral density filter with a default optical density of $d = 10$, the average photon number drops to just one nominal photon per time segment.  If the \texttt{power} parameter is set to zero, the total power reduces to that of the vacuum.
\begin{figure}[ht]
\centering
\scalebox{0.8}{\includegraphics[width=\columnwidth]{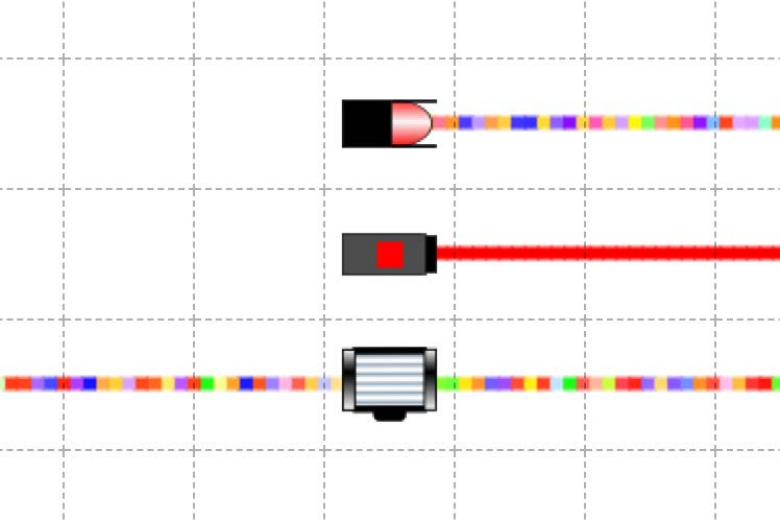}}
\caption{(Color online) Illustration of light sources in VQOL.  Shown above are an LED (top), a laser (middle), and an entanglement source (bottom).}
\label{fig:sources}
\end{figure}

Interestingly, VQOL also offers the ability to classically simulate multi-modal squeezed light and, hence, entangled states.  A dual-spatial-mode \textbf{entanglement source} based on parametric downconversion is used as a notional source of entangled photons.  Entanglement sources can be type-I or type-II and are parameterized by a squeezing strength parameter $r$ and relative phase $\varphi$.  The Jones vectors for an entanglement source are a direct translation of the Bogoliubov transformations, with the annihilation operators replaced by complex Gaussian random variables.  Since the multi-modal squeezed vacuum states are Gaussian, this provides a faithful representation of the joint statistical distribution.

In VQOL, the Jones vector for a type-I entanglement source is given by
\begin{equation}
\begin{pmatrix} a_{H} \\ a_{V} \\ b_{H} \\ b_{V} \end{pmatrix}
= \sigma_0
\begin{pmatrix}
z_{1H} \cosh r + z_{2H}^* \sinh r \\ z_{1V} \cosh r + e^{i\varphi} z_{2V}^* \sinh r \\ z_{2H} \cosh r + z_{1H}^* \sinh r \\ z_{2V} \cosh r + e^{i\varphi} z_{1V}^* \sinh r
\end{pmatrix} \; ,
\end{equation}
where $z_{1H}, z_{1V}, z_{2H}, z_{2V}$ are independent standard complex Gaussian random variables.  The Jones vectors $\vec{a} = (a_H, a_V)^\mathsf{T}$ and $\vec{b} = (b_H, b_V)^\mathsf{T}$ correspond to the two spatial modes and notionally represent the two entangled photons.  For a type-II entanglement source we have the Jones vector
\begin{equation}
\begin{pmatrix} a_{H} \\ a_{V} \\ b_{H} \\ b_{V} \end{pmatrix}
= \sigma_0
\begin{pmatrix}
z_{1H} \cosh r + z_{2V}^* \sinh r \\ z_{1V} \cosh r + e^{i\varphi} z_{2H}^* \sinh r \\ z_{2H} \cosh r + e^{i\varphi} z_{1V}^* \sinh r \\ z_{2V} \cosh r + z_{1H}^* \sinh r
\end{pmatrix} \; .
\end{equation}

For small $r$, the type-I entanglement source may be used to approximate the entangled state
\begin{equation}
\ket{\Phi} = \frac{\ket{HH} + e^{i\varphi}\ket{VV}}{\sqrt{2}} \; ,
\label{eqn:Phi}
\end{equation}
while for a type-II source it may be used to approximate the entangled state
\begin{equation}
\ket{\Psi} = \frac{\ket{HV} + e^{i\varphi}\ket{VH}}{\sqrt{2}} \; .
\label{eqn:Psi}
\end{equation}
This correspondence will be made clearer in the examples to follow.  The default settings for an entanglement source are $r = 1$ and either $\varphi = 0^\circ$, for a type-I source, or $\varphi = 180^\circ$, for a type-II source.

An entanglement source has a fixed orientation, but the directions of the two outgoing can be configured with the parameter  \texttt{directions}.  By default, this parameter is set to LR, indicating that $\vec{a}$ travels to the left and $\vec{b}$ travels to the right.  Other options are LU (left, up), LD (left, down), UR (up, right), DR (down, right), and UD (up, down).


\subsection{Measurement Devices}

VQOL offers two types of measuring devices, a \textbf{power meter} and a \textbf{detector}.  The power meter acts as a photodetector operating in linear mode, so it reports the total incident power at each time step (including contributions from the vacuum).  Power meters register power from any direction, so they have no orientation.  However, if more than one light source illuminates a given power meter, it reports the largest power among the illuminating sources.

Given a Jones vector $\vec{a} = (a_H, a_V)^\mathsf{T}$, the power meter will report the value
\begin{equation}
P = \sqrt{|a_H|^2 + |a_V|^2} \, \frac{\hbar\omega}{\Delta t} \; ,
\end{equation}
where $\omega$ and $\Delta t$ are defined as before.  Note that $P$ represents the slowly varying, time-averaged power of the wave.  Fluctuations on a time scale below $\Delta t$ are not seen.  Power levels are displayed with up to three significant figures but recorded to within $1\,\mathrm{nW}$ of precision.  In particular, power levels below $1\,\mathrm{nW}$ will be displayed and recorded as zero.  If the user highlights the power meter while the simulation is running, it will display the current and time averaged power the configuration window.

A detector in VQOL acts as a photodetector operating in Geiger mode.  It will produce a detection count (or ``click'') when the amplitude of either the horizontal or vertical component falls above a given threshold $\gamma \ge 0$.  Detectors have an effective dead time of $\Delta t$, as there can be at most one detection per time step.  A detector is parameterized by its dark count rate (DCR), measured in counts per second.  The DCR is related to the dark count probability, $\delta$, via the formula $\mathrm{DCR} = \delta/\Delta t$.  By default, DCR is set to 1000 counts per second (or 1/ms), corresponding to a dark count probability of $\delta = 0.001$.  The detection threshold for a given dark count probability $\delta$ is found from Eqn.\ (\ref{eqn:DCP}) to be
\begin{equation}
\gamma = \sigma_0 \sqrt{-\log\left( 1 - \sqrt{1-\delta} \right) } \; .
\end{equation}
For DCR = 1/ms, this takes the default value of $\gamma = 1.95$.  The user may set the DCR value for each detector separately.

\begin{figure}[ht]
\centering
\scalebox{1.0}{\includegraphics[width=\columnwidth]{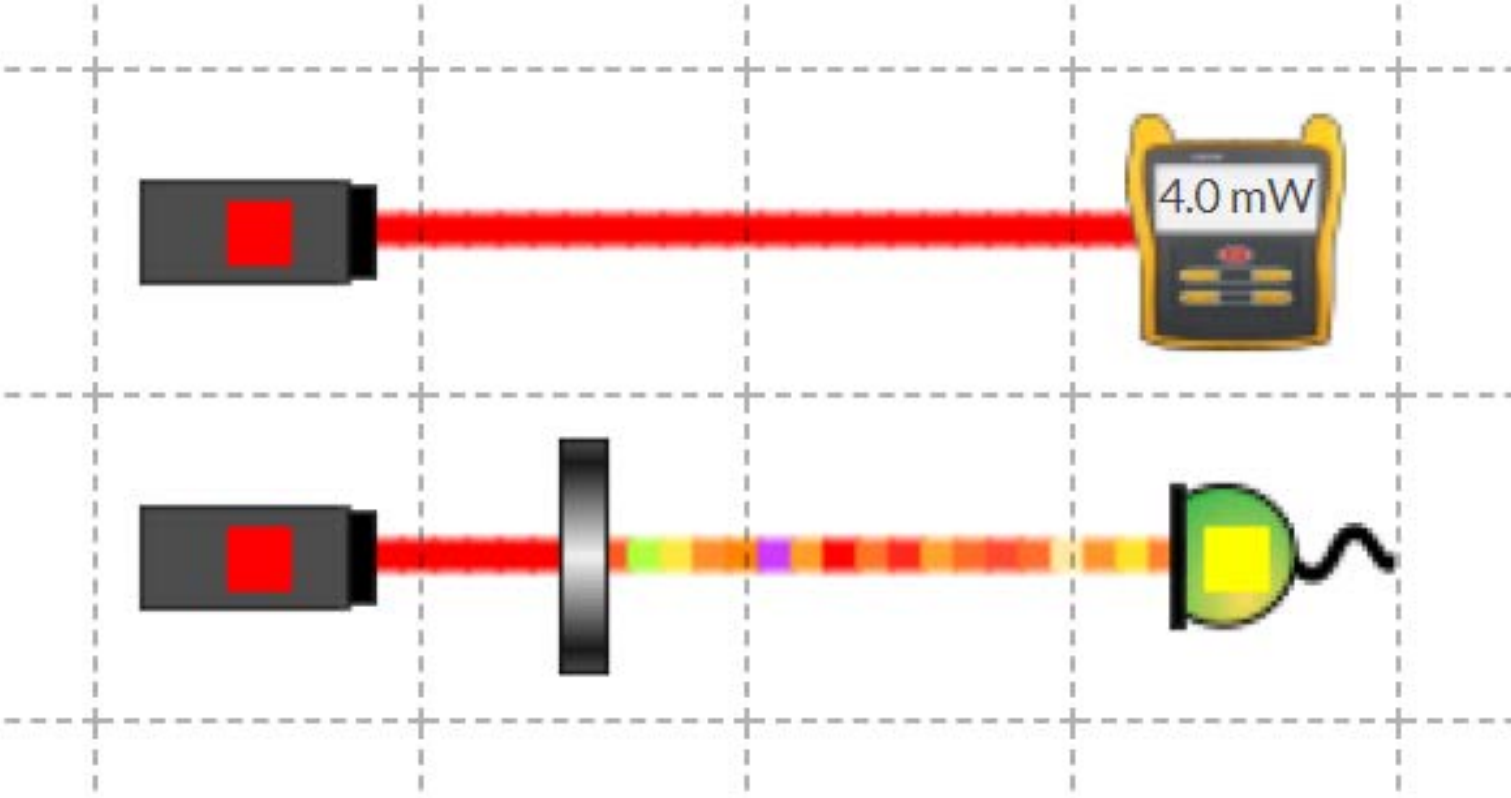}}
\caption{(Color online) Illustration of measurement devices in VQOL.  Shown above are a power meter (top) and single-photon detector (bottom), along with two lasers and a neutral density filter (gray rectangle).  The power meter reads $4.0\,\mathrm{mW}$, and the detector has just made a detection.}
\label{fig:measurement}
\end{figure}


\section{Selected Examples}
\label{sec:examples}

In this section we consider eight different examples (and a few subvariants) of experiments that can be performed in VQOL.  We begin with a balanced homodyne experiment using power meters to measure the vacuum.  Next, we examine the Born rule using several different approximations of a single-photon state.  This is followed by the use of quantum state tomography to infer the density matrix of a prepared state.  We then consider wave/particle duality in the context of a Mach-Zehnder interferometer.  Following this, we examine the use of an entanglement source to demonstrate anti-correlation and violations of Bell's inequality.  Finally, we consider the use of a beam splitter to perform a partial Bell state analysis and use this to implement a quantum teleportation scheme.  In these examples, special emphasis will be placed on the role of data analysis and post-selection to connect measured data with theoretical predictions.  All component settings are taken to be their default values unless stated otherwise.


\subsection{Homodyne Detection}

\begin{figure}[ht]
\centering
\scalebox{1.0}{\includegraphics[width=\columnwidth]{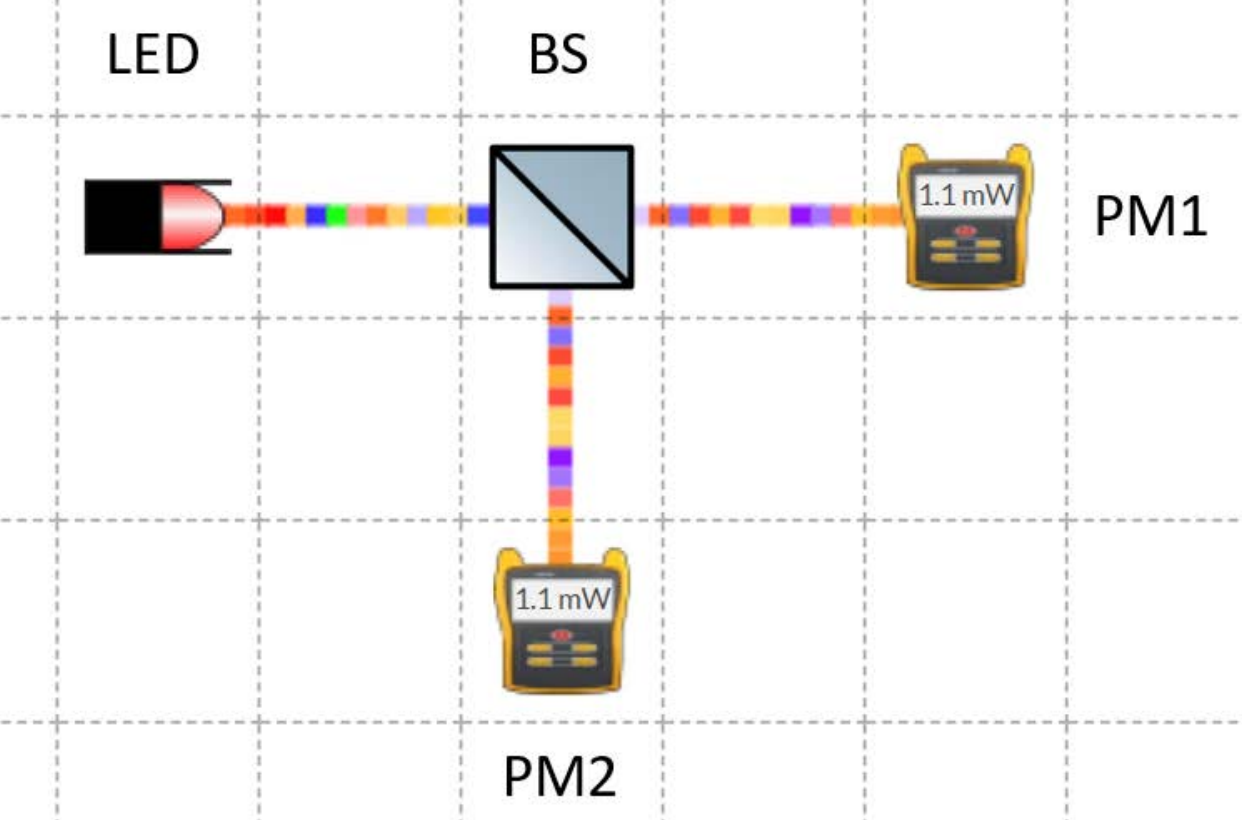}}
\caption{(Color online) Experimental setup for measuring the vacuum using balanced homodyne detection.}
\label{fig:homodyne}
\end{figure}

Even though power meters have limited resolution, we can us them to measure the much smaller power level of the vacuum, given in VQOL by $P_0 = \sigma_0^2 \hbar \omega/\Delta t$, using balanced homodyne detection.  An example of the experimental setup is shown in Fig.\ \ref{fig:homodyne}, which features an LED light source, beam splitter (BS), and two power meters (PM1 and PM2).  The Jones vector of the LED is given by
\begin{equation}
\vec{a} = \begin{pmatrix} a_H \\ a_V \end{pmatrix} = \sqrt{\sigma^2 + \sigma_0^2} \begin{pmatrix} z_{1H} \\ z_{1V} \end{pmatrix} \; ,
\end{equation}
while the Jones vector of the vacuum modes entering the (unused) top input port of the beam splitter is
\begin{equation}
\vec{b} = \begin{pmatrix} b_H \\ b_V \end{pmatrix} = \sigma_0 \begin{pmatrix} z_{2H} \\ z_{2V} \end{pmatrix} \; .
\end{equation}

The Jones vectors $\vec{a}'$ and $\vec{b}'$ of the beams exiting the beam splitter to the right and down, respectively, are given by
\begin{align}
\vec{a}' &= (\vec{a}+\vec{b})/\sqrt{2} \; ,  \\ 
\vec{b}' &= (\vec{a}-\vec{b})/\sqrt{2} \; .
\end{align}
The incident power on PM1 and PM2 is therefore
\begin{align}
P_1 &= \frac{1}{2} (|a_H+b_H|^2 + |a_V+b_H|^2) \, \hbar\omega/\Delta t \; , \\
P_2 &= \frac{1}{2} (|a_H-b_H|^2 + |a_V-b_H|^2) \, \hbar\omega/\Delta t \; ,
\end{align}
respectively.  In VQOL, these values will be recorded to within $1\,\mathrm{nW}$ of precision.

Let $x = P_1 - P_2$ and $y = P_1 + P_2$ denote the difference and sum of the two powers.  The difference may be written
\begin{equation}
\begin{split}
x &= 2 \, \mathrm{Re}( a_H b_H^* + a_V b_V^* ) \, \hbar\omega/\Delta t \\
&= 2\sqrt{(\sigma/\sigma_0)^2 + 1} \, \mathrm{Re}( z_{1H} z_{2H}^* + z_{1V} z_{2V}^* ) \, P_0 \; .
\end{split}
\end{equation}
Since $\mathsf{Var}[ \mathrm{Re}( z_{1H} z_{2H}^* + z_{1V} z_{2V}^* ) ] =  1$, we see that $x$ has a variance of $s^2 = 4(\sigma^2/\sigma_0^2 + 1) P_0^2$.  The sum, on the other hand, is
\begin{equation}
y = \left( |a_H|^2 + |a_V|^2 + |b_H|^2 + |b_V|^2 \right) \hbar\omega/\Delta t \; ,
\end{equation}
which has a mean of $\mu = 2(\sigma^2/\sigma_0^2 + 2) P_0$.  Since, for the default setting of the LED, $\sigma \gg \sigma_0$, we may estimate $P_0$ by the ratio $s^2/(2\mu)$.

We performed a 1-ms experiment in VQOL using the setup in Fig.\ \ref{fig:homodyne} and obtained measurements of $s = 57.9\,\mathrm{nW}$ and $\mu = 8.06\,\mathrm{W}$.  This resulted in an estimated value for $P_0$ of $2.08 \times 10^{-13}\,\mathrm{W}$, which, to within sampling error, is in good agreement with the true value.  Note that, due to the finite propagation in VQOL, the first few entries of the recorded power will be zero, as only vacuum states will initially be present.  These should have a negligible impact on the results but may be ignored explicitly in the analysis.  A similar experiment may be performed with a laser instead of an LED.  In this case, the variance of $x$ and mean of $y$ will be half that for the LED.  Otherwise, the same procedure may be followed.


\subsection{Detection Efficiency}

\begin{figure}[ht]
\centering
\scalebox{1.0}{\includegraphics[width=\columnwidth]{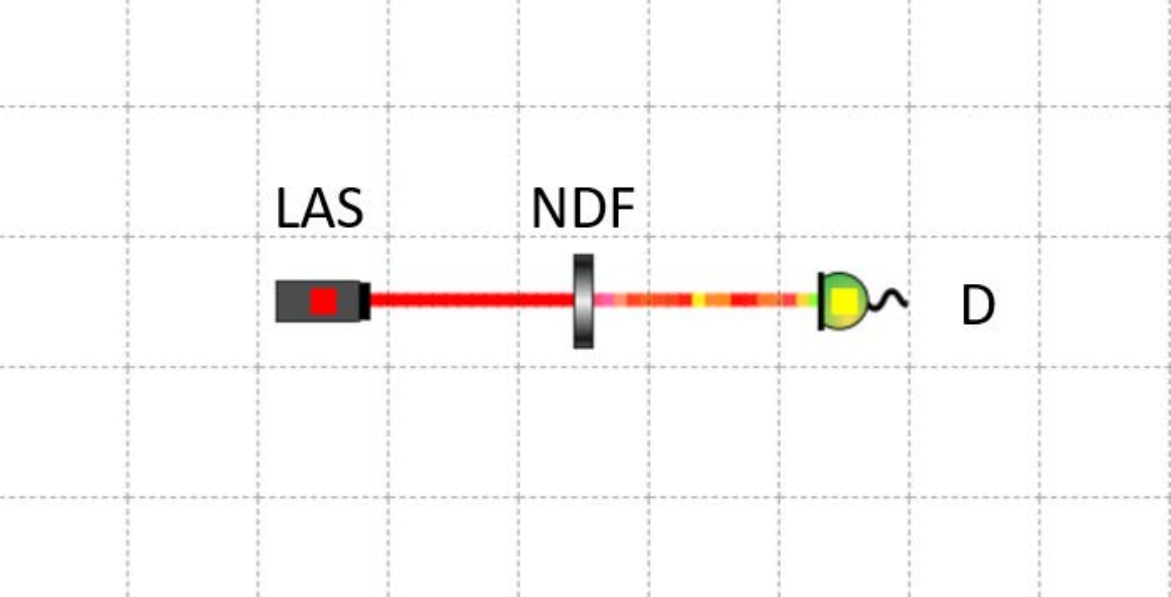}}
\caption{(Color online) Experimental setup for measuring detection efficiency using a laser (LAS), neutral density filter (NDF) and single detector (D).}
\label{fig:efficiencyLAS_setup}
\end{figure}

In Eqn.\ (\ref{eqn:eta}) we identified $\eta$ as the nominal detection efficiency.  Operationally, the efficiency of a single-photon detector is measured by counting detection events relative to some reference photon flux.  We may implement this scheme in VQOL using a laser, neutral density filter (NDF), and single detector, as shown in Fig.\ \ref{fig:efficiencyLAS_setup}).  The known power, $P$, of the laser and optical density, $d$, of the NDF allow us to estimate the photon flux as $10^{-d} P/(\hbar\omega) = 10^{-d} \|\vec{\alpha}\|^2/\Delta t$.  If we measure for a time $t$ and obtain $N$ counts, the inferred detection efficiency is
\begin{equation}
\eta_{L} = \frac{N \Delta t}{10^{-d}\|\vec{\alpha}\|^2 t} \; .
\end{equation}

\begin{figure}[ht]
\centering
\scalebox{1.0}{\includegraphics[width=\columnwidth]{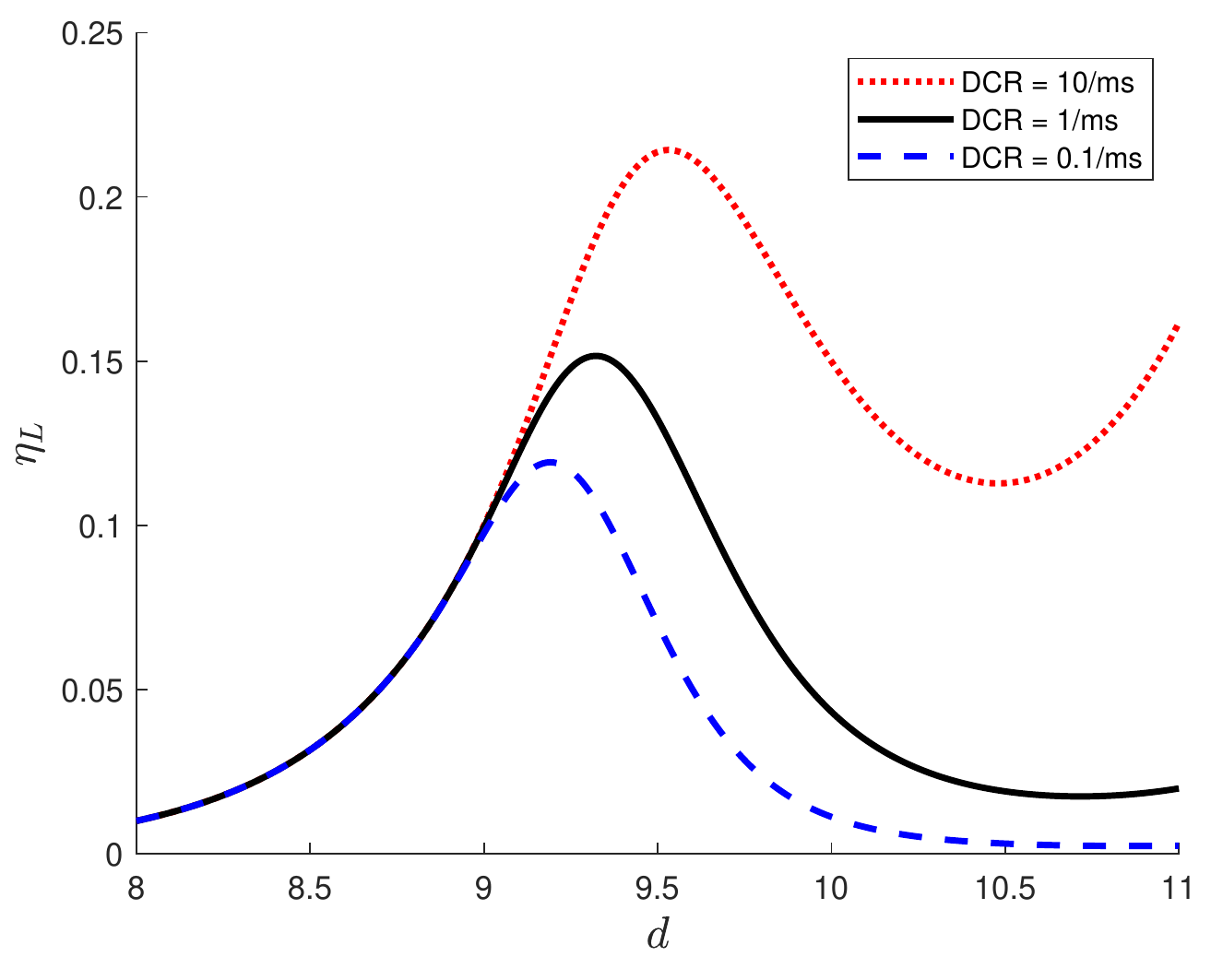}}
\caption{(Color online) Plot of inferred efficiency $\eta_{L}$ versus the optical density, $d$, for the NDF for different values of the dark count rate (DCR).}
\label{fig:efficiencyLAS_results}
\end{figure}

This experiment was implemented in VQOL and run for different values of $d$ and dark count rates.  In each case we used a default laser power of 4~mW, corresponding to $\|\vec{\alpha}\|^2 = 1$, and an experiment time of $t = 1~\mathrm{s}$.  The results are summarized in Fig.\ \ref{fig:efficiencyLAS_results}.  We observe that $\eta_{L}$ has a peak value of about 15\% at $d = 9.3$ for the default setting of DCR = 1/ms.  Higher dark count rates give larger peak efficiencies and at larger values of $d$.  For much larger values of $d$, corresponding to lower incident photon fluxes, the inferred detection efficiency goes over unity and becomes invalid, as detections are now dominated by dark counts.  Thus, the inferred detection efficiency is not an intrinsic property of the detector but, rather, is dependent upon the context of the measurement process.

\begin{figure}[ht]
\centering
\scalebox{1.0}{\includegraphics[width=\columnwidth]{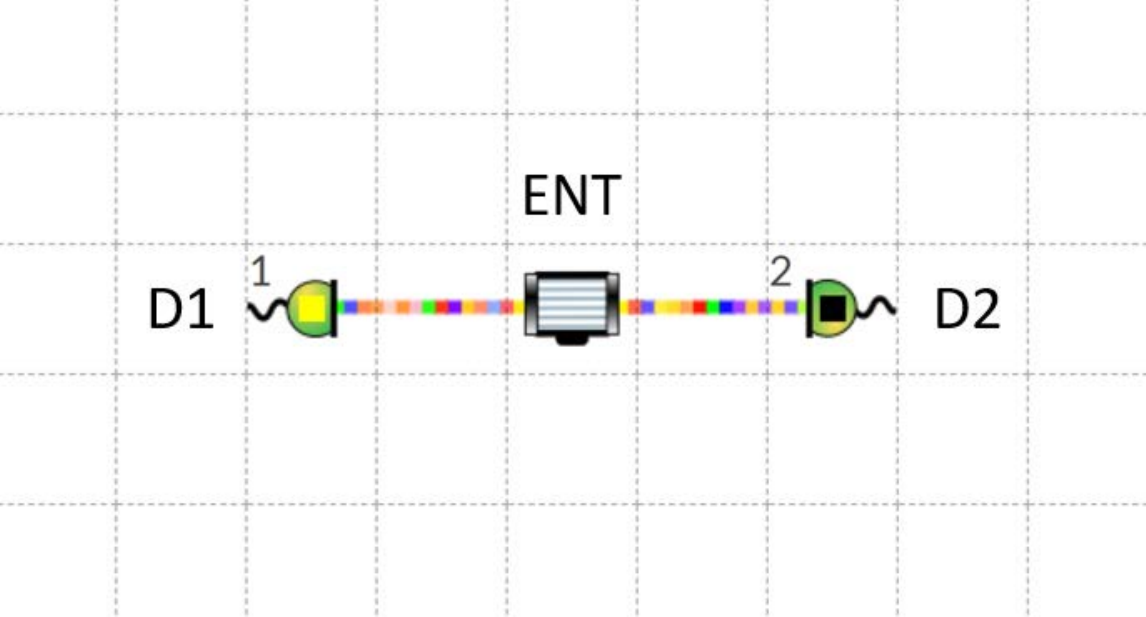}}
\caption{(Color online) Experimental setup for measuring detection efficiency using an entanglement source (ENT) and two detectors (D1 and D2).}
\label{fig:efficiencyENT_setup}
\end{figure}

An alternative scheme for measuring detection efficiency is to use a heralded entanglement source.  The modified experiment is shown in Fig.\ \ref{fig:efficiencyENT_setup}, where we have replaced the laser with an entanglement source (ENT), removed the NDF, and use two detectors, D1 and D2.  A detection on D1 is construed to herald the presence of a photon at D2.  Let $N_1$ denote the number of counts for which there is a detection on D1 but not D2.  Similarly, let $N_{12}$ denote the number of coincident counts on D1 and D2.  The inferred detection efficiency may then be defined as 
\begin{equation}
\eta_{E} = \frac{N_{12}}{N_1 + N_{12}} \; .
\end{equation}

This experiment was also implemented in VQOL and run for different values of squeezing strength $r$ and dark count rates.  In contrast to $\eta_{L}$, we find that $\eta_{E}$ approaches (but does not exceed) unity as $r$ becomes large.  There is also a dependence on DCR, similar to $\eta_{L}$ but not as dramatic.  Again, we find that the inferred detection efficiency is not an intrinsic property of the detector itself and may even appear arbitrarily close to unity under the right measurement conditions.  This, of course, is merely an artificiality born of the amplification that results from a high squeezing strength.

\begin{figure}[ht]
\centering
\scalebox{1.0}{\includegraphics[width=\columnwidth]{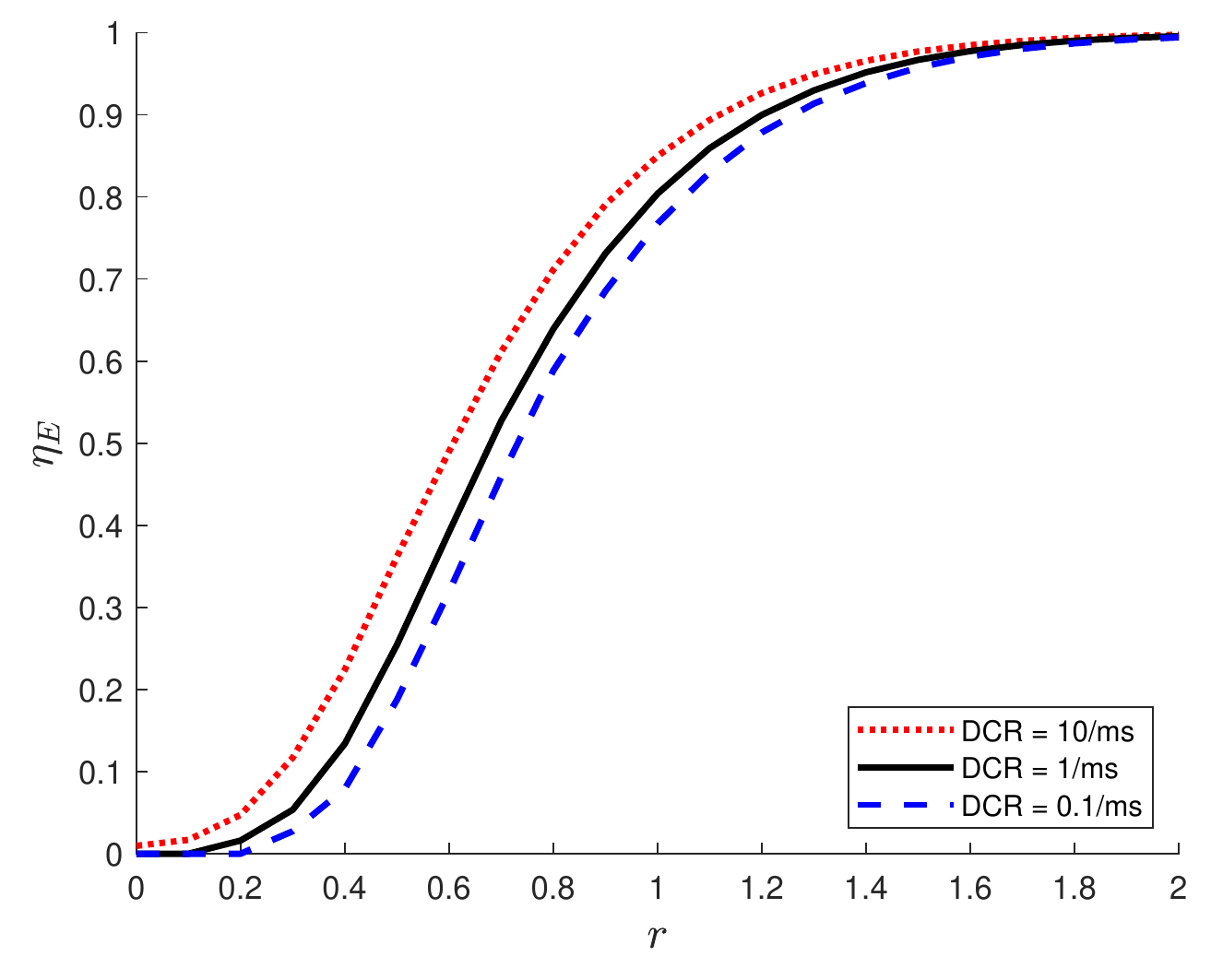}}
\caption{(Color online) Plot of inferred efficiency $\eta_{E}$ versus squeezing strength $r$ for different values of the dark count rate (DCR).}
\label{fig:efficiencyENT_results}
\end{figure}


\subsection{The Born Rule}

The Born rule provides the fundamental connection between quantum theory and observations.   Here we illustrate three experiments that may be performed in VQOL for studying the Born rule.  The first uses an attenuated laser and polarizing filter with a single detector.  The second replaces the polarizer with a polarizing beam splitter and a pair of detectors.  The third and final examples uses heralded detector with an entanglement source in place of the attenuated laser.

\subsubsection{Born Rule using a Laser and Polarizer}

\begin{figure}[ht]
\centering
\scalebox{1.0}{\includegraphics[width=\columnwidth]{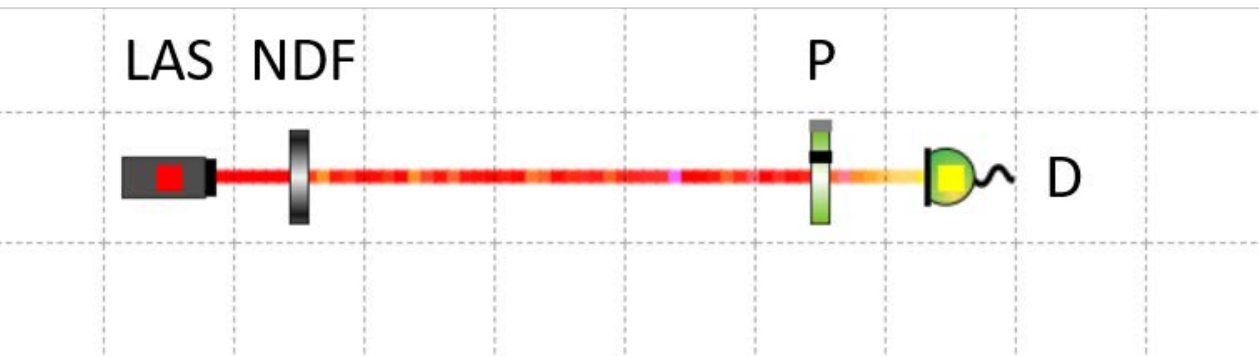}}
\caption{(Color online) Experimental setup for the Born rule using a laser (LAS), neutral density filter (NDF), polarizer (P), and detector (D).}
\label{fig:Born_setup}
\end{figure}

In this example we consider a simple experiment in which we prepare a laser (LAS) in the $\ket{H}$ polarization state and attenuate it using a neutral density filter (NDF), as shown in Fig.\ \ref{fig:Born_setup}.  A polarizer (P), set to an angle $\theta$ and phase $\phi$, is placed before a detector (D).  The polarizer has the effect of both attenuating the light and changing its polarization.  The Jones vector of the light exiting the polarizer is given by
\begin{equation}
\begin{pmatrix} a_H \\ a_V \end{pmatrix} = 10^{-d/2} \alpha \cos\theta \begin{pmatrix} \cos\theta \\ e^{i\phi} \sin\theta \end{pmatrix} + \sigma_0 \begin{pmatrix} z_H \\ z_V \end{pmatrix} \; ,
\end{equation}
where $\alpha = 10^5$ is the default amplitude of the laser and $d = 10$ is the default optical density of the NDF.  In accordance with Malus's law, the intensity varies as $\cos^2\theta$.

From the quantum perspective, we may view this experiment as the preparation of the $\ket{H}$ quantum state and subsequent measurement of the $\ket{\psi}$ state, where
\begin{equation}
\ket{\psi} = \begin{pmatrix} \cos\theta \\ e^{i\phi} \sin\theta \end{pmatrix} \; .
\end{equation}
For a single photon, the Born rule predicts that the probability of a detection is $|\braket{\psi|H}|^2 = \cos^2\theta$.  Although VQOL does not have a single-photon source, the laser and NDF provide a weak coherent state that may serve as a suitable approximation.  If the optical density $d$ is large, then only the single-photon and vacuum components will be significant.

\begin{figure}[ht]
\centering
\scalebox{1.0}{\includegraphics[width=\columnwidth]{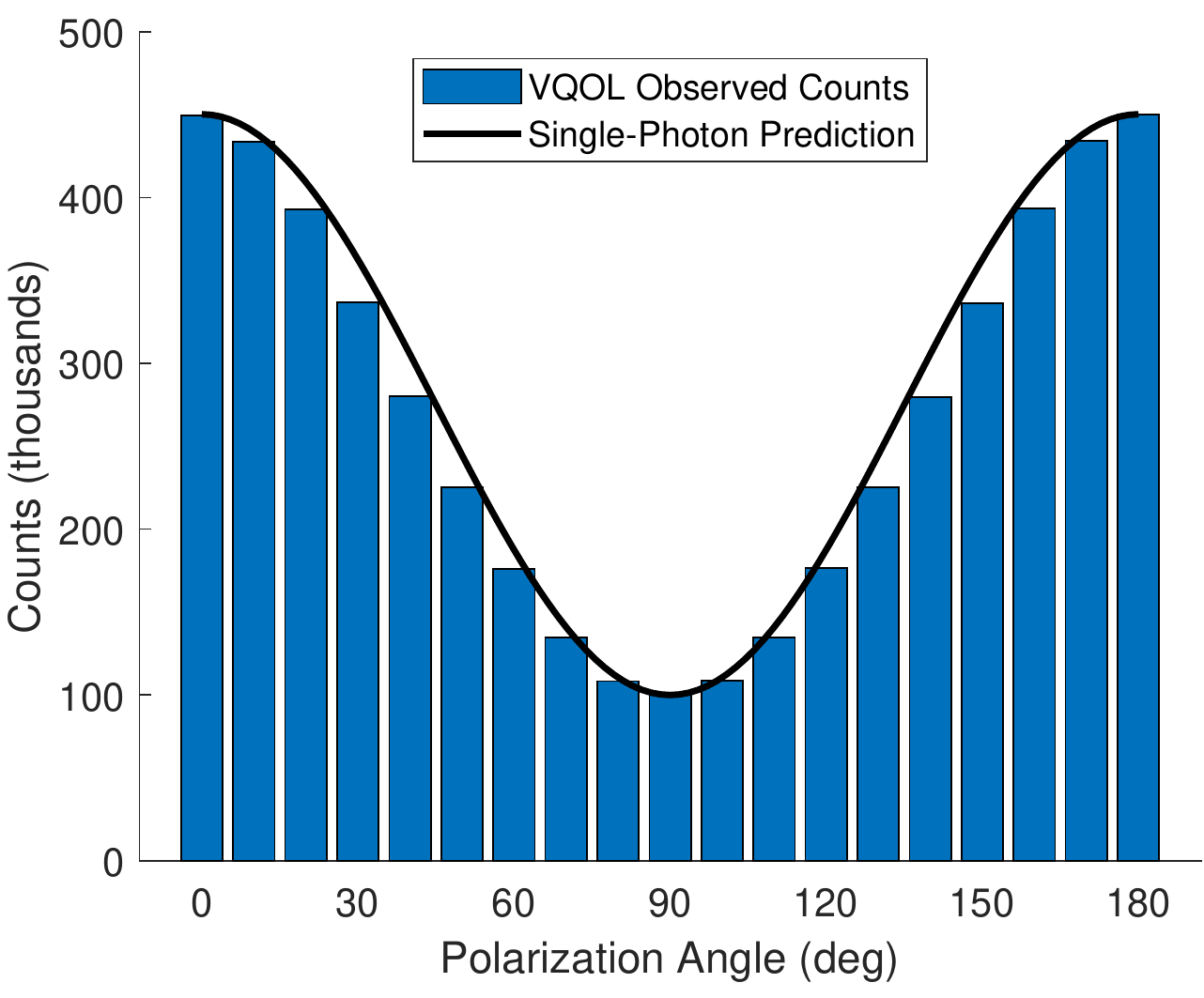}}
\caption{(Color online) Detector counts (in thousands) versus the polarization angle $\theta$ (in degrees) for a DCR of 100/ms.  The blue bars are measured results, while the black curve is the Born rule prediction for an ideal single-photon source.}
\label{fig:Born_results}
\end{figure}

In Fig.\ \ref{fig:Born_results} we have shown the detection counts for a one-second experiment using a dark count rate of 100/ms and varying the polarizer angle $\theta$ while keeping $\phi = 0^\circ$ fixed.  The results are qualitatively similar to those of Malus's law, though we note that the number of counts at $\theta = 90^\circ$ is not zero, due to the presence of dark counts.

The connection to probabilities is more subtle.  The probability of an event is generally estimated as the ratio of the number of occurrences of the event to the number of trials.  Experimentally, we measure the former, but the latter is unknown.  (Of course, in VQOL we actually \emph{do} know the number of trials, $10^6$ in this case, but this is a detail hidden in the implementation and not something experimentally available.)  A common strategy used by experimentalists is to subtract the dark counts, estimated here by the minimum number of counts, and rescale by the new maximum number of counts.  Other strategies, such as fitting to a parametric curve, work similarly.  Although this is common practice, it is important to understand that there is no ``correct'' strategy for associating counts with probabilities and that all such strategies incorporate some modeling assumptions.

In this example we have taken a rather large dark count rate.  Lower dark count rates entail higher detection thresholds, for which the approximation of Eqn.\ (\ref{eqn:approximation}) may therefore fail to hold, exhibiting deviations from the single-photon prediction.  A similar effect is observed if $d$ is too small and, hence, the laser too bright.  Although good agreement can be found with high attenuation and low dark counts, the rate of detections in this regime may be quite low.  The example we have shown here works well for a series of short, 1-ms experiments that can be run quickly and give good agreement under standard normalization.

\subsubsection{Born Rule using a Laser and PBS}

\begin{figure}[ht]
\centering
\scalebox{0.95}{\includegraphics[width=\columnwidth]{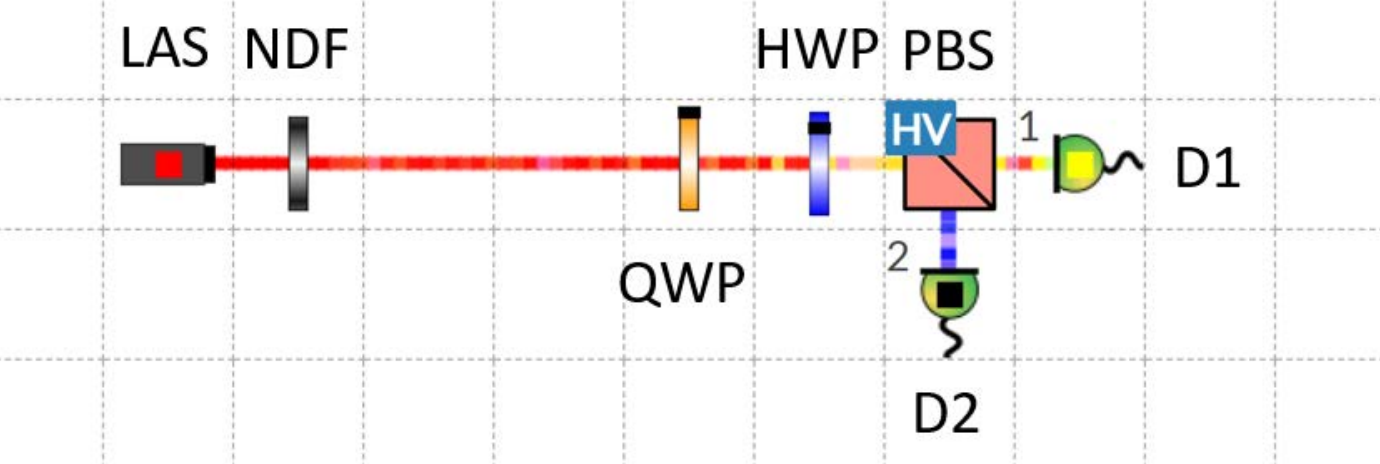}}
\caption{(Color online) Alternative experimental setup for the Born rule using a laser (LAS), neutral density filter (NDF), quarter-wave plate (QWP), half-wave plate (HWP), polarizing beam splitter (PBS) and two detectors (D1 and D2).}
\label{fig:BornAgain_setup}
\end{figure}

A variation on the previous experiment replaces the polarizer and single detector with a quarter-wave plate (QWP) and half-wave plate (HWP), followed by a polarizing beam splitter (PBS) and a pair of detectors (D1 and D2), as shown in Fig.\ \ref{fig:BornAgain_setup}.  The QWP and HWP allow one to rotate to any desired measurement basis.  (Alternatively, one may view this setup as the preparation of an arbitrary polarization state, which is subsequently measured in the $H/V$ basis.)  Let $\theta/2$ denote the fast-axis angle of the HWP.  (For this example, we shall suppose the fast-axis angle of the QWP is zero.)  The light entering the PBS is therefore linearly polarized with a polarization angle of $\theta$.  As before, we take $d = 10$ for the NDF and configure the two detectors to have a dark count rate of 100/ms.

Experimentally, we observe three possible outcomes: a single detection on D1, a single detection on D2, or a double detection on D1 and D2.  (Non-detection events are unobservable.)  Let $N_1(\theta), N_2(\theta), N_{12}(\theta)$ denote the number of counts for each of these mutually exclusive events for a given $\theta$.  A standard normalization procedure is to divide by the total number of single counts, ignoring the multi-detection events.  Thus, we define the probability
\begin{equation}
p_1(\theta) = \frac{N_1(\theta)}{N_1(\theta)+N_2(\theta)}
\end{equation}
and compare this with the single-photon Born rule prediction of $q_1(\theta) = \cos^2\theta$.

Unlike the single-detector case of the previous section, there is no guarantee that the maximum probability will be one nor that the minimum will be zero, again, due to dark counts.  A common remedy is, again, to subtract the estimated dark counts.  This leads to the alternative normalization
\begin{equation}
p'_1(\theta) = \frac{N_1(\theta) - \min N_1}{N_1(\theta) - \min N_1 + N_2(\theta) - \min N_2} \; .
\end{equation}
One can also estimate the number of dark counts directly from the dark count rate and experiment time, but for small sample sizes this may result in probability estimates that are negative or greater than one.

\begin{figure}[ht]
\centering
\scalebox{1.0}{\includegraphics[width=\columnwidth]{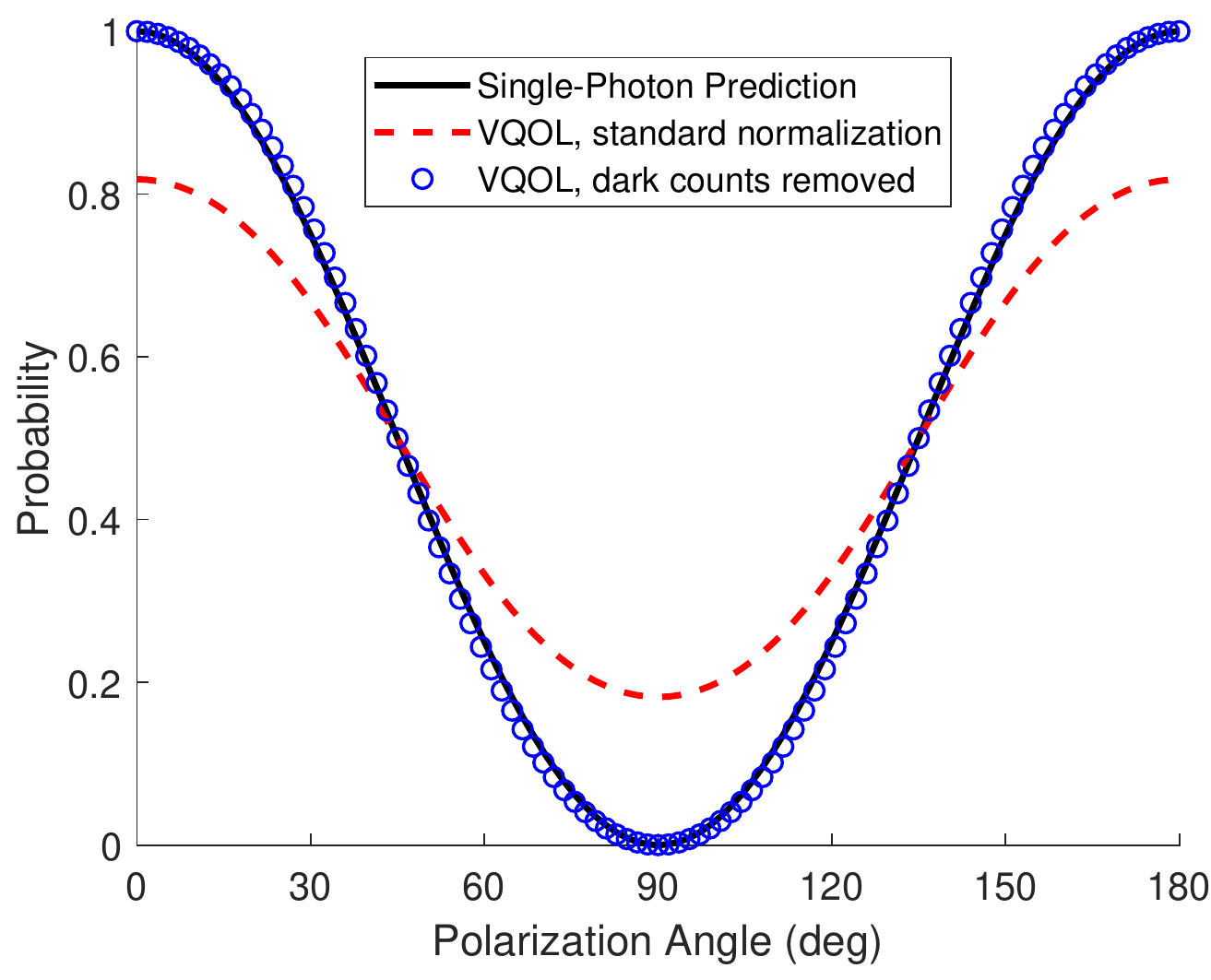}}
\caption{(Color online) Plot of probability versus the polarization angle $\theta$ (in degrees) for the experiment of Fig.\ \ref{fig:BornAgain_setup}.  The black curve is the ideal single-photon quantum prediction $q_1(\theta)$.  The red dashed curve is the VQOL result $p_1(\theta)$ under standard normalization, and the blue circles are the VQOL results $p'_1(\theta)$ with estimated dark counts removed.}
\label{fig:BornAgain_results}
\end{figure}

In Fig.\ \ref{fig:BornAgain_results} we show VQOL results for $p_1(\theta)$, using the standard normalization, and $p'_1(\theta)$, with dark counts removed, and compare these against $q_1(\theta)$.  Under the standard normalization, the visibility (i.e., the relative difference between minimum and maximum values) is rather low.  Subtracting the estimated dark counts does, however, give quite good agreement.  Using larger values of $d$ gives better agreement for $p'_1(\theta)$ but poorer visibility for $p_1(\theta)$.  On the other hand, using lower dark count rates will give better visibility but poorer agreement with $q_1(\theta)$.

\subsubsection{Born Rule using an Entanglement Source}

\begin{figure}[ht]
\centering
\scalebox{1.0}{\includegraphics[width=\columnwidth]{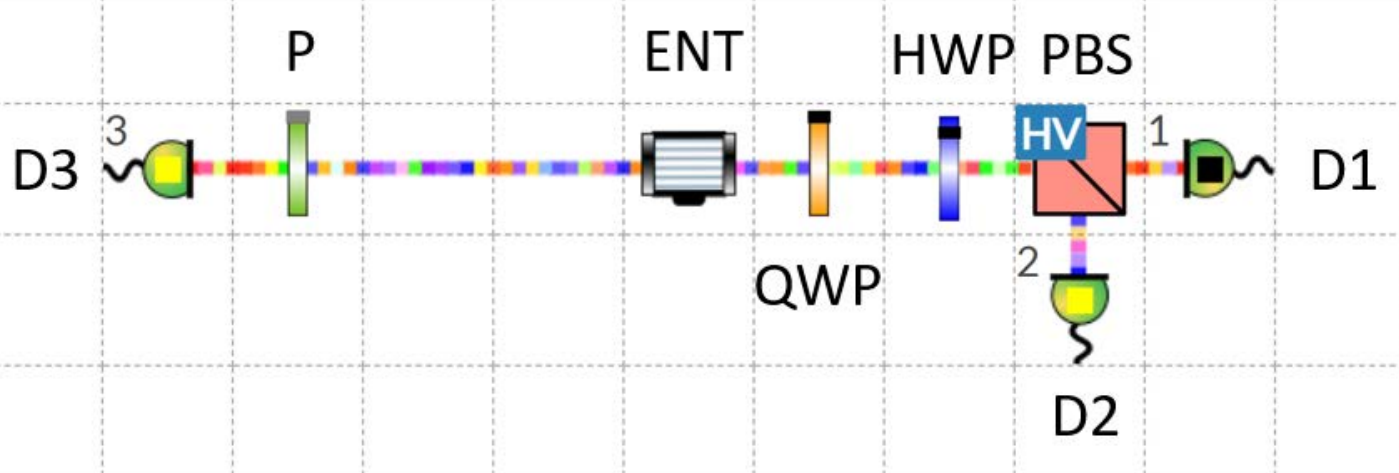}}
\caption{(Color online) Experimental setup for the Born rule using heralding of an entanglement source (ENT) by a horizontal polarizer (P) and third detector (D3).}
\label{fig:BornEntangled_setup}
\end{figure}

Our third example uses a type-I entanglement source (ENT), a polarizer (P), and a third detector (D3) in place of the laser and NDF, as shown in Fig.\ \ref{fig:BornEntangled_setup}.  The polarizer is set to transmit horizontal light, so a detection on D3 should herald the presence (or statistical propensity) of horizontally polarized light on the other side.  Thus, we post-select on events for which there is a detection on D3 \emph{and} a single detection on either D1 or D2.  Note that the distance from the entanglement source to each detector must be equal in order to have proper coincident detections.  Denoting the number of such detections by $N_{13}(\theta)$ and $N_{23}(\theta)$, the estimated probability under standard normalization is
\begin{equation}
p_1(\theta) = \frac{N_{13}(\theta)}{N_{13}(\theta)+N_{23}(\theta)} \; .
\end{equation}
As before, we may improve agreement by removing the estimated dark counts to obtain $p'_1(\theta)$.

\begin{figure}[ht]
\centering
\scalebox{1.0}{\includegraphics[width=\columnwidth]{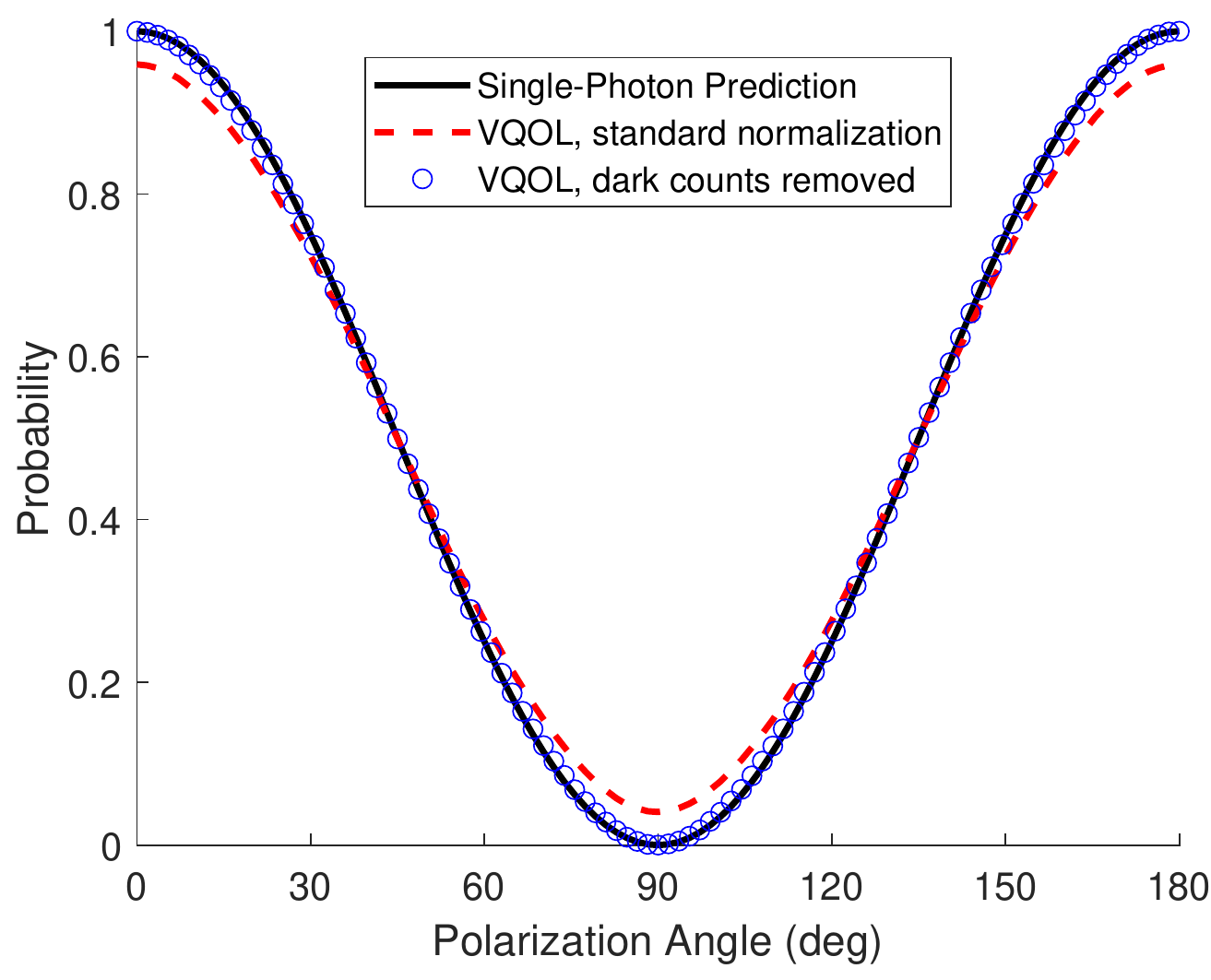}}
\caption{(Color online) Plot of probability versus the polarization angle $\theta$ (in degrees) using the entanglement source of Fig.\ \ref{fig:BornEntangled_setup}.  The black curve is the ideal single-photon quantum prediction $q_1(\theta)$.  The red dashed curve is the VQOL result $p_1(\theta)$ under standard normalization, and the blue circles are the VQOL results $p'_1(\theta)$ with estimated dark counts removed.}
\label{fig:BornEntangled_results}
\end{figure}

The results are shown in Fig.\ \ref{fig:BornEntangled_results}, where we have reverted to the default dark count rate of 1/ms for each of the detectors when using the entanglement source.  As can be seen, the heralded entanglement source provides a much better approximation to a single-photon state than does a weak coherent state.  Although a default value of $r = 1$ was used for the squeezing strength of the entanglement sources, somewhat smaller values may provide a better approximation to a single-photon state, as this suppresses higher photon number states in the corresponding squeezed state.  Similarly, somewhat lower dark count rates may provide better suppression of dark counts from the vacuum states.

\subsection{Quantum State Tomography}

A single measurement cannot uniquely determine a quantum state.  If, however, one has access to a method of preparing the same state multiple times (or multiple instances of the same prepared state), then the state may be uniquely determined by performing quantum state tomography (QST).

QST provides an estimate of the quantum state in terms of a density matrix $\mtx{\rho}$, which can be used to represent either a pure or mixed state.  Given a polarization state $\ket{\psi}$, represented as a column vector, the corresponding (pure) density matrix is the projection matrix $\ket{\psi}\bra{\psi}$, where $\bra{\psi}$ is the Hermitian conjugate of $\ket{\psi}$.  The density matrix for a general mixed state may be represented as a probabilistically weighted sum of pure-state density matrices.  As such, the density matrix $\mtx{\rho}$ must have a trace of one (i.e., $\mathrm{Tr}[\mtx{\rho}] = 1$) and be positive semi-definite (i.e., have no negative eigenvalues).  By construction it is also Hermitian.

QST may be performed using the Hilbert-Schmidt decomposition of a density matrix onto a complete, orthonormal matrix basis.  For polarization states, a common choice is to use the Pauli matrices $\mtx{I}, \mtx{X}, \mtx{Y}, \mtx{Z}$, where $\mtx{I}$ is the $2\times2$ identity and
\begin{equation}
\mtx{X} = \begin{pmatrix} 0 & 1 \\ 1 & 0 \end{pmatrix} , \;
\mtx{Y} = \begin{pmatrix} 0 & -i \\ i & 0 \end{pmatrix} , \;
\mtx{Z} = \begin{pmatrix} 1 & 0 \\ 0 & -1 \end{pmatrix} .
\end{equation}
The density matrix $\mtx{\rho}$ may then be written
\begin{equation}
\mtx{\rho} = \frac{1}{2} \mtx{I} + \frac{\mathrm{Tr}[\mtx{\rho}\mtx{X}]}{2} \, \mtx{X} + \frac{\mathrm{Tr}[\mtx{\rho}\mtx{Y}]}{2} \, \mtx{Y} + \frac{\mathrm{Tr}[\mtx{\rho}\mtx{Z}]}{2} \, \mtx{Z} \; .
\end{equation}
The traces correspond to expectation values and may be estimated experimentally by sample means.  With these three numbers, an estimate $\tilde{\mtx{\rho}}$ of $\mtx{\rho}$ may be constructed.  Note that $\tilde{\mtx{\rho}}$ is guaranteed to have unit trace and be Hermitian, but, depending upon the sample means one obtains, it may fail to be positive semi-definite.  This is an important consideration when performing QST and one that is familiar to experimenters \cite{Schmied2016}.

To measure $\mtx{X}$, $\mtx{Y}$, and $\mtx{Z}$ we measure in the $D/A$, $R/L$, and $H/V$ bases, respectively.  This can easily be done in VQOL using a polarizing beam splitter set to each of the three bases.  Alternatively, an $H/V$ PBS may be used in combination with a HWP and QWP to actively measure in the desired basis, which a more practical approach in real experiments.  Suppose we apply a QWP with fast-axis angle $\phi$, followed by a HWP with fast-axis angle $\theta$.  To measure in the $H/V$ basis, we may use $\theta = \phi = 0^\circ$.  The product of the two components acts like a phase retarder (with a phase factor of $-i$ applied to the vertical component), but this does not affect the outcomes of the detectors, which are only sensitive to amplitude.  To measure in the $D/A$ basis we may use $\theta = 22.5^\circ$ and $\phi = 45^\circ$, while for the $R/L$ basis we may take $\theta = 22.5^\circ$ and $\phi = 90^\circ$.  Alternatively, we may use one detector and a polarizing filter set to transmit each of the six polarization states to be measured.  Specifically, one may use $\mtx{P}(0^\circ,0^\circ)$, $\mtx{P}(90^\circ,0^\circ)$, $\mtx{P}(45^\circ,0^\circ)$, $\mtx{P}(-45^\circ,0^\circ)$, $\mtx{P}(45^\circ,90^\circ)$, and $\mtx{P}(45^\circ,-90^\circ)$ to measure $H$, $V$, $D$, $A$, $R$, and $L$, respectively.

\begin{figure}[ht]
\centering
\scalebox{1.0}{\includegraphics[width=\columnwidth]{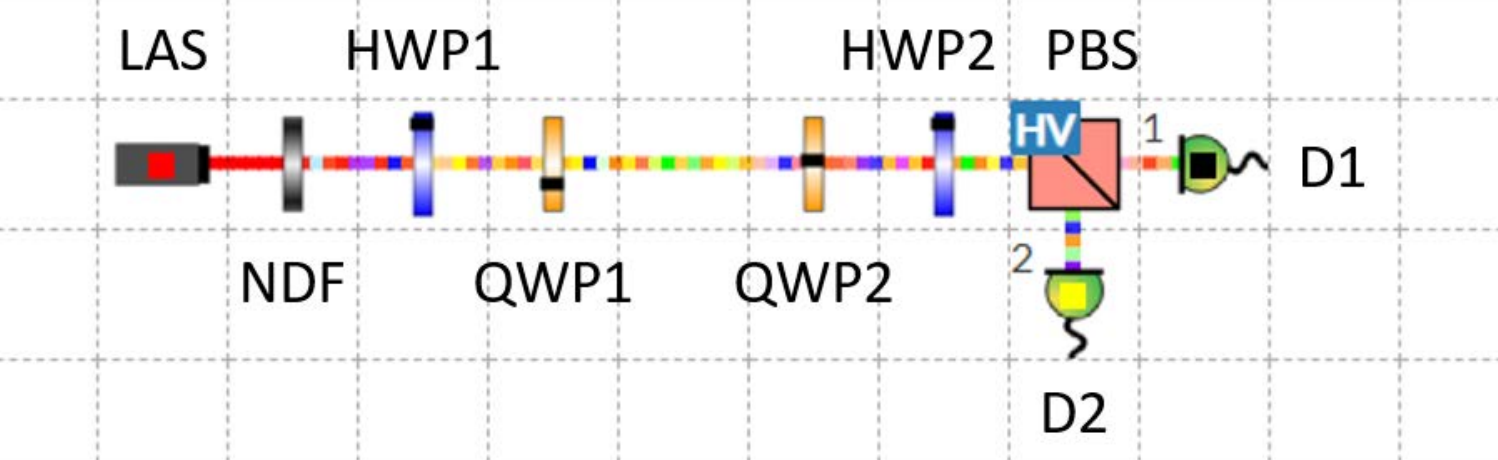}}
\caption{(Color online) Experimental setup for quantum state tomography experiment.}
\label{fig:QST}
\end{figure}

A typical experimental setup is show in Fig.\ \ref{fig:QST}.  A laser (LAS), neutral density filter (NDF), half-wave plate (HWP1) and quarter-wave plate (QWP1) are used to prepared the desired polarization state.  For measurement, QWP2 and HWP2 are used to rotate to one of the three bases, and the polarizing beam splitter (PBS) and two detectors (D1 and D2) are used to perform the final measurements.  For a given basis selection, the expectation value of the corresponding Pauli matrix observable may be estimated by $(N_1-N_2)/(N_1+N_2)$, where $N_1$ is the number of single detections on D1 and similarly for $N_2$.  (Optionally, one may first subtract the estimated number of dark counts from $N_1$ and $N_2$.)

The quality of the estimated density matrix varies greatly with the level of attenuation, as parameterized by the optical density $d$ of the NDF.  This may be quantified by the fidelity, defined to be
\begin{equation}
F = \bra{\psi} \tilde{\mtx{\rho}} \ket{\psi} \; ,
\end{equation}
where $\ket{\psi}$ is the ideal prepared quantum state.  Generally, larger values of $d$ result in poorer fidelity (i.e., smaller values of $F$) due to contamination from dark counts.  Smaller values of $d$, corresponding to stronger light, yield higher fidelities but may also lead to an invalid estimate of the density matrix that is not positive semi-definite.  In such cases, the fidelity may actually be greater than one!

\begin{figure}[ht]
\centering
\scalebox{1.0}{\includegraphics[width=\columnwidth]{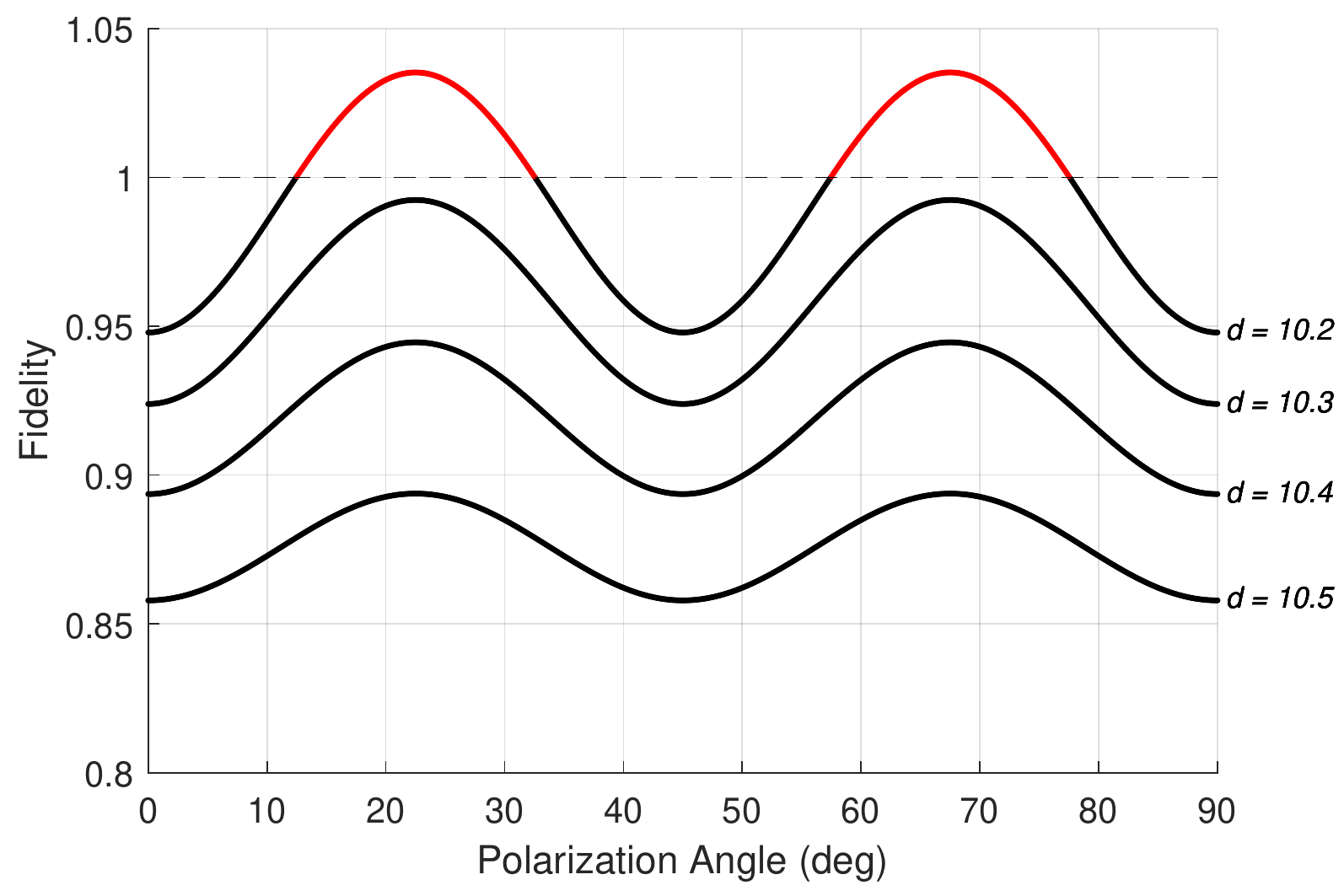}}
\caption{(Color online) Plot of fidelity versus polarization angle for several values of the optical density, $d$, of the NDF.  The horizontal axis shows twice the fast-axis angle of HWP1.  The dashed line shows the theoretical upper bound for the fidelity, and the red (gray) portions of the curve show where the fidelity exceeds unity.}
\label{fig:fidelity}
\end{figure}

These observations are summarized in Fig.\ \ref{fig:fidelity}, where we have plotted the fidelity for various prepared polarization states and optical densities.  Otherwise, all components are configured to their default settings.  We see that polarization states that fall along one of the measured bases tend to have the lowest fidelity but yet also tend to yield valid density matrices.  (However, changes in $\phi$ will translate these curves horizontally.)  For intermediate polarizations and strong light, the fidelity may rise above unity, corresponding to cases in which the estimated density matrix fails to be positive semi-definite and is therefore invalid.

\subsection{Mach-Zehnder Interferometer}

\begin{figure}[ht]
\centering
\scalebox{1.0}{\includegraphics[width=\columnwidth]{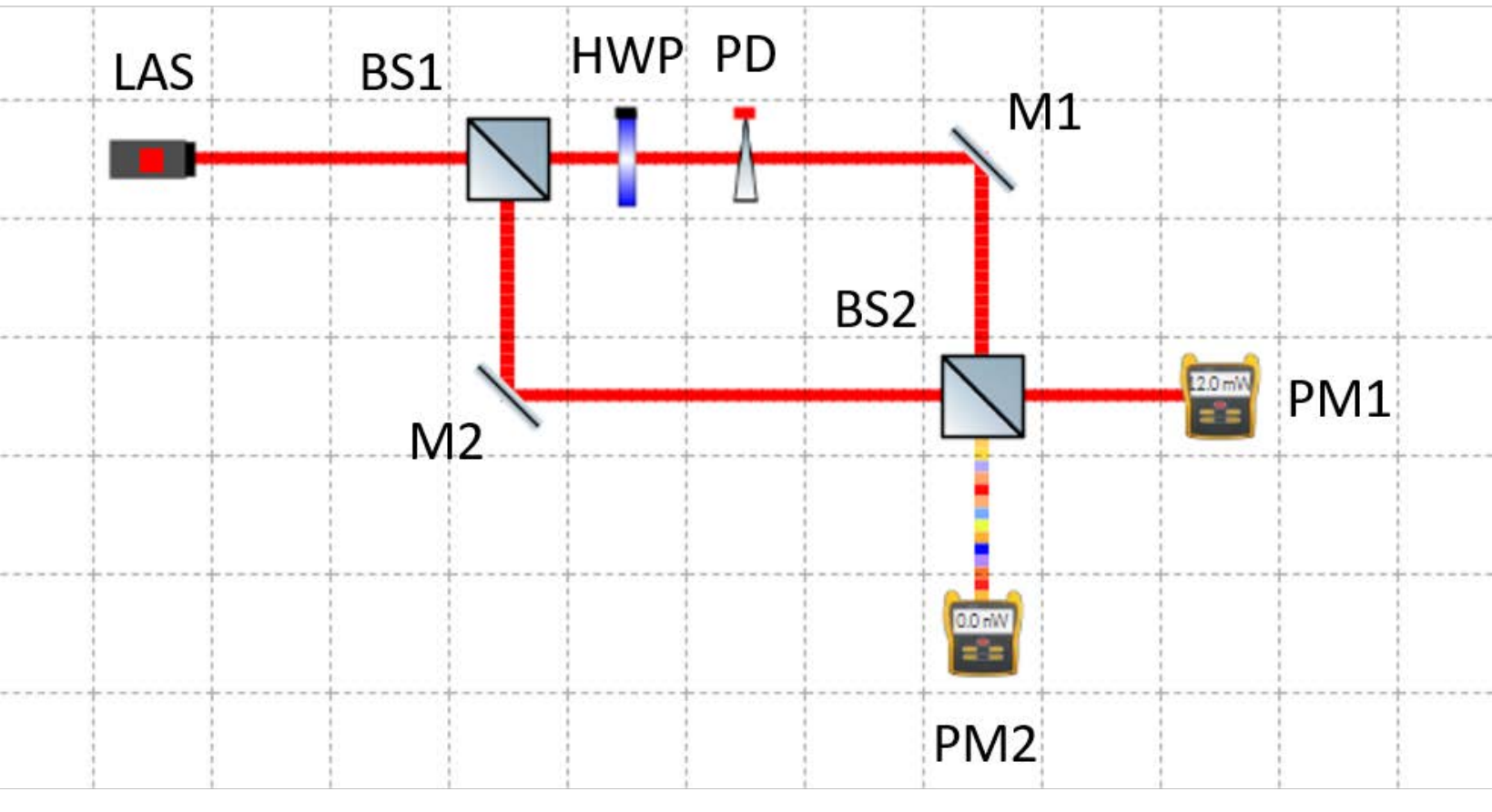}}
\caption{(Color online) Experimental setup for classical Mach-Zehnder interferometer experiment with a laser (LAS), two 50/50 beam splitters (BS1, BS2), two mirrors (M1, M2), a half-wave plate (HWP), phase delay (PD), and two power meters (PM1, PM2).}
\label{fig:CMZI}
\end{figure}

A Mach-Zehnder interferometer is a simple optical device that may be used to illustrate the dual wave/particle behavior of light.  A typical setup is shown in Fig.\ \ref{fig:CMZI}.  The interferometer consists of a pair of mirrors (M1, M2) and 50/50 beam splitters (BS1, BS2).  Along one path of the interferometer are placed a half-wave plate (HWP), with a fast-axis angle of $\theta$, and a phase delay (PD), with a phase angle $\phi$.  We will consider three variants of this experiment: one with classical light, one with quantum light, and a third in which the second beam splitter is removed.

\subsubsection{Classical Mach-Zehnder Interferometer}

In the classical version of the experiment, we have a laser providing horizontally polarized light as input to the interferometer and a pair of power meters (PM1 and PM2) to measure the output.  The Jones vectors for the light entering BS1 are therefore
\begin{equation}
\vec{a} = \begin{pmatrix} a_H \\ a_V \end{pmatrix} = \alpha \begin{pmatrix} 1 \\ 0 \end{pmatrix}  + \sigma_0 \begin{pmatrix} z_{1H} \\ z_{1V} \end{pmatrix} \; ,
\end{equation}
for the laser light, and
\begin{equation}
\vec{b} = \begin{pmatrix} b_H \\ b_V \end{pmatrix} = \sigma_0 \begin{pmatrix} z_{2H} \\ z_{2V} \end{pmatrix} \; ,
\end{equation}
for the downward vacuum modes.  Exiting the first beam splitter, BS1, these become
\begin{equation}
\begin{bmatrix} \vec{a} \\ \vec{b} \end{bmatrix}
\xrightarrow{\sf BS1}
\frac{1}{\sqrt{2}} \begin{bmatrix} \vec{a}+\vec{b} \\ \vec{a}-\vec{b} \end{bmatrix} \; .
\end{equation}
Following HWP, PD, M1, and M2 we have
\begin{equation}
\frac{1}{\sqrt{2}}
\begin{bmatrix} a_H - b_H \\ a_V - b_V \\ e^{i\phi} (a_H+b_H) \cos2\theta + e^{i\phi} (a_V+b_V) \sin2\theta \\ e^{i\phi} (a_H+b_H) \sin2\theta - e^{i\phi} (a_V+b_V) \cos2\theta \end{bmatrix} .
\label{eqn:openconfig}
\end{equation}
Finally, following the second beam splitter, BS2, we have
\begin{multline}
a'_H = \left( \frac{1+e^{i\phi} \cos2\theta}{2} \right) a_H
- \left( \frac{1-e^{i\phi} \cos2\theta}{2} \right) b_H \\
+ \frac{e^{i\phi}}{2} (a_V + b_V) \sin2\theta
\end{multline}
and
\begin{multline}
b'_H = \left( \frac{1-e^{i\phi} \cos2\theta}{2} \right) a_H 
- \left( \frac{1+e^{i\phi} \cos2\theta}{2} \right) b_H \\
- \frac{e^{i\phi}}{2} (a_V + b_V) \sin2\theta \; .
\end{multline}
Similar expressions hold for $a'_V$ and $b'_V$, namely,
\begin{multline}
a'_V = \left( \frac{1-e^{i\phi} \cos2\theta}{2} \right) a_V
- \left( \frac{1+e^{i\phi} \cos2\theta}{2} \right) b_V \\
+ \frac{e^{i\phi}}{2} (a_H + b_H) \sin2\theta
\end{multline}
and
\begin{multline}
b'_V = \left( \frac{1+e^{i\phi} \cos2\theta}{2} \right) a_V 
- \left( \frac{1-e^{i\phi} \cos2\theta}{2} \right) b_V \\
- \frac{e^{i\phi}}{2} (a_H + b_H) \sin2\theta \; .
\end{multline}

For $\theta = 0^\circ$ and $\phi = 0^\circ$ we see that $a'_H = a_H$, with the other three components being negligible.  Thus, PM1 reads the full power (i.e., 4.0 mW) and PM2 reads zero, due to constructive and destructive inteference, respectively, in the interferometer.  Changing $\phi$ alters this interference and, hence, the readings on each power meter.  Taking $\phi = 90^\circ$, for example, gives a balanced result, with each power meter reading 2.0 mW, while setting $\phi = 180^\circ$ reverses the roles of the two output beams.

Changing $\theta$ provides ``which-way'' path information, thereby also changing the interference and, hence, the readings on the two power meters.  In particular, when $\theta = 45^\circ$ the polarization in the upper arm flips to vertical and the two power meters become balanced.  With $\theta = 90^\circ$, the roles are reversed, with PM2 reading 4.0 mW and PM1 reading zero.  Note that although vacuum modes are present in the simulation they play no significant role relative to the much stronger laser light.  Hence, this may be viewed as a purely classical effect.


\subsubsection{Quantum Mach-Zehnder Interferometer}

\begin{figure}[ht]
\centering
\scalebox{1.0}{\includegraphics[width=\columnwidth]{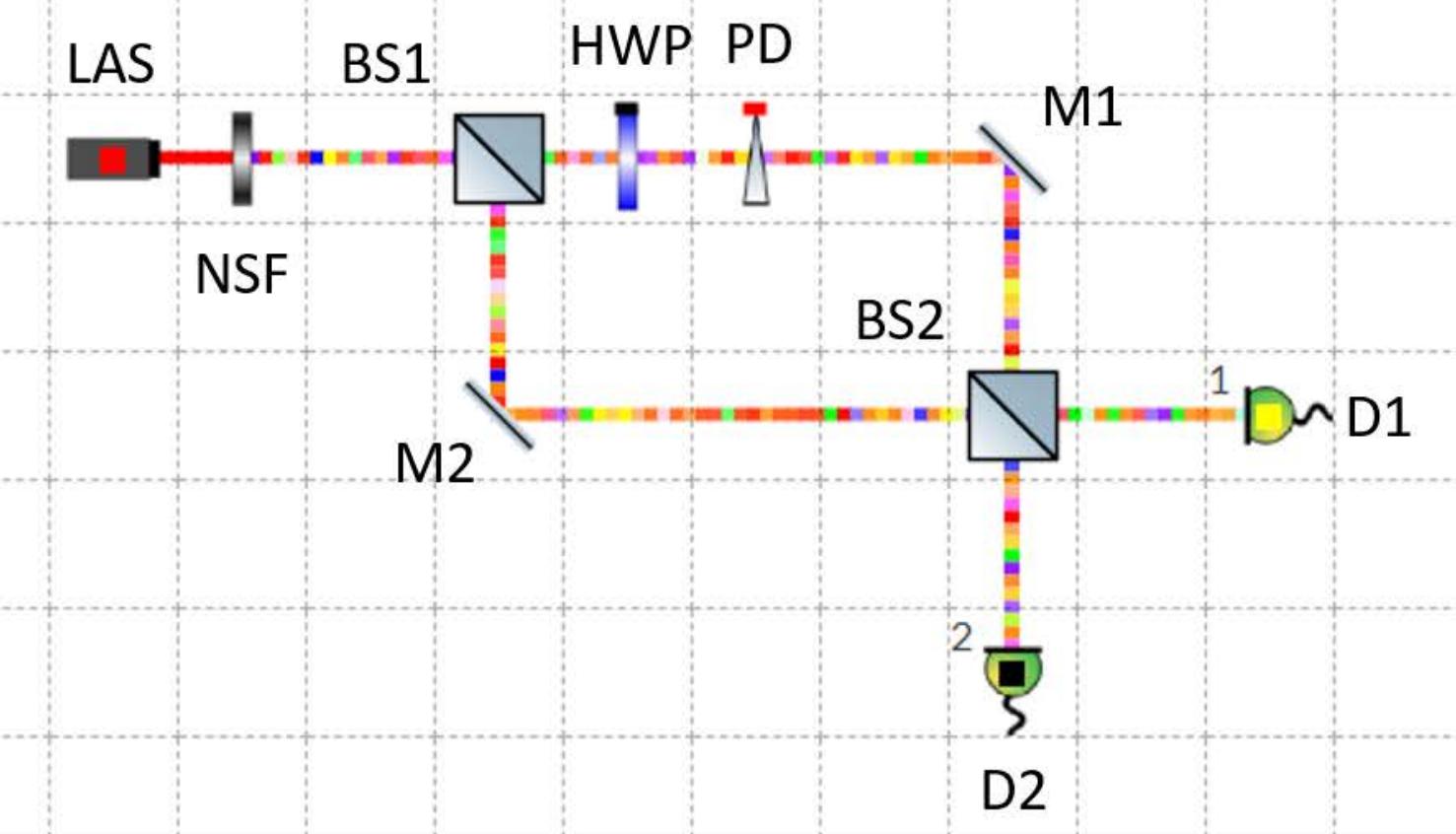}}
\caption{(Color online) Experimental setup for quantum Mach-Zehnder interferometer experiment.}
\label{fig:QMZI}
\end{figure}

To explore the quantum regime, we simply add a neutral density filter (NDF) and replace the power meters with a pair of detectors (D1 and D2), as shown in Fig.\ \ref{fig:QMZI}.  Qualitatively, the behavior is quite similar to that of the classical experiment, as the detector counts for weak coherent light are proportional to the classical intensity, once dark counts are subtracted.  The introduction of an NDF changes $\vec{a}$ to 
\begin{equation}
\vec{a} = \begin{pmatrix} a_H \\ a_V \end{pmatrix} = 10^{-d/2} \alpha \begin{pmatrix} 1 \\ 0 \end{pmatrix}  + \sigma_0 \begin{pmatrix} z_{1H} \\ z_{1V} \end{pmatrix} \; ,
\end{equation}
where $d$ is the optical density.  Otherwise, the expressions for $\vec{a}'$ and $\vec{b}'$ remain valid.

Let $D_1$ denote the set of events for which detector D1 clicks (irrespective of D2), and let $D_2$ be defined similarly, so
\begin{subequations}
\begin{align}
D_1 &= \Bigl\{ |a'_H| > \gamma ~\text{ or }~ |a'_V| > \gamma \Bigr\} \; , \\
D_2 &= \Bigl\{ |b'_H| > \gamma ~\text{ or }~ |b'_V| > \gamma \Bigr\} \; .
\end{align}
\end{subequations}
Let $p_1(\phi,\theta)$ and $p_2(\phi,\theta)$ denote the probabilities for the single-detection events $D_1 \cap \bar{D}_2$ and $\bar{D}_1 \cap D_2$, respectively.  Because $D_1$ and $D_2$ are independent,
\begin{subequations}
\begin{align}
p_1(\phi,\theta) &= \Pr[D_1] \bigl( 1 - \Pr[D_2] \bigr) \; , \\
p_2(\phi,\theta) &= \Pr[D_2] \bigl( 1 - \Pr[D_1] \bigr) \; .
\end{align}
\end{subequations}

The probabilities $\Pr[D_1], \Pr[D_2]$ can be computed directly from Eqn.\ (\ref{eqn:PrD}) using the expectation values of $a'_H, b'_H, a'_V, b'_V$, which are given, respectively, by
\begin{subequations}
\begin{align}
\alpha'_H &= \left( \frac{1+e^{i\phi}\cos2\theta}{2} \right) 10^{-d/2} \alpha \; , \\
\beta'_H &= \left( \frac{1-e^{i\phi}\cos2\theta}{2} \right) 10^{-d/2} \alpha \; , \\
\alpha'_V &= \frac{e^{i\phi}}{2} \sin2\theta \, 10^{-d/2} \alpha \; , \\
\beta'_V &= -\frac{e^{i\phi}}{2} \sin2\theta \, 10^{-d/2} \alpha \; .
\end{align}
\end{subequations}
These results may be compared to the ideal quantum predictions for a single-photon state, which are
\begin{subequations}
\begin{align}
q_1(\phi,\theta) &= \frac{1}{2} (1 + \cos\phi \cos2\theta) \; , \\
q_2(\phi,\theta) &= \frac{1}{2} (1 - \cos\phi \cos2\theta) \; .
\end{align}
\end{subequations}

\begin{figure}[ht]
\centering
\scalebox{1.0}{\includegraphics[width=\columnwidth]{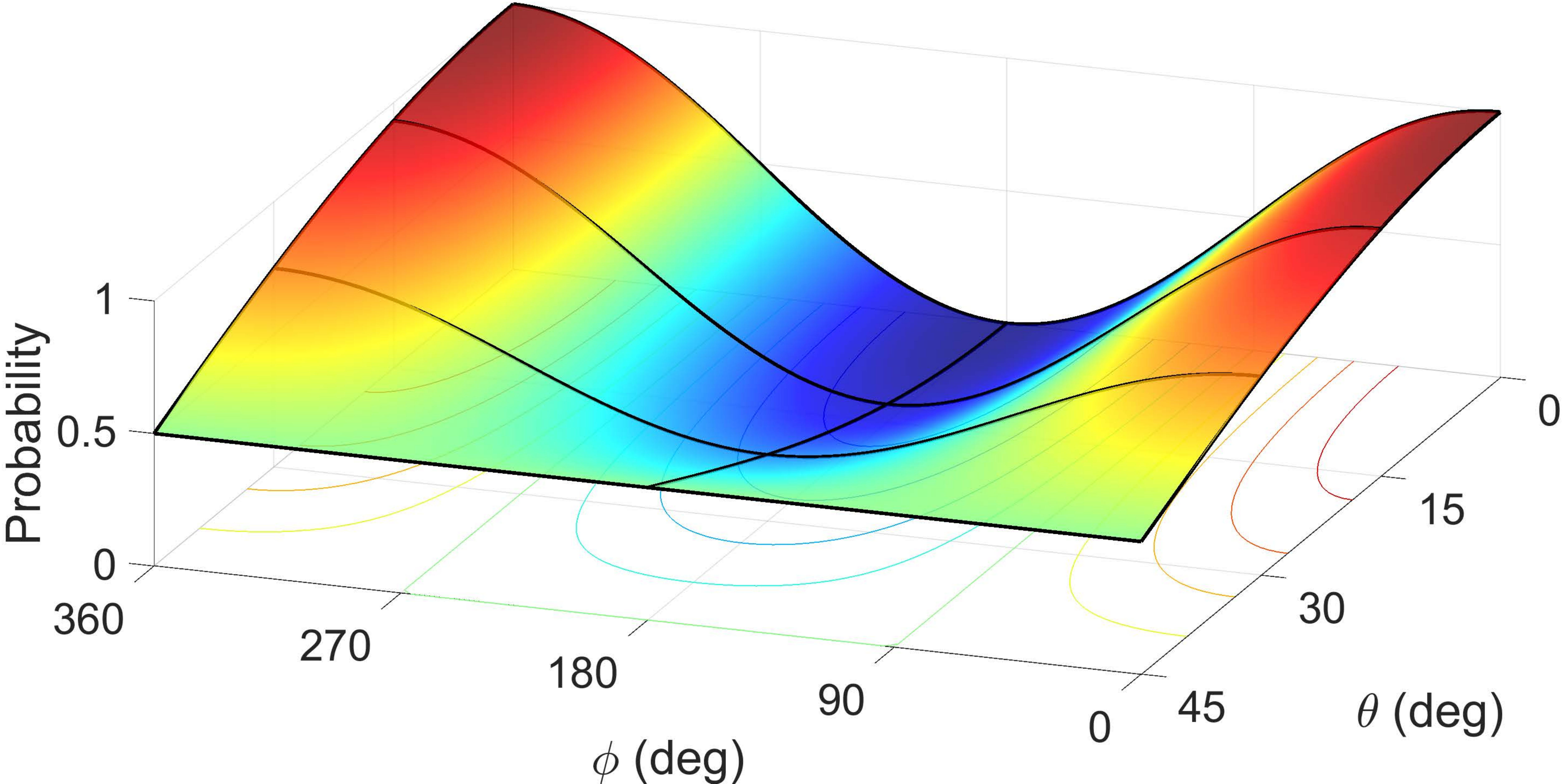}}
\caption{(Color online) Surface plot of $p'_1(\phi,\theta)$ versus $\phi$ and $\theta$ using $d = 12$ .  The color scale ranges from dark blue (zero) to dark red (one).  The black curves are $q_1(\phi,\theta)$, the ideal quantum prediction for a single photon, for fixed values of either $\phi \in \{ 0^\circ, 180^\circ, 360^\circ \}$ or $\theta \in \{ 0^\circ, 15^\circ, 30^\circ, 45^\circ \}$.}
\label{fig:wave-particle}
\end{figure}

The probabilities $p_1, p_2$ will not generally match the single-photon quantum probabilities $q_1, q_2$ due to the presence of vacuum and higher photon number modes in the weak coherent state used in VQOL.  A better correspondence can be achieved, as before, by subtracting the estimated dark counts and rescaling by the total number of single detection events.  Specifically, we define
\begin{equation}
p'_1(\phi,\theta) = \frac{p_1(\phi,\theta) - \delta_1}{p_1(\phi,\theta)  - \delta_1 + p_2(\phi,\theta) - \delta_2}  \; ,
\end{equation}
where $\delta_1$ and $\delta_2$ are the dark count probabilities for D1 and D2, respectively, and $p'_2(\phi,\theta)$ is defined similarly.  An example is shown in Fig.\ \ref{fig:wave-particle} for the case $d = 12$ and a default dark count rate of 1/ms.  We note that the gradual introduction of which-way information as $\theta$ increases results in a commensurate reduction in the interference pattern until, at $\theta = 45^\circ$, it vanishes completely.  One can see the gradual morphing between wave-like and particle-like behavior exhibited in the experiment.

Using the default parameter setting of $d = 10$ will also result in a well-defined interference pattern, although it will not match the ideal quantum predictions quite as well.  Typical maximum count rates for this case are 43/ms.  On the other hand, using $d = 12$ results in small samples (about 1.08/ms) and poor visibility, as the maximum number of counts is not much larger than the dark count rate.  The best parameter settings to use will therefore depend on the pedagogical or investigative purposes of the experiment.


\subsubsection{Wheeler's Delayed-Choice Experiment}

Conceptual difficulties can arise when one thinks of light as a particle rather than a wave.  Common descriptions of the Mach-Zhender experiment are often phrased in particle language such as, ``a photon enters the first beam splitter and is either transmitted or reflected'' or perhaps the more nuanced but no less confusing, ``a photon enters the first beam splitter and is both transmitted and reflected.''  Although the two descriptions attempt to capture the particle and wave aspects of light, respectively, both are couched in a particle-centric language.

In VQOL there are no single-photon sources.  Although each time step may appear to represent a single photon, it more properly represents a short pulse of light over which the vacuum modes are coherent.  Setting the experiment time to a single time bin (i.e., 1 $\mu$s), one sees a single pulse split into two at the first beam splitter, which is not the behavior of a single particle.  It may be surprising, then, that what emerges from the second beamsplitter is not a single, recombined pulse but, rather, \emph{two} pulses.  Although not apparent in the visual animation, this is due to a second, vacuum state being introduced in the top input port of the first beam splitter.

The contrast between particle-like and wave-like behavior is often illustrated using John Wheeler's delayed-choice \emph{gedanken} experiment \cite{Ma2016}.  In the experiment, the choice of whether to use BS2 is delayed until after the light passes through BS1.  If one has in mind the notion that the light must ``decide'' to behave either as a particle or a wave before entering BS1, then it must somehow nonlocally reconsider its decision when faced with the presence or absence of BS2.  This odd state of affairs can be understood more clearly in the context of VQOL, for which light is always a wave.  If BS2 is absent, the two beams interfere briefly at the point of intersection then resume their original polarizations.  The particle-like behavior that results from this open configuration can be seen in the symmetry of the four modes in Eqn.\ (\ref{eqn:openconfig}).  If BS2 is present, the two beams interfere in such a manner as to alter the amplitude or polarization of each beam.  Each pulse therefore interacts locally with the components and other beams it encounters.


\subsection{Photon Anticorrelation}

\begin{figure}[ht]
\centering
\scalebox{1.0}{\includegraphics[width=\columnwidth]{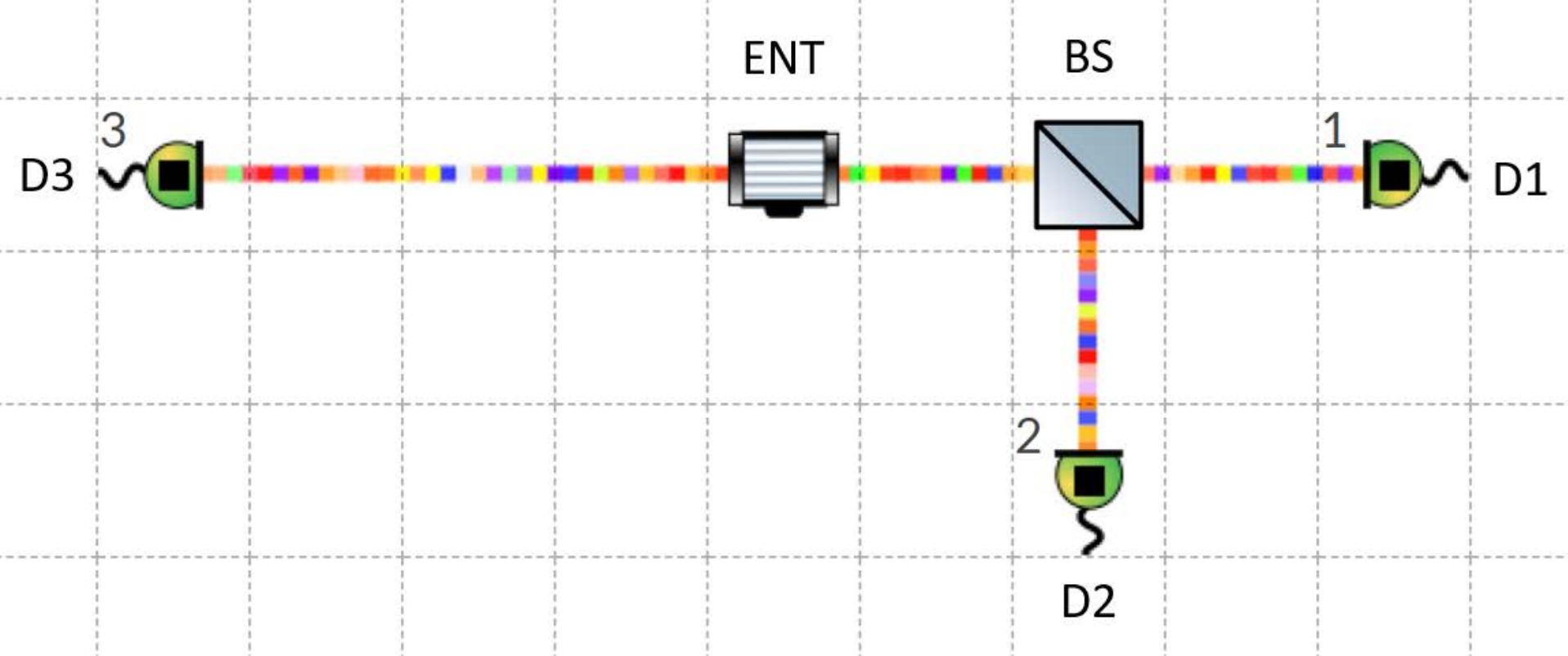}}
\caption{(Color online) Experimental setup for the photon anticorrelation experiment.}
\label{fig:Grangier_setup}
\end{figure}

This experiment examines the peculiar particle-like behavior of light and is modeled after the 1986 experiment by Grangier, Roger, and Aspect \cite{Grangier1986}, a modern version of which is described in Ref.\ \cite{Thorn2004}.  A typical experimental setup is shown in Fig.\ \ref{fig:Grangier_setup}.  An entanglement source (ENT) produces two beams of light, traveling left and right.  The right beam enters a beam splitter (BS), and the outputs are directed to detectors D1 and D2.  The left beam is directed to a third detector, D3, and is used for heralding.

If we run the experiment, we typically see detections on either D1 or D2, when there is one at all, but rarely on both.  This suggests particle-like behavior, but it may also be consistent with wave-like behavior such as from an attenuated LED.  The question at hand is whether these few coincident detections are merely accidental (i.e., the result of independent random events that occasionally just happen to coincide) or actually too few to be explained in this manner.  Specifically, let $D_1, D_2, D_3$ represent the events that a detection occurs on D1, D2, D3, respectively, irrespective of the others.  If $\vec{a}$ and $\vec{c}$ are the random Jones vectors describing the right and left output beams, respectively, of ENT and $\vec{b}$ is the vacuum mode entering BS from above, then these events are precisely
\begin{align}
D_1 &= \left\{ \tfrac{1}{\sqrt{2}} |a_H + b_H| > \gamma \mbox{ or } \tfrac{1}{\sqrt{2}} |a_V + b_V| > \gamma \right\} \\
D_2 &= \left\{ \tfrac{1}{\sqrt{2}} |a_H - b_H| > \gamma \mbox{ or } \tfrac{1}{\sqrt{2}} |a_V - b_V| > \gamma \right\} \\
D_3 &= \bigl\{ |c_H| > \gamma \mbox{ or } |c_V| > \gamma \bigr\} \; .
\end{align}

The events $D_1$ and $D_2$ will be statistically independent if (and only if) $\Pr[D_1 \cap D_2] = \Pr[D_1] \Pr[D_2]$.  Furthermore, since $D_1 \supseteq D_1 \cap \bar{D}_2$ and $D_2 \supseteq \bar{D}_1 \cap D_2$, we have the following inequality relating the probability of coincident detections to single detections:
\begin{equation}
\Pr[D_1 \cap D_2] \ge \Pr[D_1 \cap \bar{D}_2] \Pr[\bar{D}_1 \cap D_2] \; .
\label{eqn:indep}
\end{equation}
Any violation of this inequality would entail statistical dependence between $D_1$ and $D_2$ and, hence, coincident detections that are not ``merely accidental.''  (The converse is not true: A bright light source will make $D_1$ and $D_2$ strongly correlated.)

The probabilities in Eqn.\ (\ref{eqn:indep}) cannot be measured directly, as the number of trials cannot be determined.  Instead, we condition on detections at D3 since, for an ideal two-photon entangled state, a detection on D3 would notionally herald the presence of a single photon on the other side.  This leads to the anti-correlation coefficient
\begin{equation}
\alpha_{12|3} = \frac{\Pr[D_1 \cap D_2|D_3]}{\Pr[D_1 \cap \bar{D}_2|D_3] \Pr[\bar{D}_1 \cap D_2|D_3]} \; .
\end{equation}
Thus, $\alpha_{12|3} < 1$ implies that $D_1$ and $D_2$ are statistically dependent with respect to the conditional distribution $\Pr[\,\cdot\,|D_3]$.  In particular, this means that a detection at D1, say, implies that a detection at D2 is \emph{less} likely to occur, conditioned on $D_3$.

To estimate $\alpha_{12|3}$ experimentally, we use
\begin{equation}
\tilde{\alpha}_{12|3} = \frac{N_{123} (N_3 + N_{13} + N_{23} + N_{123})}{N_{13} \, N_{23}} \; ,
\end{equation}
where $N_{123}$ is the number of triple coincident counts on D1, D2, and D3, $N_{13}$ is the number of double coincident counts on D1 and D3, $N_{23}$ is the number of double coincident counts on D2 and D3, and $N_3$ is the number of single counts on just D3.  The sum in the numerator is therefore the total number of counts on D3, regardless of the other two detectors.

\begin{figure}[ht]
\centering
\scalebox{1.0}{\includegraphics[width=\columnwidth]{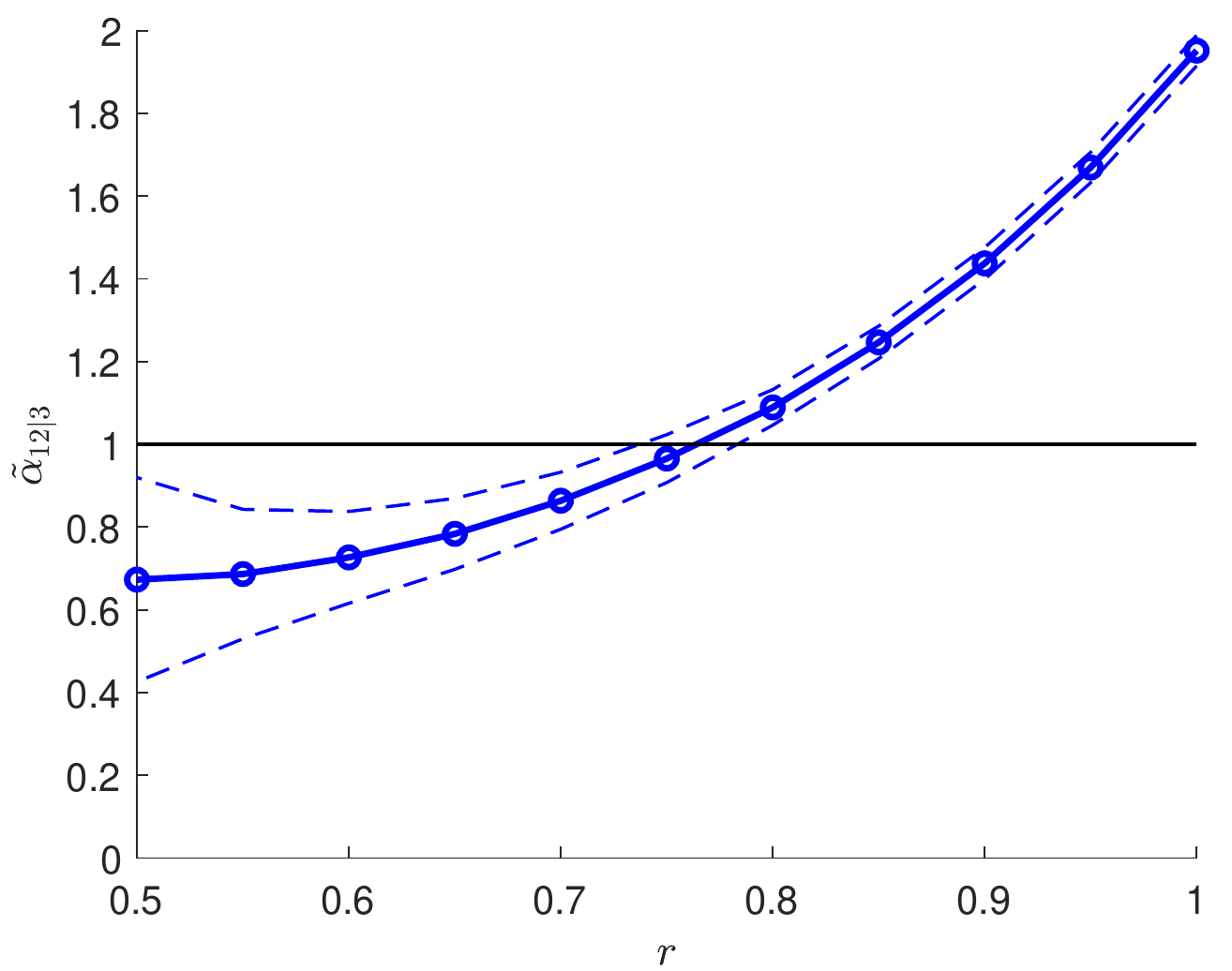}}
\caption{(Color online) Plot of $\tilde{\alpha}_{12|3}$ for various values of $r$.  The blue circles are the mean values over 100 1-s runs.  The dashed lines show the range of values that may be expected in approximately 95\% of the runs.  Values below the black horizontal line indicate anti-correlation behavior.}
\label{fig:Grangier_results}
\end{figure}

Table \ref{tbl:Grangier} shows the results of a single, one-second run in VQOL for a type-I entanglement source with $r = 0.7$.  For this particular run, we obtained $\tilde{\alpha}_{12|3} = 0.89$.  Over 100 such runs we obtained a mean value of $0.864 \pm 0.004$, which violates the inequality by 34 standard deviations.  We may therefore conclude that the number of coincidences at D1 and D2, conditioned on a detection at D3, is indeed much smaller that would be expected by mere chance.  Smaller values of $r$ may yield lower $\tilde{\alpha}_{12|3}$ values but will show greater variability.  In Fig.\ \ref{fig:Grangier_results} we show the results for a similar set of runs, with $r$ values ranging from $0.5$ to $1$.  Although much smaller values have been observed experimentally, the results obtained in VQOL can produce statistically significant violations.

\begin{table}[ht]
\centering
\begin{tabular}{ccc}
\hline
Event & Variable & Counts \\
\hline
$D_1 \cap \bar{D}_2 \cap \bar{D}_3$ & $N_1$ & 8401 \\
$\bar{D}_1 \cap D_2 \cap \bar{D}_3$ & $N_2$ & 8373 \\
$\bar{D}_1 \cap \bar{D}_2 \cap D_3$ & $N_3$ & 43\,407 \\
\hline
$D_1 \cap D_2 \cap \bar{D}_3$ & $N_{12}$ & 63 \\
$D_1 \cap \bar{D}_2 \cap D_3$ & $N_{13}$ & 6710 \\
$\bar{D}_1 \cap D_2 \cap D_3$ & $N_{23}$ & 6772 \\
\hline
$D_1 \cap D_2 \cap D_3$ & $N_{123}$ & 703
\end{tabular}
\caption{Table of measured counts obtained in VQOL for the experiment shown in Fig.\ \ref{fig:Grangier_setup} using a type-I entanglement source with a squeezing strength of $r = 0.7$.}
\label{tbl:Grangier}
\end{table}

\subsection{Bell Inequality Violations}

Bell's inequality is an important result used both for fundamental tests of quantum mechanics and to verify the security of entanglement-based quantum communication networks \cite{Ekert1991,Vazirani2014,Bierhorst2018}.  Violations of the inequality are used to certify that the system cannot have been spoofed by a classical counterfeit, but only when performed in a loophole-free manner.  Here, we show how VQOL may be used to perform this well-known experiment.

\begin{figure}[ht]
\centering
\scalebox{1.0}{\includegraphics[width=\columnwidth]{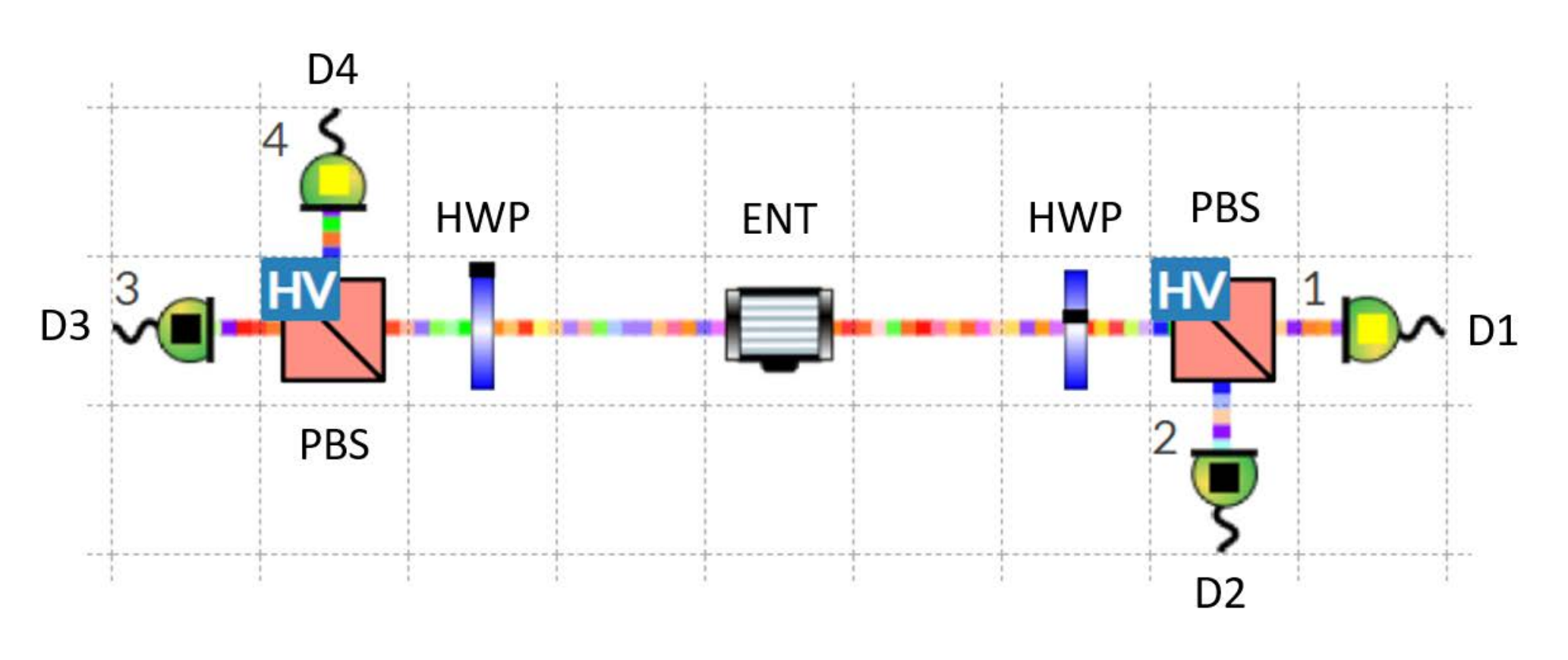}}
\caption{(Color online) Experimental setup for testing violations of the Bell-CHSH inequality.}
\label{fig:Bell_setup}
\end{figure}

We will use the Clauser, Horne, Shimony, and Holt (CHSH) variant of Bell's inequality commonly used in experiments \cite{Bell1964,CHSH1969}.  A typical experimental setup is shown in Fig.\ \ref{fig:Bell_setup}.  A type-II entanglement source is used to generate multi-mode squeezed light that, when restricted to a two-photon subspace, produces the singlet Bell state
\begin{equation}
\ket{\Psi^-} = \frac{\ket{HV} - \ket{VH}}{\sqrt{2}}
\end{equation}
given by Eqn.\ (\ref{eqn:Psi}) with $\varphi = 180^\circ$.  The left and right output beams are sent to Alice and Bob, respectively, who each control a half-wave plate, a polarizing beam splitter, and a pair of detectors.  HWP1, controlled by Alice, is set to a fast-axis angle of either $\alpha_1 = 0^\circ$ or $\alpha_2 = 22.5^\circ$, thereby measuring in either the $H/V$ or $D/A$ basis, respectively.  HWP2, controlled by Bob, is set to a fast-axis angle of either $\beta_1 = 11.25^\circ$ or $\beta_2 = 78.75^\circ$.  Once the experiment is run, Alice and Bob compare their results and compute a set of correlations between their measurement outcomes for the four different choices of measurement settings.

At each time step, there are 16 possible outcomes: no detections, four possible single detections, six possible double detections, four possible triple detections, and a quadruple detection.  As is commonly done in such experiments, we post-select on the four possible double-detection events in which exactly one of D1 or D2 clicks and exactly one of D3 or D4 clicks.  In such cases, if Alice gets a detection on, say, D3 she assigns the outcome $+1$, while if she gets a detection on D4 she assigns the outcome $-1$.  Bob makes similar assignments for D1 and D2, respectively.  The outcome of their joint measurement is the product of their individual measurements, since the observables are separable, so if Alice gets a single detection on D4 and Bob gets a single detection on D1, the outcome of their measurement is $(-1)(+1) = -1$.

In Table \ref{tbl:Bell} we have tabulated the full detector counts for each of the four possible measurement settings $(\alpha_i,\beta_j)$ for a 1-ms experiment with a default type-II entanglement source and default detector settings.  Of particular note are the counts corresponding to valid coincident detection events by Alice and Bob.  Let $N_{ij}^{13}$ denote the number of coincident counts for Alice on D3 and Bob on D1 when measurement setting $(\alpha_i,\beta_j)$ is used, with $N_{ij}^{14}$, $N_{ij}^{23}$, and $N_{ij}^{24}$ defined similarly.

\begin{table}[ht]
\begin{tabular}{ccccc}
\hline
Detectors & \; $\alpha_1,\beta_1$ & \; $\alpha_1,\beta_2$ & \; $\alpha_2,\beta_1$ & \; $\alpha_2,\beta_2$ \\
\hline
1 & 39 & 26 & 30 & 27 \\
2 & 29 & 32 & 29 & 33 \\
3 & 19 & 32 & 37 & 38 \\
4 & 25 & 32 & 27 & 27 \\
\hline
1,2 & 0 & 0 & 0 & 0 \\
1,3 & \bf 11 & \bf 11 & \bf 8 & \bf 32 \\
1,4 & \bf 34 & \bf 37 & \bf 45 & \bf 9 \\
2,3 & \bf 52 & \bf 44 & \bf 36 & \bf 5 \\
2,4 & \bf 5 & \bf 8 & \bf 7 & \bf 43 \\
3,4 & 0 & 0 & 0 & 0 \\
\hline
1,2,3 & 4 & 3 & 5 & 4 \\
1,2,4 & 2 & 1 & 2 & 0 \\
1,3,4 & 4 & 1 & 4 & 4 \\
2,3,4 & 6 & 4 & 2 & 1 \\
\hline
1,2,3,4 & 2 & 1 & 2 & 1
\end{tabular}
\caption{Full list of detector counts for all four measurement settings of the Bell inequality experiment.  Only the counts in boldface are used in the analysis.}
\label{tbl:Bell}
\end{table}

Following the standard practice of normalizing by the total number of coincident counts, the correlation $C_{ij}$ is defined by
\begin{equation}
C_{ij} = \frac{N_{ij}^{13} - N_{ij}^{14} - N_{ij}^{23} + N_{ij}^{24}}{N_{ij}^{13} + N_{ij}^{14} + N_{ij}^{23} + N_{ij}^{24}} \; .
\end{equation}
Using the values from Table \ref{tbl:Bell}, we find $C_{11} = -0.6863$, $C_{12} = -0.6200$, $C_{21} = -0.6875$, $C_{22} = +0.6854$, which yields a Bell statistic of
\begin{equation}
S = \bigl| C_{12} + C_{12} \bigr| + \bigl| C_{21} - C_{22} \bigr| = 2.679 \; .
\end{equation}
This is well above the upper bound of 2 given by the CHSH inequality, thus demonstrating a violation.

\begin{figure}[ht]
\centering
\scalebox{1.0}{\includegraphics[width=\columnwidth]{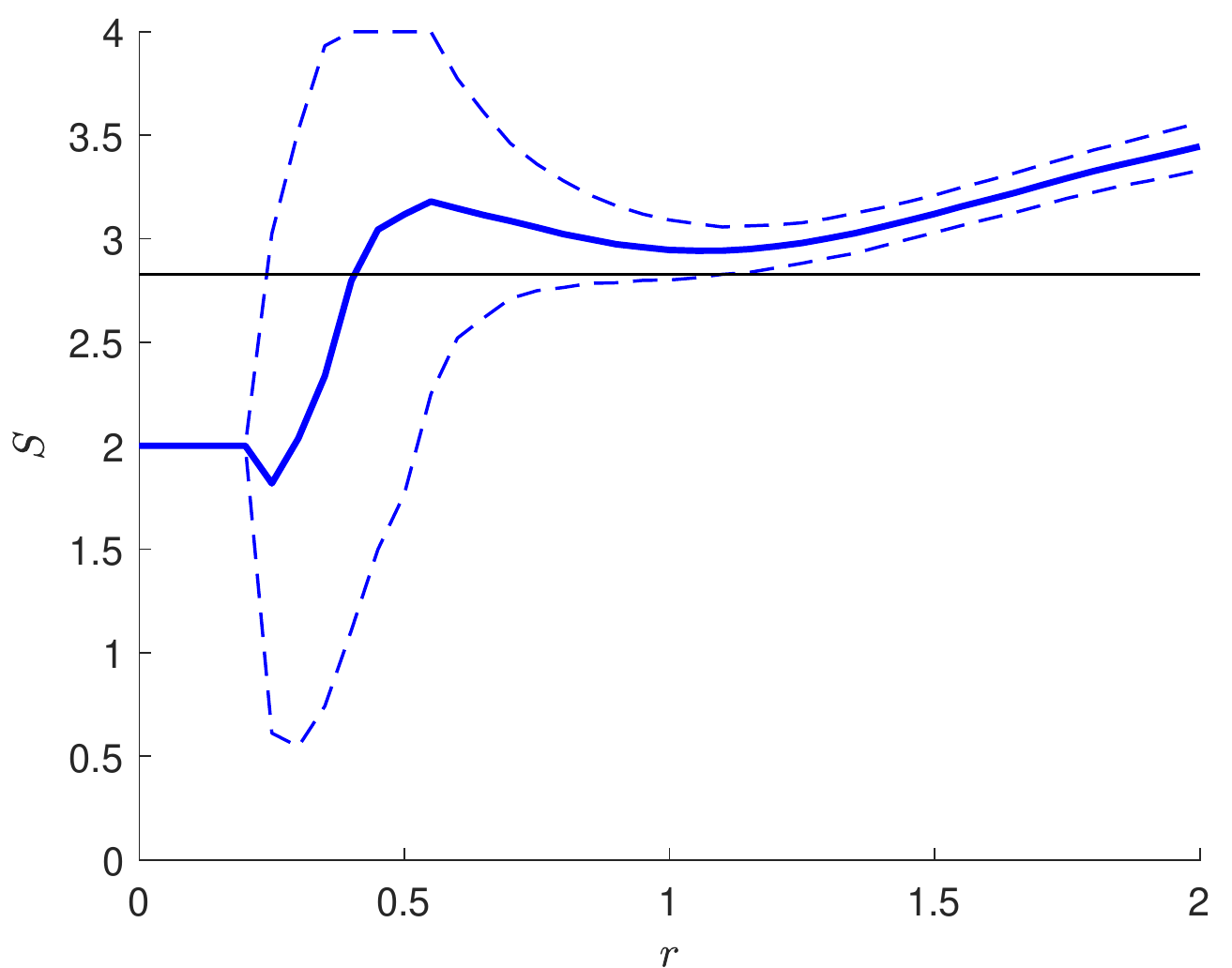}}
\caption{(Color online) Plot of the Bell statistic, $S$, as a function of the squeezing parameter, $r$.  The dashed lines show the range of values that may be expected in approximately 95\% of the 1-ms runs.  The horizontal line indicates the Tsirelson bound of $2\sqrt{2}$.}
\label{fig:Bell_results}
\end{figure}

 In Fig.\ \ref{fig:Bell_results} we have plotted the Bell statistic averaged over 1000 1-ms experiments as a function of $r$.  Above $r = 0.2$, we see that violations of the classical bound are common although with considerable variance.  We note that it is possible within VQOL to achieve values of the Bell statistic falling above the Tsirelson bound of $2\sqrt{2}$, which is the upper bound for all quantum states and measurement settings.  Although uncommon, this effect has been observed in real experiments \cite{Tasca2009}.   Such large values may be understood as arising from the post-selection of coincident detection events used to calculate $S$ and may therefore be construed as simply an artifact of the data analysis.

As VQOL is a fully classical model, it may seem suprising that a violation of the Bell inequality is even possible.  To understand this, we can estimate the coincidence detection efficiency from the data in Table \ref{tbl:Bell}.  For example, under measurement setting $(\alpha_1,\beta_1)$, when Bob makes a single detection on D1, Alice sees a corresponding single detection of either D3 or D4 about $(11+34)/(39+11+34) = 53.6\%$ of the time.  This is well below the efficiency bound of $2/(\sqrt{2}+1) \approx 82.8\%$ needed to close the detection loophole, which explains why a violation is possible.  The physical explanation of this is that VQOL violates the fair-sampling assumption: instances in which there is a valid coincident detection turn out to be statistically distinct from the entire ensemble of random realizations.  This provides a physical understanding for how the fair-sampling hypothesis may be violated.


\subsection{Hong-Ou-Mandel Effect}

The Hong-Ou-Mandel (HOM) effect is often used to illustrate two-photon interference.  As it is commonly described, two identical photons enter a 50/50 beam splitter, with one in each input port, and a single pair comes out of one output port or the other.  First demonstrated experimentally in 1987 \cite{HOM1987}, the HOM effect serves as a cornerstone for applications such as quantum teleportation, entanglement swapping, and quantum networks \cite{KLM2001}.

\begin{figure}[ht]
\centering
\scalebox{0.9}{\includegraphics[width=\columnwidth]{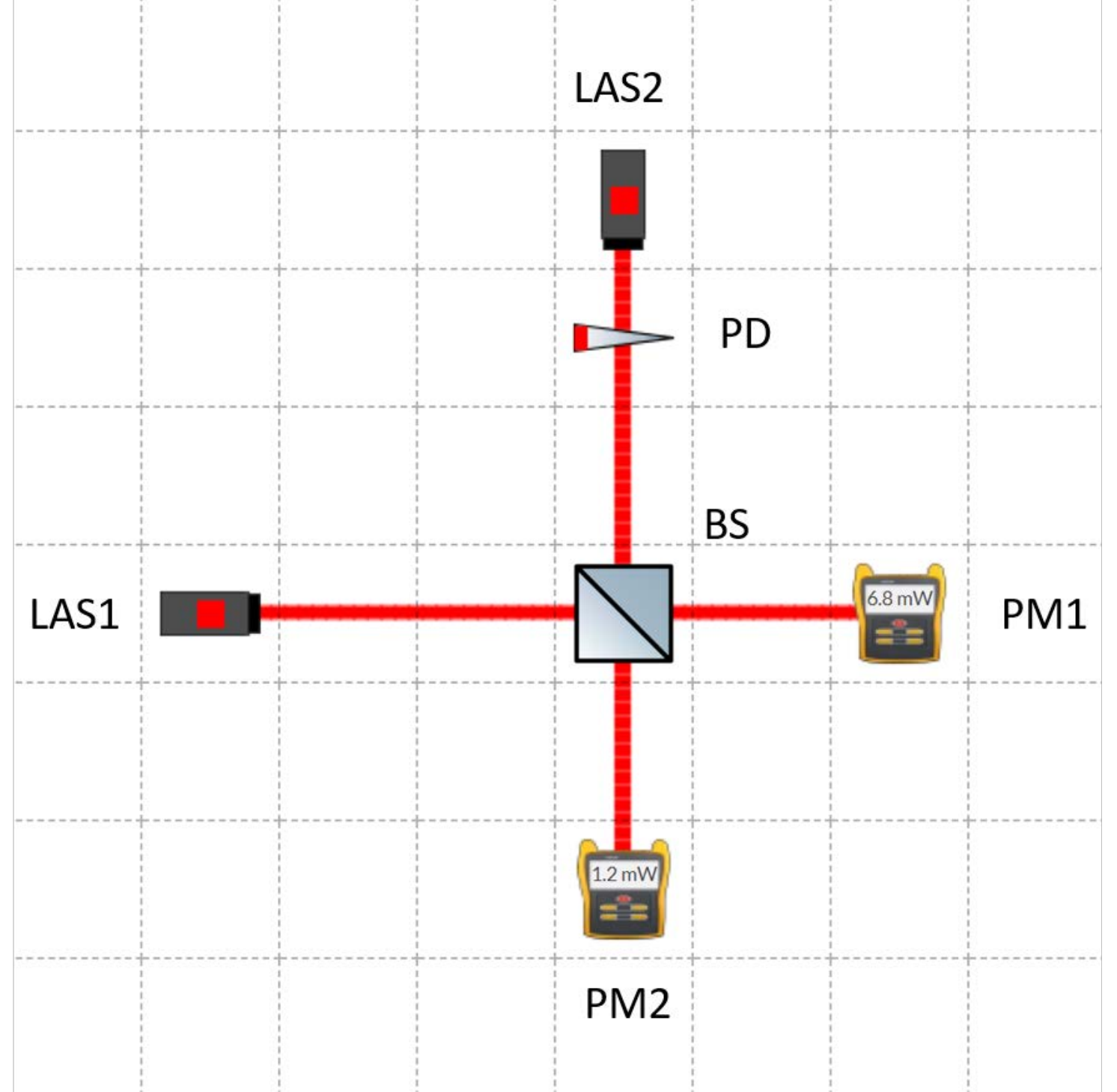}}
\caption{(Color online) Experimental setup for classical HOM experiment, including two lasers (LAS1 and LAS2), a phase delay (PD) set to $45^\circ$, a beam splitter (BS), and two power meters (PM1 and PM2).}
\label{fig:CHOM}
\end{figure}

We can begin to understand the HOM effect classically in VQOL by considering two lasers, a phase delay component, a beam splitter, and two power meters, as illustrated in Fig.\ \ref{fig:CHOM}.  The lasers produce coherent light with matching amplitudes and phases.  (In a real laser, these phases would be random.)  After the phase delay component, which artificially introduces a relative phase between the two spatial modes, the complex amplitudes are $\alpha_H = 10^5$ for the right ($\rightarrow$) mode and $\beta_H = \alpha_H e^{i\phi}$ for the down ($\downarrow$) mode.  (The vertical components are zero.)  After the beam splitter, these become
\begin{subequations}
\begin{align}
\alpha'_H &= \frac{\alpha_H}{\sqrt{2}}(1 + e^{i\phi}) \; , \\
\beta'_H &= \frac{\alpha_H}{\sqrt{2}}(1 - e^{i\phi}) \; .
\end{align}
\end{subequations}

When there is no relative phase delay (i.e., $\phi = 0^\circ$), the right mode has a power reading of 8.0 mW, while the down mode shows a power of 0.0 nW.  With $\phi = 90^\circ$, the two readings are identical, and with $\phi = 180^\circ$ they are reversed.  More generally, the power readings are $\mbox{(8.0 mW)} \cos^2(\phi/2)$ and (8.0 mW) $\sin^2(\phi/2)$ for a relative phase delay of $\phi$.  If the phase is random, for example by the introduction of a dephaser, then the power will fluctuate according to an arcsine distribution that is sharply peaked at the maximum and minimum values.  This crudely mimics the behavior of a pair of identical photons appearing at either one detector or the other.

The quantum version of the experiment may be used to perform Bell-state analysis: given one of four Bell states prepared using an entanglement source, determine which of the four was prepared.  The four Bells states are
\begin{align}
\ket{\Phi^{\pm}} &= \frac{\ket{HH}\pm\ket{VV}}{\sqrt{2}} \; , \\
\ket{\Psi^{\pm}} &= \frac{\ket{HV}\pm\ket{VH}}{\sqrt{2}} \; .
\end{align}
Using linear optical components, one can only determine three classes of Bell states: two particular Bell states and a set containing the other two \cite{Calsamiglia2001}.  Fortunately, this is sufficient for most applications.  In this example, we will prepare one of the four Bell states and determine whether it is $\ket{\Psi^-}$, $\ket{\Psi^+}$, or in the set $\{\ket{\Phi^+}, \ket{\Phi^-}\}$.

\begin{figure}[ht]
\centering
\scalebox{1.0}{\includegraphics[width=\columnwidth]{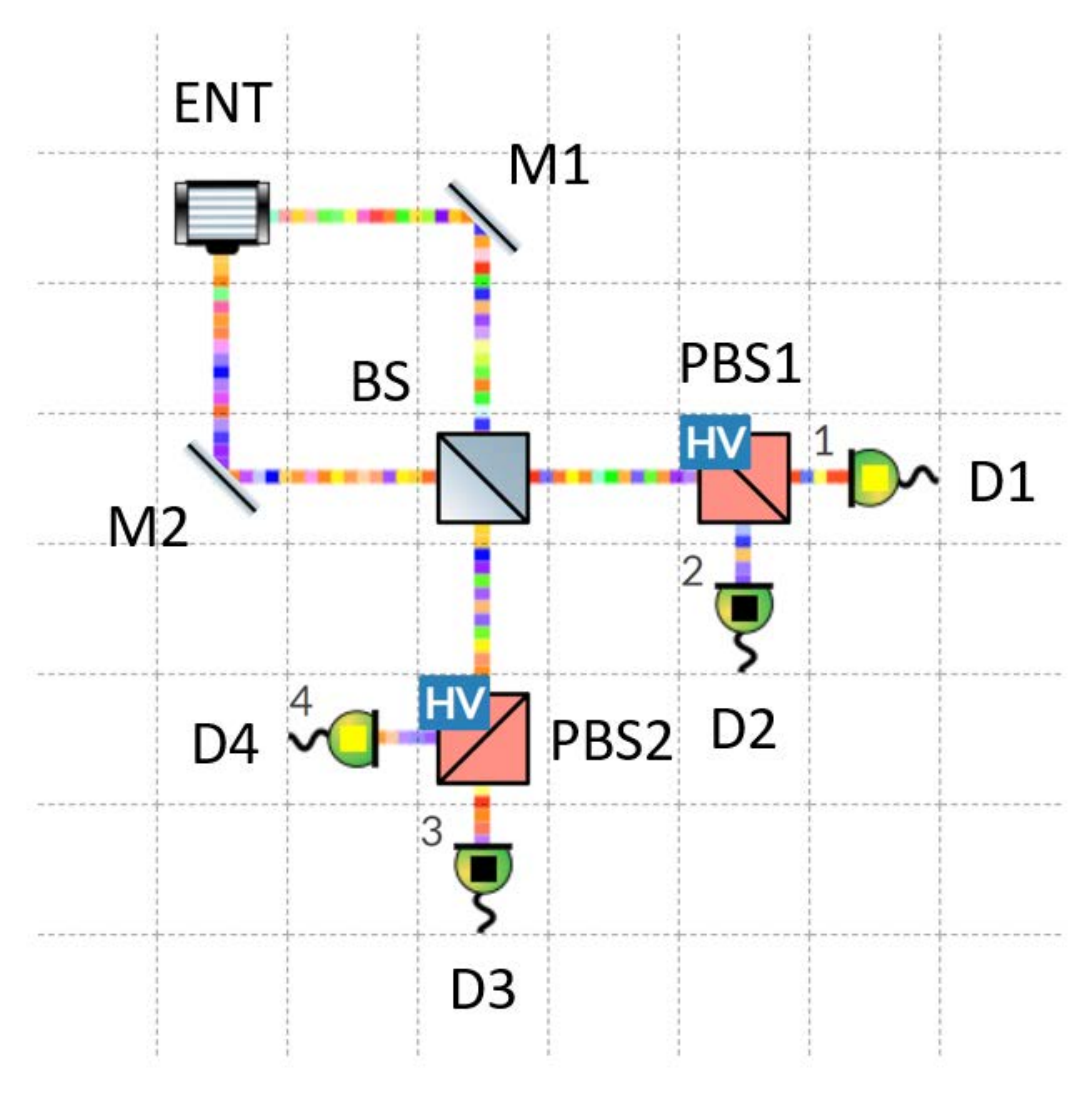}}
\caption{(Color online) Experimental setup for Bell state analysis showing confirmation of a $\ket{\Psi^-}$ state.}
\label{fig:BSA_setup}
\end{figure}

We have illustrated a typical setup for Bell-state analysis in Fig.\ \ref{fig:BSA_setup}.  An entanglement source (ENT) is configured to produce output beams in the right and down spatial modes.  A pair of mirrors redirect these beams to a beam splitter (BS) whose outputs are sent to a pair of $H/V$ polarizing beam splitters, PBS1 and PBS2, corresponding to the right and down spatial modes, respectively.  The $H$ and $V$ output beams of PBS1 (PBS2) are sent to detectors D1 and D2 (D3 and D4), respectively.

To prepare $\ket{\Psi^-}$ we use a type-II entanglement source with phase $\varphi = 180^\circ$, while for $\ket{\Psi^+}$ we use $\varphi = 0^\circ$.  Similarly, $\ket{\Phi^+}$ and $\ket{\Phi^-}$ are prepared using a type-I entanglement source with $\varphi = 0^\circ$ and $\varphi = 180^\circ$, respectively.  The results of a 1-ms experimental run with a default squeezing strength of $r = 1$ are shown in Table \ref{tbl:BSA}.  We see that when $\ket{\Psi^-}$ is prepared we tend to get coincident detections on either D1 and D4 \emph{or} D2 and D3.  When $\ket{\Psi^+}$ is prepared we tend to get coincident detections on either D1 and D2 \emph{or} D3 and D4.  Preparing $\ket{\Phi^{\pm}}$ we see only that there is a significant increase in the number of single-detection events.

\begin{table}[ht]
\begin{tabular}{ccccc}
\hline
Detectors & \quad $\ket{\Psi^-}$ & \quad $\ket{\Psi^+}$ & \quad $\ket{\Phi^+}$ & \quad $\ket{\Phi^-}$ \\
\hline
1 & 56 & 55 & \bf 104 & \bf 102 \\
2 & 59 & 64 & \bf 96 & \bf 100 \\
3 & 54 & 48 & \bf 91 & \bf 92 \\
4 & 49 & 49 & \bf 111 & \bf 101 \\
\hline
1,2 & 7 & \bf 75 & 18 & 19 \\
1,3 & 4 & 2 & 24 & 19 \\
1,4 & \bf 68 & 0 & 23 & 22 \\
2,3 & \bf 76 & 6 & 17 & 18 \\
2,4 & 8 & 9 & 21 & 19 \\
3,4 & 5 & \bf 69 & 20 & 19 \\
\hline
1,2,3 & 6 & 7 & 7 & 1 \\
1,2,4 & 10 & 5 & 7 & 5 \\
1,3,4 & 8 & 9 & 4 & 9 \\
2,3,4 & 10 & 6 & 6 & 6 \\
\hline
1,2,3,4 & 4 & 2 & 2 & 1
\end{tabular}
\caption{Table of detector counts for each of the four prepared Bell states.  The bold faced numbers highlight the  possible outcomes for an ideal quantum state.}
\label{tbl:BSA}
\end{table}

These results comport with the quantum predictions, as we will now show.  Suppose we prepare the $\ket{\Psi^-}$ state.  This may be written in terms of creation operators as
\begin{equation}
\ket{\Psi^-} = \frac{1}{\sqrt{2}} \left( \hat{a}_H^\dagger \hat{b}_V^\dagger - \hat{a}_V^\dagger \hat{b}_H^\dagger \right) \ket{\vec{0}} \; .
\end{equation}
where $\ket{\vec{0}}$ is the quantum vacuum state.  Applying the beam splitter transformation to the four creation operators, we obtain the new state
\begin{equation}
\ket{\Psi'^-} = \frac{1}{\sqrt{2}} \left( \hat{a}_V^\dagger \hat{b}_H^\dagger - \hat{a}_H^\dagger \hat{b}_V^\dagger \right) \ket{\vec{0}} \; ,
\end{equation}
for which we expect coincident detections on D2 and D3 (corresponding to the $\hat{a}_V^\dagger \hat{b}_H^\dagger$ term) \emph{or} D1 and D4 (corresponding to the $\hat{a}_H^\dagger \hat{b}_V^\dagger$ term), each with equal probability.  Similarly, the state $\ket{\Psi^+}$ transforms to
\begin{equation}
\ket{\Psi'^+} = \frac{1}{\sqrt{2}} \left( \hat{a}_H^\dagger \hat{a}_V^\dagger - \hat{b}_H^\dagger \hat{b}_V^\dagger \right) \ket{\vec{0}} \; ,
\end{equation}
for which we expect detections on D1 and D2 \emph{or} D3 and D4.  Finally, the states $\ket{\Phi^{\pm}}$ transform to
\begin{equation}
\ket{\Phi'^{\pm}} \propto \left[ (\hat{a}_H^\dagger)^2 \pm (\hat{a}_V^\dagger)^2 \pm (\hat{b}_H^\dagger)^2 \pm (\hat{b}_V^\dagger)^2 \right] \ket{\vec{0}} \; ,
\end{equation}
for which we would expect single detections on the four detectors, each with equal probability.

As with any experiment, perfect classification will be confounded by errors.  In a single shot, failures to achieve valid coincident detections will be common, and misclassifications will occasionally occur.  Performance can be improved using different squeezing strengths or dark count rates, with associated tradeoffs in sample size and other metrics of performance.  In the next section we will investigate such considerations in the context of a particular application.


\subsection{Quantum Teleportation}

The ability to perform a partial Bell-state analysis using a beam splitter allows for an optical implementation of the quantum teleportation protocol.  First proposed in 1993 \cite{Teleportation1993} and demonstrated experimentally in 1997 \cite{Bouwmeester1997}, quantum teleportation has been performed across 143 km of terrestrial atmosphere, between a ground observatory and an orbiting satellite 1400 km away, and through 600 m of sewer line \cite{Canary2012,Satellite2017,Danube2004}.  Practical applications include secure quantum networks and measurement-based quantum computing \cite{Repeaters1998,Pirandola2015,Gisin2007,Ladd2010}.

\begin{figure}[ht]
\centering
\scalebox{1.0}{\includegraphics[width=\columnwidth]{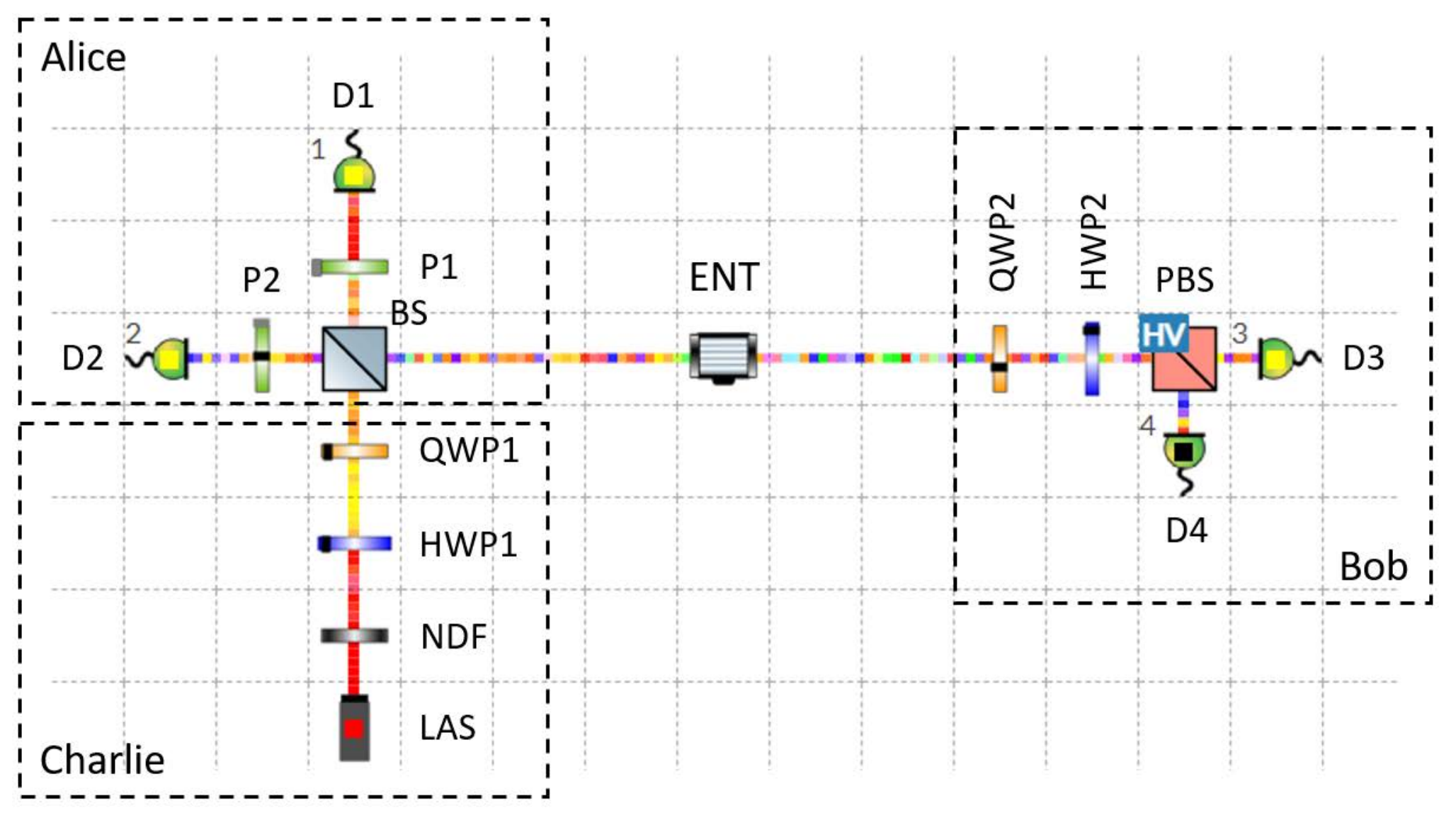}}
\caption{(Color online) Experimental setup for quantum teleportation experiment.}
\label{fig:teleportation_setup}
\end{figure}

A simple example of a quantum teleportation experiment can be described in terms of three notional agents: Alice, Bob, and Charlie.  As shown in Fig.\ \ref{fig:teleportation_setup}, Charlie prepares a single-qubit polarization state to be teleported using a laser (LAS), a neutral density filter (NDF), a half-wave plate (HWP1) with a fast-axis angle of $\theta$, and a quarter-wave plate (QWP1) with a fast-axis angle of $\phi$.  The state is teleported to Bob, who analyzes it with a quarter-wave plate (QWP2) set to a fast-axis angle of $\phi'$, a half-wave plate (HWP2) set to a fast-axis angle of $\theta'$, along with a polarizing beam splitter (PBS), and two detectors at the horizontal (D3) and vertical (D4) output ports.

To effect the teleportation, an entanglement source (ENT) configured to prepare the $\ket{\Psi^-}$ Bell state is placed between them.  The right beam travels to Bob, while the left beam travels to Alice, who performs the Bell-state analysis.  The left beam from ENT and the laser light from Charlie enter Alice's beam splitter (BS).  The beam exiting upward passes through a horizontal polarizer (P1) before hitting detector D1, while the left output beam passes through a vertical polarizer (P2) before hitting detector D2.  A valid teleportation will occur when D1 and D2 both click, but even this will not guarantee a valid detection by Bob, as we shall see.  Note that it is important to ensure that the distance from the sources to the detectors is equal so that proper coincidences are detected.  If this is not done, the teleportation scheme will not work!

We performed the experiment with all default parameters except for the optical density of the NDF, which we took to be $d = 9$.  This was done to better match the relatively bright entanglement source, as the default value of ten yields very few samples.  Four prepared states were considered: $\ket{H}$ ($\theta = 0^\circ, \phi = 0^\circ$), $\ket{D}$ ($\theta = 22.5^\circ, \phi = 45^\circ$), $\ket{R}$ ($\theta = 22.5^\circ, \phi = 0^\circ$), and a fourth, ``typical'' state $\ket{\psi}$ ($\theta = 22.5^\circ, \phi = 30^\circ$).  To verify the teleported state, Bob sets $\theta' = \theta$ and $\phi' = \phi - 90^\circ$, which inverts Charlie's preparation and, ideally, should produce a single detection on D3.

The results are summarized in Table \ref{tbl:teleport}, where we have displayed only the detections corresponding to valid teleportation events (i.e., those for which there are detections on both D1 and D2).  The four remaining possibilities are (1) Bob gets no detections, (2) Bob detects on D3 only, (3) Bob detects on D4 only, or (4) Bob detects on both D3 and D4.  The first and fourth possibilities are disregarded as invalid, leaving only two numbers to be considered: $N_H$, the number of counts on D1, D2, D3, and not D4, and $N_V$, the number of counts on D1, D2, D4, and not D3.  We define the fidelity of the teleported state as follows:
\begin{equation}
F = \frac{N_H}{N_H + N_V} \; .
\label{eqn:fidelity}
\end{equation}

For example, when Charlie prepares the state $\ket{H}$, Alice gets a total of 37 coincident detections on D1 and D2, each indicating a valid teleportation event.  Of these, 27 give a single detection on D3, indicating a successful teleportation of Charlie's state.  Of the remaining events, five go undetected by Bob, four are invalid double detections, and one indicates the wrong state was teleported.  With $N_H = 27$ and $N_V = 1$, we estimate a fidelity of 0.96 with a 95\% Clopper-Pearson confidence interval of $(0.82, 1.0]$.  For this experimental setup, horizontal and vertical polarizations give the best fidelity, while diagonal or circularly polarized states give the worst fidelity.  The state $\ket{\psi}$ is typical of most, with a fidelity of 0.76.

\begin{table}[ht]
\begin{tabular}{ccccc}
\hline
Detectors & \quad $\ket{H}$ \quad & \quad $\ket{D}$ \quad & \quad $\ket{R}$ \quad & \quad $\ket{\psi}$ \quad \\
\hline
1,2       & 5   & 92 & 99 & 72 \\
1,2,3    & \bf 27 & \bf 41  & \bf 36   & \bf 32 \\
1,2,4    & \bf 1   & \bf 16  & \bf 14  & \bf 10 \\
1,2,3,4 & 4   & 1   & 5   & 9 \\
\hline
Fidelity & 0.96 & 0.72 & 0.72 & 0.76
\end{tabular}
\caption{Table of detector counts for each of four different teleported states.  The bold faced numbers correspond to valid detection events.  The bottom row shows the fidelity computed from Eqn.\ (\ref{eqn:fidelity}) for each state.}
\label{tbl:teleport}
\end{table}

The classical fidelity limit is considered to be 2/3, which these examples all violate \cite{Massar1995}. Experimentally, teleportation fidelities as high as 0.90 have been observed, though values between 0.80 and 0.90 are more common \cite{Danube2004,Valivarthi2020}.  By tuning the parameters $d$ and $r$, it is possible to get higher fidelities than those reported in Table \ref{tbl:teleport}.  Taking larger values of $d$ and smaller values of $r$ tends to suppress higher photon number states and, hence, to give a closer approximation to the ideal single-photon state to be teleported as well as the ideal two-photon entangled Bell state used to effect the teleportation.  Post-selection on valid detection events, furthermore, serves to suppress the vacuum states.  Of course, values of $d$ that are too large or values of $r$ that are too small will result in states that are nearly identical to the vacuum and, hence, will give poor fidelity.  Furthermore, this regime will yield very few valid detections, resulting in either poor sample statistics or the necessity to perform longer experiments.

\begin{figure}[ht]
\centering
\scalebox{0.9}{\includegraphics[width=\columnwidth]{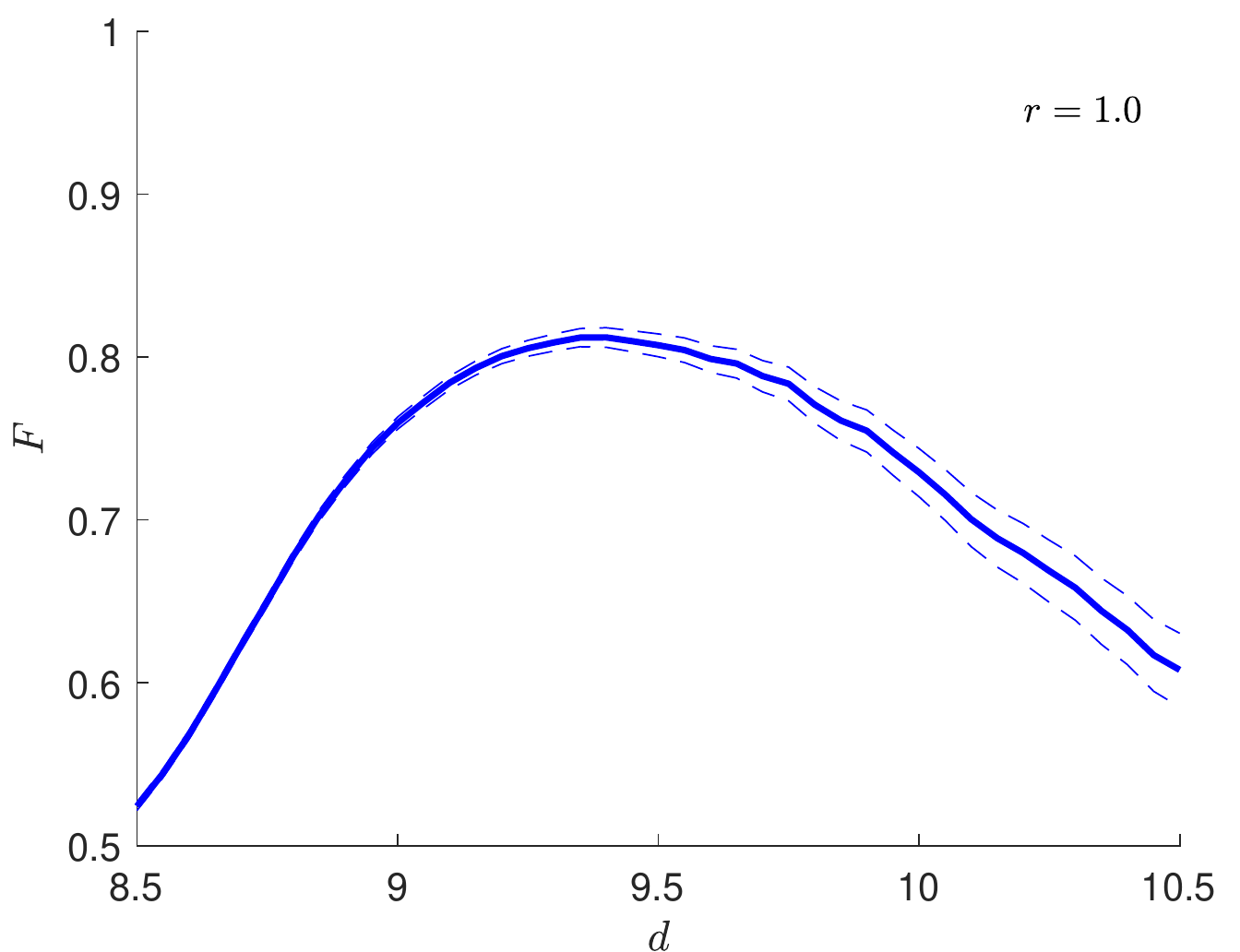}}
\centerline{$(a)$}
\scalebox{0.9}{\includegraphics[width=\columnwidth]{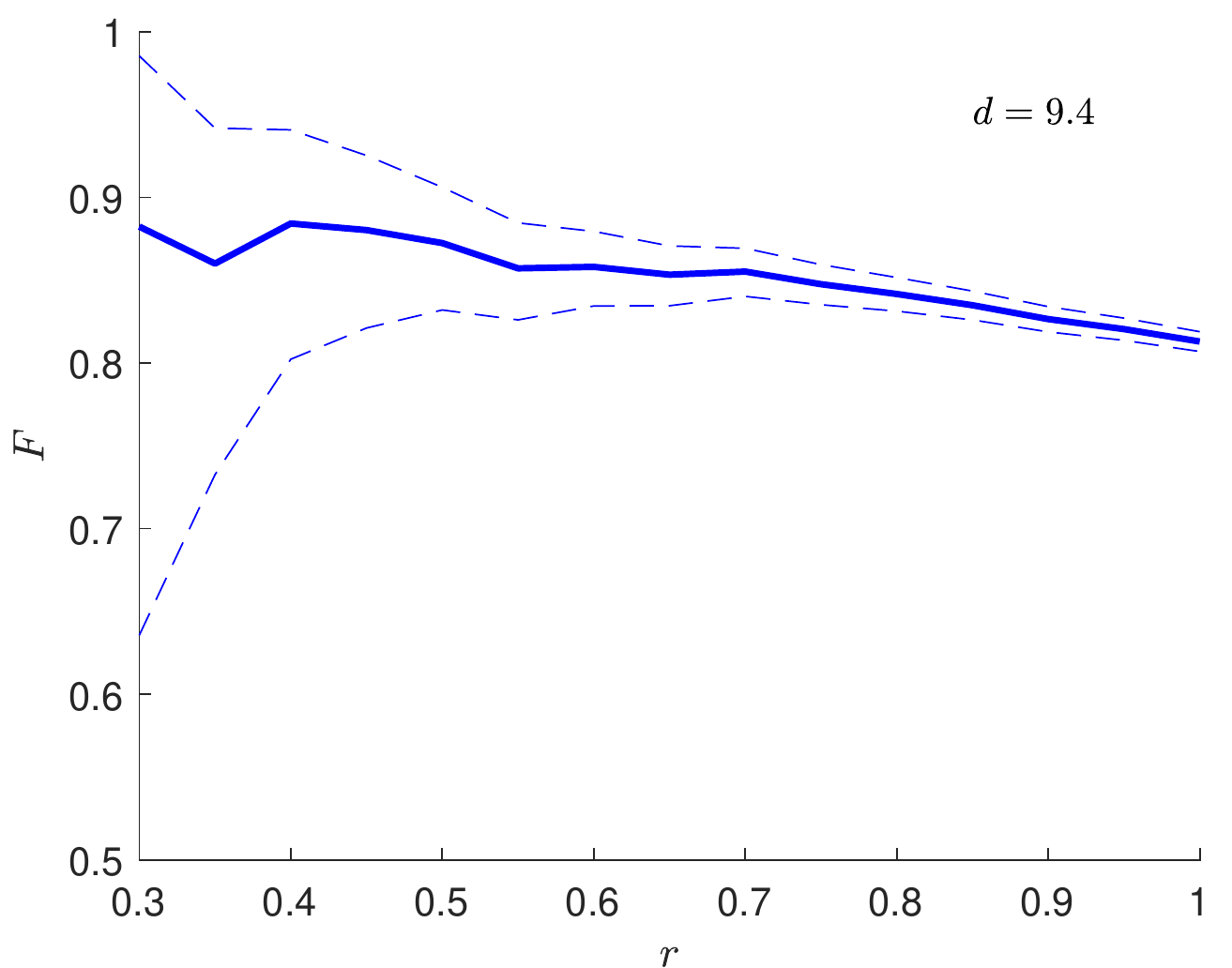}}
\centerline{$(b)$}
\caption{(Color online) Plots of fidelity $F$ versus optical density $d$ [top, subfigure (a)] and squeezing strength $r$ [bottom, subfigure (b)].  For subfigure (a), $r = 1.0$, while, for subfigure (b), $d = 9.4$.  The dashed lines are the upper and lower bounds for 95\% Clopper-Pearson confidence intervals.}
\label{fig:teleport_fidelity}
\end{figure}

The tradeoffs in these various parameter selections are summarized in Fig.\ \ref{fig:teleport_fidelity}.  These results are for the teleportation of the $\ket{\psi}$ state and for 1-s experiments.  Taking $r = 1$ to be fixed, we see that a peak fidelity of about 0.82 can be achieved with $d = 9.4$.  As $d$ increases the sample size decreases, resulting in greater uncertainty and statistical error.  Dependence on $r$ is rather more flat, with fidelity generally increasing with smaller values of $r$.  However, values below about $r = 0.5$ result small sample sizes and, hence, high variability.  Across both parameters, we find that taking $d = 9.4$ and $r = 0.65$ gives a near-optimal fidelity of about 0.87, with a 95\% confidence interval of $(0.86, 0.89)$ for a single, 1-s run.

Although this experiment was performed using a single laser and entanglement source, a similar experiment may be performed with two entanglement sources, one of which Charlie uses to prepare a heralded single-photon state by performing a measurement on the extra spatial mode.  (This, in effect, becomes an entanglement swapping experiment.)  Also, although Bob uses knowledge of Charlie's prepared state in order to validate the teleportation procedure, he could just as easily remain ignorant of the teleported state and perform quantum state tomography to infer the quantum state directly.  Finally, we note that a slight modification of this experiment can be used to realize a superdense coding experiment \cite{Mattle1996}.


\section{Discussion}
\label{sec:discussion}

We have illustrated several examples of experiments in VQOL exhibiting characteristically quantum phenomena, such as wave/particle duality and entanglement.  In most cases, agreement between VQOL and the corresponding ideal quantum predictions is only approximate.  This is due, in part, to our restriction to using only Gaussian states of light, such as coherent laser light or entangled squeezed light, which contain contributions from vacuum modes as well as higher photon-number states.  It is also due, in part, to our treatment of detectors as simple amplitude threshold crossing devices.  Departures from ideal behavior can occur when the light source is too weak, in which case vacuum contributions may dominate, or when the light source is too strong, in which case multi-photon components may dominate.  Similarly, different detection thresholds can provide varying levels of agreement.  Thresholds that are too high will result is small sample sizes or necessitate long experiment runs; those that are too low may suffer from an excess of dark counts.  The best agreement will depend on the experiment and the method of analysis but will usually be for intermediate values.  In the examples discussed we have tried to identify parameter regimes that provide a good compromise between these competing factors.

We have also highlighted the important role of data analysis is comparing experimental observations to theoretical predictions.  Quantum mechanics is fundamentally a theory of probabilities, and the meaning and estimation of these probabilities depends critically on the context and interpretation of the experiment.  Although in principle one has access to the true underlying number of random trials in VQOL, we have emphasized how data analysis in real experiments must be performed without this information.  In many cases, this leads to the necessity of performing post-selection on particular outcomes, such as the use of heralding to approximate single-photon sources or the use of coincident detections in performing quantum teleportation.  Even in the absence of post-selection, renormalization may be required to obtain probabilities from raw counts.  Furthermore, the removal of dark counts and ``accidental'' coincidences, both of which are common data analysis techniques, may also be required.  The seemingly mundane process of comparing raw observations to theoretical predictions can be a critical step in understanding and interpreting quantum theory.

The restriction to 1-$\mu$s time bins is an artificiality intended to represent the typical dead time of an avalanche photodiode as well as the coherence times of the light sources.  Common photodetection timing devices have a sampling resolution of about 1 ns, although timing jitter can be resolved below 10 ps \cite{Zadeh2020}  This restriction makes VQOL ill suited for experiments focusing on high-resolution timing with small, nonzero time delays.  Likewise, the subtle effects of detector afterpulsing and saturation are also ignored \cite{Eisaman2011}.  The choice of using a 1-$\mu$s coherence time is an artificiality born of expediency.  At the beginning of each time step a new and independent random realization of the ZPF is drawn.  At this time, any light incident upon a detector will either be such as to trigger a detection or not.  If a detection is triggered, the detector remains dead until the next time step.  Otherwise, the Jones vector remains unchanged, and hence cannot trigger a detection, until the next time step.  In other words, the detectors are such that they do not accumulate energy over time but, rather, trigger immediately when adequate conditions are realized at the beginning of a time step.  This is consistent with the rather short response times of avalanche photodiodes (typically less than a nanosecond) compared with their relatively long recovery times (typically around a microsecond).


\section{Conclusion}
\label{sec:conclusion}

The Virtual Quantum Optics Laboratory (VQOL) is a versatile simulation tool that can be used to design and execute a wide array of experiments in classical and quantum optics.  The simplicity of the user interface makes it a valuable classroom resource that can be used to introduce and explore quantum concepts.  The novelty and sophistication of the underlying models also make it an excellent tool for exploratory theoretical research or as a complement to actual quantum optics experiments.

We have illustrated several examples of experiments in VQOL exhibiting characteristically quantum phenomena, such as wave/particle duality and entanglement.  Agreement with quantum theory is only approximate and depends critically on both the choice of parameter settings and the manner in which the data are post-selected and analyzed.  In this regard, VQOL is very different from traditional simulators in that it is intended to more closely match observations than theory.  Despite its simplicity, we believe VQOL can be a powerful tool for learning and understanding key quantum phenomena.


\ifarxiv
\section*{Acknowledgments}
\else
\begin{acknowledgments}
\fi
This work was supported in part by Applied Research Laboratories, The University of Texas at Austin (ARL:UT), the National Science Foundation (NSF), under Grant No.\ 1842086, and the Office of Naval Research (ONR), under Grant Nos.\ N00014-18-1-2233 and N00014-17-1-2107.
\ifarxiv
\else
\end{acknowledgments}
\fi





\end{document}